\documentclass[12pt]{book}
\usepackage{amsmath,amssymb,amsfonts,ifthen,graphicx,epsfig,booktabs,caption}
\usepackage{hyperref,slashed,multirow,wrapfig}                 
\usepackage{axodraw4j}

\makeindex

\newcommand{\lsim}{\mbox{\raisebox{-.6ex}{~$\stackrel{<}{\sim}$~}}}
\newcommand{\gsim}{\mbox{\raisebox{-.6ex}{~$\stackrel{>}{\sim}$~}}} 

\newcommand{\be}{\begin{equation}}
\newcommand{\ee}{\end{equation}}
\newcommand{\bea}{\begin{eqnarray}}
\newcommand{\eea}{\end{eqnarray}}

\newcommand{\K}{K\"{a}hler\hspace{0.1cm}}
\newcommand{\OR}{O'Raifeartaigh\hspace{0.1cm}}
\newcommand{\g}{\widetilde{G}}
\newcommand{\Hub}{\mathrm{H}}
\newcommand{\mpl}{m_{P}}
\newcommand{\Mpl}{M_{P}}
\newcommand{\mg}{m_{\widetilde{G}}}

\newcommand{\neutr}{\widetilde{\chi}^0}
\newcommand{\mw}{m_{\widetilde{W}}}
\newcommand{\w}{\widetilde{W}}
\newcommand{\gl}{\tilde{g}}
\newcommand{\half}{\dfrac{1}{2}}
\newcommand{\mgl}{m_{\tilde{g}}}

\newcommand{\qcd}{\mathrm{SU(3)_C}}

\newcommand{\ew}{\mathrm{SU(2)_L}}
\newcommand{\pho}{\widetilde{\gamma}}
\newcommand{\zi}{\tilde{z}}
\newcommand{\chp}{\widetilde{W}^+}
\newcommand{\chm}{\widetilde{W}^-}
\newcommand{\charg}{\widetilde{\chi}}

\def\e{{\,\rm eV}}
\def\G{{\,\rm GeV}}
\def\M{{\,\rm MeV}}
\def\k{{\,\rm keV}}
\def\T{{\,\rm TeV}}

\title{Gravitino phenomenology and cosmological implications of supergravity}

\author{Andrea Ferrantelli\\ \ Draft date \today }

 \date{ }
\begin{document}

\begin{center}

HELSINKI INSTITUTE OF PHYSICS $\qquad$ INTERNAL REPORT SERIES

\vskip 10mm
HIP-2010-01

\vskip 30mm

\textbf{GRAVITINO PHENOMENOLOGY AND COSMOLOGICAL IMPLICATIONS OF SUPERGRAVITY}

\vskip 10mm

ANDREA FERRANTELLI

\vskip 5mm
{\it University of Helsinki and Helsinki Institute of Physics, \\P.O.Box 64 (Gustaf H\"allstr\"omin katu 2)\\
FIN-00014 University of Helsinki, Finland}


\vskip 20mm

\textit{ACADEMIC DISSERTATION}\\
\textit{To be presented, with permission of the Faculty of Science}\\
\textit{of the University of Helsinki, for public criticism}\\
\textit{in the Lecture Hall A129 of Chemicum, A. I. Virtasen aukio 1,}\\
\textit{on the 27th of January 2010 at 12 O'Clock.}

\vskip 20mm

Helsinki 2010

\end{center}

~

\vfill

\begin{center}

ISBN 978-952-10-6056-4 (paperback)\\
ISSN 1455-0563\\
ISBN 978-952-10-6057-1 (pdf)\\
http://ethesis.helsinki.fi\\
Yliopistopaino\\
Helsinki 2010

\end{center}


\newpage

\begin{flushright}
\textit{Dedicated to my mother Vilma,\\
She who taught me to think,\\
and to my father Alessandro,\\
He who taught me to dream.
}
\end{flushright}

\frontmatter

\pagestyle{plain}



\section*{Acknowledgements}

	Certainly, this work would not have been possible without the guidance of my supervisor Kari Enqvist. I owe my maturation as a physicist to him. He provided my work with freedom and independence, and stimulated new perspectives with his deep intuitions.
	I am deeply grateful to him for always believing in me, and for being a paternal figure even in the difficult moments, when my discomfort would easily disappear after his encouraging words.

	Emidio Gabrielli and Anca Tureanu traced the path of the young lad at the beginning of his journey. With great patience and devotion, they have shown me how to organise my work, both operatively and strategically. I feel extremely indebted to them. I am also honoured to acknowledge John McDonald as my collaborator. He has enhanced my knowledge of cosmology exponentially, with an impressive amount of notions and insights. He helped me to refine significantly my writing style as well. I am grateful also to Jukka Maalampi and Iiro Vilja, for reading this manuscript carefully and providing me with useful comments and insights. I also thank Antonio Riotto for honouring me with his presence as my opponent.

	Masud Chaichian, Katri Huitu, Kimmo Kainulainen, Kazunori Khori, Anupam Mazumdar and Santosh Kumar Rai are highly acknowledged for their interesting comments, which in several occasions revealed to be very important. I also thank my officemates and colleagues at the Helsinki Institute of Physics for providing always with an extremely relaxed, yet productive atmosphere.

	I felt my father Alessandro and my mother Vilma always very close, even though they were physically so far away. I am extremely grateful to them for never imposing anything to me, and for letting me fulfil my dreams in complete freedom.

	My friends and bandmates, both in Finland and all around Europe, have been so important during these years of hard work. My attitude towards enjoying life properly and my growth as a musician are very much due to their support.

\begin{flushright}

Andrea Ferrantelli\\
Helsinki, January 2010

\end{flushright}

\addcontentsline{toc}{section}{Preface}

\newpage

A. Ferrantelli: Gravitino phenomenology and cosmological implications of supergravity, University of Helsinki, 2010, 121 p., Helsinki Institute of Physics Internal Report Series, HIP-2010-01, ISBN 978-952-10-6056-4 (paperback), ISSN 1455-0563, ISBN 978-952-10-6057-1 (pdf).\\
INSPEC classification: A9880B, A9880D, A9530C.
Keywords: cosmology, early universe, inflation, supergravity, unitarity, gauge mediation, MSSM.

\section*{Abstract}

	In this thesis we consider the phenomenology of supergravity, and in particular the particle called "gravitino". We begin with an introductory part, where we discuss the theories of inflation, supersymmetry and supergravity. Gravitino production is then investigated into details, by considering the research papers here included.

	First we study the scattering of massive W bosons in the thermal bath of particles, during the period of reheating. We show that the process generates in the cross section non trivial contributions, which eventually lead to unitarity breaking above a certain scale. This happens because, in the annihilation diagram, the longitudinal degrees of freedom in the propagator of the gauge bosons disappear from the amplitude, by virtue of the supergravity vertex. Accordingly, the longitudinal polarizations of the on-shell W become strongly interacting in the high energy limit. By studying the process with both gauge and mass eigenstates, it is shown that the inclusion of diagrams with off-shell scalars of the MSSM does not cancel the divergences.

	Next, we approach cosmology more closely, and study the decay of a scalar field $S$ into gravitinos at the end of inflation. Once its mass is comparable to the Hubble rate, the field starts coherent oscillations about the minimum of its potential and decays pertubatively. We embed $S$ in a model of gauge mediation with metastable vacua, where the hidden sector is of the \OR type. First we discuss the dynamics of the field in the expanding background, then radiative corrections to the scalar potential $V(S)$ and to the \K potential are calculated. Constraints on the reheating temperature are accordingly obtained, by demanding that the gravitinos thus produced provide with the observed Dark Matter density. We modify consistently former results in the literature, and find that the gravitino number density and $T_R$ are extremely sensitive to the parameters of the model. This means that it is easy to account for gravitino Dark Matter with an arbitrarily low reheating temperature.

\addcontentsline{toc}{section}{Abstract}

\newpage

\section*{List of included papers}

The two research papers included in this thesis are:

	 \cite{WW} A.~Ferrantelli,
``Scattering of massive W bosons into gravitinos and tree unitarity in broken supergravity'',\\  
  JHEP {\bf 0901} (2009) 070  [arXiv:0712.2171 [hep-ph]].

   \cite{Infl} A. Ferrantelli, J. McDonald, "Cosmological evolution of scalar fields and gravitino dark matter in gauge mediation at low reheating temperatures",
  \\ e-Print: arXiv:0909.5108 [hep-ph] (\textit{Accepted for publication in JCAP}).

\section*{Author's contribution}

In \cite{Infl} the main idea came from John, who also sketched the overall \emph{modus operandi} and considered the dynamics of the $S$ field in Sect.\ref{dynamics}, which I have cross-checked. I calculated all the Coleman-Weinberg potentials and the supergravity corrections in Sect.\ref{sect:cw}. I have studied the constraints on the gravitino mass range and on the GMSB model in Sect.\ref{section:constr} (which all came from my own original idea). By comparison with the literature, I have recalculated the $S$ field decay rates, and found a correction factor in the coefficient for the cutoff Eq.(\ref{Lambdaconstr}). I have written the first draft of the paper, which was then substantially modified by both me and John.

	The study of the free streaming lenght of the gravitino, and the discussion in Sect.\ref{cosmo} about the term $\mg\Mpl^2$ in the superpotential, were suggested to me by Kazunori Khori during our frequent conversations at Lancaster University.

\addcontentsline{toc}{section}{List of included papers}





\tableofcontents

\addcontentsline{toc}{section}{Contents}

\mainmatter


\pagestyle{headings}

\chapter{Inflationary cosmology}\label{chapt:intro}

\begin{flushleft}
\textit{"Tyger! Tyger! Burning bright\\
In the forests of the night,\\
What immortal hand or eye\\
Could frame thy fearful symmetry?"\\}
- William Blake, "The Tiger"

\textit{"E quindi uscimmo a riveder le stelle."}\\
- Dante Alighieri; Inferno XXXIV, 139
\end{flushleft}

Cosmology, (from the Greek $\kappa o\sigma\mu o\lambda o\gamma\acute{\iota}\alpha$, namely -$\kappa\acute{o}\sigma\mu o\varsigma$, kosmos, "Universe" and -$\lambda o\gamma\acute{\iota}\alpha$, -logia, "study") has extremely ancient origins, which can be dated back to the prehistoric era. Though the word Cosmology is recent (first used in 1730 in Christian Wolff's Cosmologia Generalis), the study of the Universe has a long and complex history. It involves science, religion, philosophy and esoterism.

	Thousands years ago the Babylonians were skilled in Astronomy, building up a tradition that was further developed by the Ancient Greeks. The latter were the first to build a cosmological model within which to interpret the motions of planets and stars. In the fourth century BC, they believed that the stars were fixed on a celestial sphere which rotated about the spherical Earth every 24 hours. The planets, the Sun and the Moon, would have then moved in the ether between the Earth and the stars.

	This model was refined, and in the second century AD it lead to Ptolemy's system. It was based on religious and philosophical fundaments, as the circular motion of stars and planets was justified by the belief that perfect motion should be in circles. The earth was put at the centre of the Universe, by choosing a reference frame which turned out not to be practical. Four geometric devices, the deferent, epicycle, eccentric and equant had to be introduced to describe the planetary motions correctly. The resulting system was extremely complicated.

	However, Ptolemy's model was successful in describing the celestial motions with great accuracy. Much later, in the 16th century, Copernicus assumed that the Earth was not at the centre of the Universe, but rotated and moved in a circular orbit about the Sun. Unfortunately, his model could not match the accuracy of Ptholemy's system, which was still regarded as the standard cosmological model.

	Nevertheless, after a few decades Galileo discovered that there are moons orbiting the planet Jupiter. This was a first observational evidence against the Earth-centred system. The final strike of genius came from Johannes Kepler. Kepler postulated that the planets moved in ellipses, not perfect circles, about the Sun. The Heliocentric model soon replaced Ptolemy's system, but it still was an entirely phenomenological model. The theoretical justification for Kepler's laws came after about one century, when Newton showed that elliptical motion could be explained by his law for the gravitational force.

	After these fundamental discoveries, the interest in the solar system was then pushed forward to larger scales. The philosopher Immanuel Kant, among the others, proposed that the Milky Way was just one "island Universe" or galaxy, and that beyond it must be other galaxies.

	Since the formulation of the theory of General Relativity by Einstein in 1915, it became possible to discuss the evolution of the Universe from a fully consistent physical viewpoint. 
The Russian mathematician and meteorologist Alexander Friedman found in 1917 that the Einstein equations could describe an expanding Universe. This implied an instantaneous birth of the Universe, about ten thousand million years ago in the past, with creation of all the matter at just one instant. The British astronomer Fred Hoyle called it the "Big Bang".

	In 1929, Edwin Hubble discovered through observations of the redshift of galaxies \cite{Hubble} that the Universe is expanding. George Gamow and his collaborators found in 1946 that nucleosynthesis requires a hot and extremely dense state at the origin of the Universe. This statement is now regarded as the standard Big Bang scenario.
	Remarkably, they also predicted that the actual Universe is filled with a background radiation of the black body-type. This was confirmed by the experimental discovery of the Cosmic Microwave Background Radiation (CMBR or simply CMB) by Penzias and Wilson in 1965 \cite{Penzias}.

	In the standard Big Bang scenario, the state of the Universe (either radiation- or matter-dominated) consists of a decelerated expansion, due to the negative second derivative of the scale factor. However, several cosmological problems such as flatness and horizon problem plague this description. The theory of \emph{inflation} was originally proposed in 1980 by Alan Guth in \cite{Guth} and by Katsuhiko Sato in \cite{Sato}, as a mechanism for resolving these problems.
	Contemporary with Guth, Alexei Starobinsky argued that quantum corrections to gravity would replace the initial singularity of the Universe with an exponentially expanding state \cite{Staro}. Einhorn and Sato published a model similar to Guth's and showed that it would resolve the puzzle of the magnetic monopole abundance in Grand Unified Theories \cite{ES}. Like Guth, they concluded that such a model (which is now called old inflation) not only required fine tuning of the cosmological constant, but also would very likely lead to a much too granular Universe, namely to large density variations resulting from bubble wall collisions.

	The bubble collision problem was solved in 1982 by Andrei Linde in \cite{Linde1} and independently by Andreas Albrecht and Paul Steinhardt in \cite{AS}, in a revised version that is named new inflation.

	The basic idea is slow-roll inflation, where instead of tunneling out of a false vacuum state, inflation occurred by a scalar field rolling down a potential energy hill. When the field rolls very slowly compared to the expansion of the Universe, inflation occurs and eventually ends when the potential becomes steeper. However, there was still a problem of fine tuning also in this model, since the scalar field would need to spend enough time in the false vacuum to provide with a sufficient amount of inflation. Eventually, in 1983 Linde \cite{Lindechao} introduced a variant version of the slow-roll in which initial conditions of the scalar fields are chaotic.
	This corresponds to chaotic inflation. It predicts that our homogeneous and isotropic Universe may be produced in the regions where inflation occurs sufficiently. The basic difference between chaotic and old (or new) inflation, is that the latter requires a Universe in a state of thermal equilibrium from the beginning, whereas the former does not necessarily need this assumption. Moreover, chaotic inflation powerfully solves also the problem of initial conditions, since it can start at Planckian densities.

	Since the definition of cosmic inflation, many different models have been constructed \cite{LL}. In particular, the trend is now to incorporate particle physics in the model, in order to create a fully consistent physical theory \cite{Linde,LR}. In fact, the detailed particle physics mechanism responsible for inflation still needs to be understood.
	The issues of preheating, reheating, the generation of density perturbations and the creation of structures, involve a whole class of hard-core physics, from supergravity to string theory. Also the origin and the nature of the inflaton field are still unclear and problematic, as in chaotic inflation it is just a scalar particle with no defined structure. Attempts to embed this field within a particle physics theory are object of intense discussion nowadays. For instance, there are models where the inflaton is defined by linear combinations of flat directions in the MSSM \cite{Anupam}.

	The inflationary paradigm not only solves the flatness and horizon problems, but it also generates density perturbations which then concur to the formation of cosmological structures in the Universe. This is very important for observations, as it was first noticed in the early 80s by Stephen Hawking \cite{Hawking}, Starobinsky \cite{Staro2} and several others \cite{Density}.
	Nearly scale-invariant spectra of cosmological perturbations are generated by quantum fluctuations of the inflaton, whose scales are frozen during the expansion. Long after the ending of inflation, the scales re-enter the Hubble radius. Accordingly, the perturbations which originated during inflation can create the structures of the Universe which correspond to a large scale. Such scale-invariant spectra were observed by the COBE satellite in 1992 \cite{COBE}, and the recent WMAP5 data \cite{WMAP5} confirm this interpretation.

	In this thesis we cover some aspects of this fascinating, yet problematic subject. We shall attempt to describe a number of cosmological issues from the point of view of particle physics. In the introductory part we provide with some basics which are relevant to our discussion, which then focuses on the research papers here included.

	In Chapter 1 we discuss the theory of cosmological inflation, that provides the background for the physical mechanisms which are studied in this work. After considering the standard Big Bang cosmology, we review old and new inflation, chaotic inflation and reheating.

	Chapter 2 introduces supersymmetry (SUSY) and supergravity (SUGRA). We discuss the spontaneous breaking of global and local SUSY, and the gravitino as the gauge field of supergravity. We obtain from the general SUGRA Lagrangian the interactions of the gravitino.

	Chapter 3 refers to the paper \cite{WW}. We investigate the scattering of W bosons into gravitino and gaugino in the broken phase, by using both gauge and mass eigenstates. Differently from what is obtained for unbroken gauge symmetry, we find in the scattering amplitudes new structures, which can lead to violation of unitarity above a certain scale. We show that the longitudinal polarizations of the on-shell W become strongly interacting at high energies, and show that the inclusion of diagrams with off-shell scalars of the MSSM does not cancel the divergences.

	Chapter 4 considers the production of Dark Matter gravitinos via the decay of a scalar field at the end of inflation. We discuss the dynamics of the field and the model of gauge mediation in which it is embedded. It is shown that the gravitino density is extremely sensitive to the parameters of the hidden sector. For the case of an O'Raifeartaigh hidden sector, the observed Dark Matter density can be explained by gravitinos even for low reheating temperatures $T_R \lsim 10 \G$. This chapter examines the contents and results of the paper \cite{Infl}.

In the Appendix we summarise our conventions for spinors, propagators and spin sums. We provide also with a list of the Feynman rules for the MSSM and for supergravity which are relevant for this thesis.

\newpage

\section{The history of the Universe in a nutshell}
\begin{center}
\begin{figure}[ht]
\begin{center}
	\includegraphics[width=0.7\textwidth]{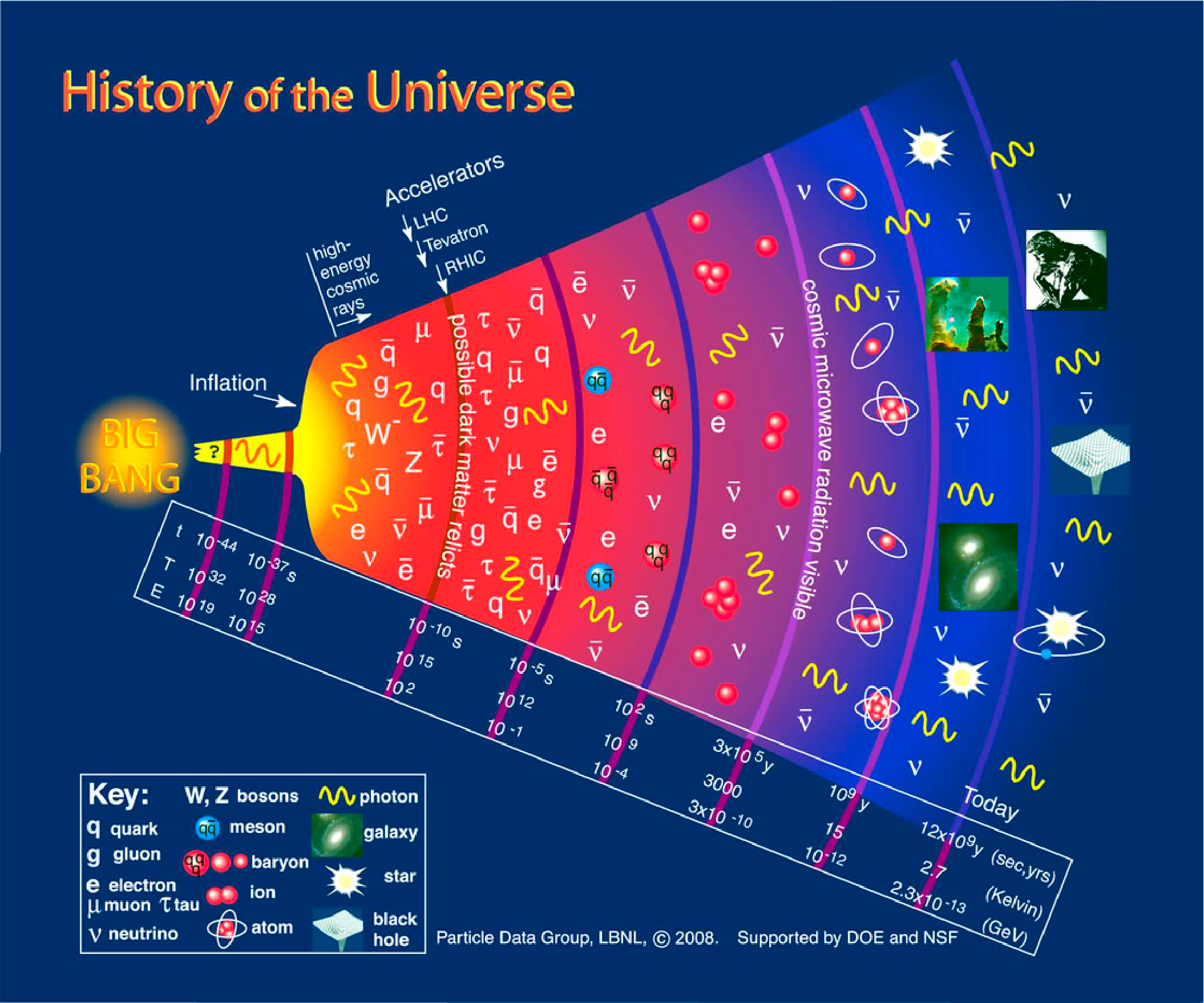}
\end{center}
	\caption{Schematic history of the Universe.}
	\label{fig:history}
\end{figure}
\end{center}

	Since the beginning of time (provided it did have a beginning), the Universe has passed through several epochs, which are summarised in Fig \ref{fig:history} and listed chronologically below. Some of them are described by established physics that is verified by experimental data, others by hypotheses and conjectures based on physics beyond the Standard Model. Some are still completely obscure to our understanding \cite{Dolgovcosmo}.

\begin{itemize}
	
\item 
	Nobody actually knows anything certain about the beginning. Was it a period that can be described by quantum gravity? What about time, did it exist or not? These are still open questions, which belong to the so-called pre-inflationary cosmology.

\item 
	Inflation, namely a period of exponential expansion of the Universe. This model is nowadays widely accepted by the scientific community, since it cures the problems of the standard big bang cosmology in agreement with the observational data. The quantum fluctuations during inflation were indeed imprinted on the metric. They can be observed as CBR fluctuations in the power spectrum, namely as deviations from homogeneity and isotropy in the matter distribution.

\item 
	End of inflation, particle production. Once that the exponential expansion has come to an end, the inflaton field releases the energy of the Universe. Photons and other elementary particles are produced via several mechanisms, which are addressed in the next sections. 

\item 
	Baryogenesis. The confinement of quarks takes place and baryons are formed. The excess of matter with respect to antimatter is created.

\item 
	The Universe reaches thermal equilibrium and it is cooled down adiabatically. Several phase transitions take place in this epoch: breaking of grand unified theories (GUT), electroweak (EW) symmetry, supersymmetry (SUSY), phase transition from free quark-gluon phase to confinement in quantum chromodynamics (QCD) and so on. Possible formation of topological defects.

\item 	Decoupling of neutrinos from the cosmological plasma, when the Universe is 1 sec old and $T\approx 1 \M$.

\item 	Big Bang Nucleosynthesis (BBN) is the epoch when the light elements $^2 H$, $^3 He$, $^4 He$, $^7 Li$ were formed. It occurs in the time interval from $1 \, s$ to $200\, s$, and $T\approx 1-0.07\M$. It is probably the only mechanism in the early Universe for which we have a very good agreement of theory with data.

\item 
	Structure formation starts when the cosmological matter turns from relativistic to non-relativistic. This corresponds to the epoch of equivalence between radiation and matter domination. It takes place at $T\sim\e$, with redshift $z_{eq}\approx 10^4$.

\item 
	Decoupling from matter of the Cosmic Microwave Background (CMB), which after that has propagated almost freely in the Universe. This is called the period of hydrogen recombination, at $T\approx 0.2\e$ or $z_{rec}\approx10^3$. Baryons began to fall into already evolved seeds of structures generated by Dark Matter.

\item 
	A variegated cosmological jungle of stars, planets, black holes, worm holes and other cosmological objects is formed. The present time corresponds to $t\approx 12-15\times 10^9\,\rm{yr}$ and $T=2.3\times 10^{-13}\G=2.7\,\rm K$. The redshift $z$ is defined by $z+1\equiv a(t_0)/a(t)$, where $a(t_0)$ is the scale factor of the universe at the actual time $t_0$. Accordingly, $z_0=z(t_0)=0$.

\end{itemize}

\newpage

\section{The standard Big Bang cosmology}\label{sect:friedman}

	Let us assume the cosmological principle \cite{LL}: at large scales, the Universe is homogeneous and isotropic. Technically speaking, this is achieved by using the Friedman-Robertson-Walker (FRW) metric
\be
ds^2=g_{\mu\nu}dx^\mu dx^\nu=-dt^2+a^2(t)\left[\dfrac{dr^2}{1-kr^2}+r^2(d\theta^2+\sin^2\theta d\phi^2) \right],
\label{FRW}
\ee
where $a(t)$ is the scale factor, which measures the radius of the Universe according to the flow of the cosmic time $t$. The constant $k$ is the cosmological curvature, which can take the values 0 (flat Universe), 1 (closed of spherical Universe) or -1 (open of hyperbolic Universe).

	Besides the geometrical structure of the Universe, the matter content is primarily important for its evolution. This is defined by the equation of state between the energy density $\rho(t)$ and the pressure $p(t)$. Namely, for a radiation-dominated Universe we have $p=\rho/3$, whereas if it is matter (or dust) dominated, $p=0$. Given the metric and the matter content, the dynamics of the Universe is determined by solutions of the Einstein equations in General Relativity. With a cosmological constant\footnote{The cosmological constant $\Lambda$ was originally postulated by Einstein to achieve a stationary Universe. Even though it is now clear that our Universe is expanding (Einstein called it the "biggest blunder" of his life), since $\Lambda$ is directly related to dark energy, this was clearly another strike of genius of his.}, they read as
\be
G_{\mu\nu}+\Lambda g_{\mu\nu}\equiv R_{\mu\nu}-\dfrac{1}{2}Rg_{\mu\nu}+\Lambda g_{\mu\nu}=8\pi GT_{\mu\nu}\,.
\label{Einstein}
\ee
Here $R_{\mu\nu}$ is the Ricci tensor, $R$ is the Ricci (curvature) scalar, $T_{\mu\nu}$ is the energy momentum tensor and $G_{\mu\nu}$ is the Einstein tensor. $G$ is Newton's gravitational constant and it is related to the Planck mass $\mpl=1.2211\times 10^{19}\G$ through $\mpl=(\hbar c/G)^{1/2}$ (or either $\mpl=(1/G)^{1/2}$ in natural units). Throughout this thesis, we will use extensively also the reduced Planck mass, defined as $\Mpl\equiv\mpl/\sqrt{8\pi}=2.43\times 10^{18}\G$.

	Remarkably, the Einstein equations (\ref{Einstein}) summarise the main challenges of contemporary cosmology. The left hand side contains the geometric structure of the Universe, which can be modified by either using the cosmological constant or by properly modifying the Einstein tensor. The latter approach is discussed in theories of modified gravity \cite{Francaviglia}, which have been recently proposed as an alternative to dark energy. On the other hand, the energy momentum tensor $T_{\mu\nu}$ encodes the matter constituents, among which Dark Matter. Accordingly, the topics of this thesis deal with the right hand side of (\ref{Einstein}).

	By using the FRW metric (\ref{FRW}), and ignoring the cosmological constant, the Einstein equations become the two \emph{Friedman equations}. From the 00 component of the Eistein tensor we obtain
\be
\Hub^2=\dfrac{8\pi}{3\mpl^2}\rho-\dfrac{k}{a^2}\,,
\label{Fried}
\ee
where the Hubble expansion rate is defined as $\Hub\equiv\dot{a}/a$. The components $G^{11}=G^{22}=G^{33}$ give the dynamics of the scale factor $a(t)$:
\be
\dfrac{\ddot{a}}{a}=-\dfrac{4\pi}{3\mpl^2}(\rho+3p)\,.
\label{acc}
\ee
In the above, we assume the perfect fluid form for the energy-momentum tensor
\be
T^{\mu\nu}=(\rho+p)u^{\mu}u^{\nu}+pg^{\mu\nu}\,.
\ee
By combining (\ref{Fried}) and (\ref{acc}), one finds the continuity equation,
\be
\dot{\rho}+3\Hub(\rho+p)=0\,,\label{Fried2}
\ee
which is the General Relativity version of energy-momentum conservation. The Friedman equation (\ref{Fried}) can also be rewritten as
\be
\Omega-1=\dfrac{k}{a^2\Hub^2}\,, \qquad \mathrm{where} \qquad \Omega\equiv\frac{\rho}{\rho_c}=\dfrac{8\pi}{3\Hub^2\mpl^2}\rho\,.
\label{Omega}
\ee
The dimensionless parameter $\Omega$ measures the amount of deviation of the actual energy density $\rho$ of the Universe from the critical density $\rho_c=3\Hub^2\mpl^2/8\pi$, that corresponds to a spatially flat geometry. In fact, if $k=0$ then $\rho=\rho_c$ and $\Omega=1$. In this case, the solutions of the Friedman equations are
\be
\qquad a\propto t^{1/2}\,,\qquad \rho\propto a^{-4}\,,
\ee
for radiation domination and
\be
\qquad a\propto t^{2/3}\,,\qquad \rho\propto a^{-3}\,,
\ee
for matter domination. In both cases, this corresponds to an expanding Universe with a decelerating expansion, since the second time derivative of the scale factor is negative ($\ddot{a}<0$) by virtue of Eq.(\ref{acc}).

\subsection{Problems of the standard scenario}

	The cosmological models (also named "Friedman models") above defined are consistent from a theoretical viewpoint, but they are plagued by a number of phenomenological problems \cite{Linde}.

	Let us discuss them in this section. The \textbf{flatness problem} originates from Eq.(\ref{Omega}). Since the term $a^2\Hub^2$ always decreases, the matter density $\Omega$ shifts away from 1 with the expansion of the Universe. In contrast, observations seem to suggest that nowadays $\Omega \approx \mathcal{O}(1)$ \cite{WMAP5}. Accordingly, it must have been very close to 1 in the past. It can be shown that in order to have $0.1\lsim \Omega\lsim 2$, consistently with the experimental data, the early Universe must have had $|\Omega-1|\lsim 10^{-59}\dfrac{\mpl^2}{T^2}$. This implies that at the Planck epoch when $T\approx \mpl$,
\be
|\Omega-1|=\left|\dfrac{\rho}{\rho_c}-1 \right|\lsim 10^{-59}\,.
\ee
Thus if the initial density would have been larger than $\rho_c$ by e.g. a factor of $10^{-57}$, the Universe would have collapsed long ago. Conversely, if the density were smaller than $\rho_c$ by the same factor, the expansion would have been so rapid to forbid the formation of structures. Therefore we are left with an extreme fine-tuning of the initial conditions, namely with a fairly unnatural coincidence in the history of the Universe.

	The \textbf{horizon problem} deals with causality. First consider a comoving wavelength $\lambda$ and a physical wavelength $a\lambda$ such that $a\lambda\lsim\Hub^{-1}$ (namely, it is inside the Hubble radius $\Hub^{-1}$). Since in the Friedman models $a\propto t^p$ with $0<p<1$, we find that $a\lambda\propto t^p$. However, $\Hub^{-1}\propto t$ and the physical wavelength becomes much smaller than the Hubble radius, defining a causal region which is very small within the horizon.

	More into detail, with the notations of Ref.\cite{Tsujikawa}, the particle horizon $D_H(t)$ is defined as the region where the light travels from the moment of the big bang $t_*$,
\be
D_H(t)=a(t)d_H(t)\,, \qquad d_H(t)=\int^{t}_{t_*}\dfrac{d\tau}{a(\tau)}\,,
\ee
where $d_H(t)$ is the comoving distance. The photons we observe in the CMB were emitted at the time of decoupling, and $D_H(t_{dec})=a(t_{dec})d_H(t_{dec})$ corresponds to the region where photons have interacted at that time. If the comoving distance at the actual time ($t_0$) is $d_H(t_0)$, we have
\be
\dfrac{d_H(t_{dec})}{d_H(t_0)}\propto\left(\dfrac{t_0}{t_{dec}} \right)^{-1/3}\propto \left(\dfrac{10^5}{10^{10}} \right)^{1/3}\approx 2\times10^{-2}\,.
\ee
This implies that the regions where there is causal interaction of photons are small. However, in the CMB we actually see photons which thermalise to the same temperature in all regions. This is in contradiction with the standard big bang cosmology.

	The \textbf{origin of large scale structure} is also problematic. The anisotropies in the last scattering surface (LSS) have been observed by the COBE satellite. Their amplitudes are small and almost scale-invariant. In the standard cosmology, it is impossible to generate them between the big bang and the LSS, because they spread too fast and spoil the formation of structures.

	Finally, the \textbf{monopole problem} regards the creation of many unwanted relics (such as monopoles, topological defects and cosmic strings) due to supersymmetry breaking \cite{Linde}. The cosmic string, in particular, predicts the primordial formation of gravitinos and moduli fields, whose energy density decreases like matter, namely as $a^{-3}$. The radiation energy density decreases as $a^{-4}$ in the radiation-dominated era, therefore such relics would be dominant in the Universe. This is of course inconsistent with observations.

\section{The inflationary model}

	The physical mechanisms which are analysed in Chapters \ref{chapt:WW} and \ref{chapt:Infl} take place right after the end of inflation, which therefore provides a common background. Following Refs.\cite{Linde,Tsujikawa}, we here provide with an introductory description of the inflationary paradigm.

	The basic idea is simple, yet extremely powerful. Since the problems of the standard Big Bang cosmology are related to an always decelerating expansion, let us assume that in the early Universe there has been an early stage with accelerated expansion. In other words, we impose
\be
\ddot{a}>0\,,
\label{accinfl}
\ee
that by virtue of (\ref{acc}) implies
\be
\rho+3p<0\,.
\label{diseqstate}
\ee
Eq.(\ref{accinfl}) means that $\dot{a}$ (and therefore $a\Hub$) increases during inflation. Accordingly, the comoving Hubble radius $(a\Hub)^{-1}$ now decreases, and the $a^2\Hub^2$ term in
\be
\Omega-1=\dfrac{k}{a^2\Hub^2}\,,
\ee
increases during inflation and $\Omega$ becomes rapidly $\mathcal{O}(1)$.  After inflation ends, $\Omega$ starts to decrease like in the standard scenario. However, if inflation lasts long enough, $\Omega$ remains close to unity until the present day. The flatness problem is now avoided.

	Moreover, during inflation the scale factor evolves as $a\propto t^p$, with $p>1$. This implies that the physical wavelength $a\lambda$ is pushed outside the Hubble radius, which grows linearly with time. Namely, causality is now guaranteed in regions which are much larger than the horizon, and the horizon problem is solved. Also in this case, however, inflation should last for enough time, since after inflation the Hubble radius starts to grow again faster than the physical wavelength. Since we need that before decoupling the photons cover a comoving distance that is much larger than the distance after the decoupling, the following condition must be satisfied,
\be
\int^{t_{dec}}_{t_*}\dfrac{dt}{a(t)}\gg\int^{t_0}_{t_{dec}}\dfrac{dt}{a(t)}\,.
\ee
This happens if the Universe expands $e^{70}$ times during inflation \cite{LL}.

	Large scale structures can be generated after inflation, since the comoving Hubble radius decreases during the inflationary expansion. This means that the nearly scale-invariant perturbations, which are needed for the creation of cosmological structures, are causally related and small quantum fluctuations are thus generated. After the scale is pushed outside the Hubble radius during inflation, the perturbations can be described as classical. When inflation ends, the evolution of the Universe is described by the standard Big Bang model, and the comoving Hubble radius begins to increase. Then the scales of perturbations cross inside the horizon, and causality follows. The small perturbations originated during inflation appear as large-scale perturbations after this second horizon crossing. This produces the seeds of density perturbations which are observed in the anisotropies of the CMB.

	We conclude with a comment on particle production. During the inflationary epoch the energy density decreases very slowly. Indeed, $a\propto t^p$, where $p>1$, implies $\Hub\propto t^{-1}\propto a^{-1/p}$, and $\rho\propto a^{-2/p}$. At the same time, the energy density of massive particles decreases much faster ($\propto a^{-3}$), thus the particles are red-shifted away during inflation and diluted. Also the monopole problem is thus solved.

\subsection{Slow-roll conditions and inflationary models}

	After considering the main idea of inflation and its cosmological effects, we now discuss the theory more closely.
Let us consider a homogeneous scalar field $\varphi$, which we will call the \emph{inflaton}. The potential $V(\varphi)$ makes the Universe expand exponentially. The energy density and the pressure density of this particle are written as:
\be
\rho=\dfrac{1}{2}\dot{\varphi}^2+V(\varphi)\,,\qquad p=\dfrac{1}{2}\dot{\varphi}^2-V(\varphi)\,.
\ee
Substituting this into (\ref{Fried}) and (\ref{Fried2}), we get
\be
\Hub^2=\dfrac{8\pi}{3\mpl^2}\left[ \dfrac{1}{2}\dot{\varphi}^2+V(\varphi) \right]\,,
\label{friedinfl1}
\ee
and
\be
\ddot{\varphi}+3\Hub\dot{\varphi}+V^{\prime}(\varphi)=0\,,
\label{friedinfl2}
\ee
where we have neglected the curvature term $k/a^2$. The above means that during inflation, Eq.(\ref{diseqstate}) provides with $\dot{\varphi}^2 < V(\varphi)$, namely the potential energy of the inflaton dominates over its kinetic energy. Accordingly, since we need a very flat potential to guarantee a sufficient amount of inflation, we impose the following slow-roll conditions:
\bea
&&\dot{\varphi}^2 \ll V(\varphi)\,,\label{slow1}\\
&&|\ddot{\varphi}|\ll |3\Hub\dot{\phi}|\,.\label{slow2}
\eea
By introducing the fundamental slow-roll parameters,
\be
\epsilon\equiv\dfrac{\mpl^2}{16\pi}\left(\dfrac{V^{\prime}}{V} \right)^2,\qquad \eta\equiv\dfrac{\mpl^2}{8\pi}\dfrac{V^{\prime\prime}}{V}\,,
\ee 
it can be shown that Eqs.(\ref{slow1}) and (\ref{slow2}) imply $\epsilon\ll 1$, $|\eta|\ll 1$. In this limit, Eqs.(\ref{friedinfl1}) and (\ref{friedinfl2}) can be rewritten respectively as follows,
\bea
&\Hub^2\approx\dfrac{8\pi}{3\mpl^2}V(\varphi)\,,\label{sr1}\\
&3\Hub\dot{\varphi}\approx -V^{\prime}(\varphi)\,.
\label{sr2}
\eea
The above are called the \emph{slow-roll equations}. Accordingly, inflation ends when either $\epsilon$ or $\eta$ grow enough so that they approach order unity.
We remark that the conditions $\epsilon\ll 1$ and $|\eta|\ll 1$ are just constraints on the shape of the potential, and that they do not necessarily imply the slow-roll equations.

 To measure the amount of inflation, the standard quantity is the number of e-foldings, defined as
\be
N\equiv \ln\dfrac{a_f}{a_i}=\int^{t_f}_{t_i}{\Hub dt}\,,
\ee
where the subscripts $i$ and $f$ denote respectively the quantities at the beginning and at the end of inflation. In order to solve the flatness problem, $|\Omega_f-1|\lsim 10^{-60}$ right after the end of inflation. The ratio of this quantity between the initial and final phase of inflation is
\be
\dfrac{|\Omega_f-1|}{|\Omega_i-1|}\lsim\left(\dfrac{a_i}{a_f} \right)^2=e^{-2N}\,,
\ee
if one assumes that $\Hub$ is nearly constant in that period. From the above equation, if $|\Omega_i-1|\approx 1$, we find that the number of e-foldings has to be $N\gsim 70$ to solve both the flatness problem and the horizon problem.

	In the next subsections we will sketch some inflationary models, which differ by the particle content and the scalar potential $V(\varphi)$. In particular, we will consider chaotic and hybrid inflation. There exists a number of other models, for instance natural inflation \cite{PNGBs}, but we will not describe them here.

\subsubsection{Chaotic inflation}
	This model was proposed in 1983 by Andrei Linde \cite{Lindechao}. It is defined by chaotically distributed initial conditions and by a potential that can be either quadratic,
\be
V(\varphi)=\dfrac{1}{2}m^2\varphi^2\,,
\ee
or quartic,
\be
V(\varphi)=\dfrac{1}{4}\lambda\varphi^4\,,
\ee
thus provided with a self-interaction term.
 With a quadratic potential, the slow-roll equations (\ref{sr1}) and (\ref{sr2}) can be rewritten as
\be
\Hub^2\approx\dfrac{4\pi m^2\varphi^2}{3\mpl^2}\,,\qquad 3\Hub\dot{\varphi}+m^2\varphi\approx 0\,,
\ee
which provide the following solutions,
\be
\varphi\approx\varphi_i- \dfrac{m\mpl}{2\sqrt{3\pi}}t\,,
\ee
and
\be
a\approx a_i\exp{\left[2\sqrt{\dfrac{\pi}{3}}\dfrac{m}{\mpl}t\left(\varphi_i-\dfrac{m\mpl}{4\sqrt{3\pi}}t \right) \right]}\,.
\ee
$\varphi_i$ is the initial value of the inflaton. What happens physically, is that the inflaton finds itself displaced from the true vacuum, to which it rolls back, as in Fig.(1.2).
	In the case of a quartic potential, when $\varphi > \lambda^{-1/4}\mpl$, $\varphi$ has a greater energy density than the Planck density, and classical physics cannot describe the evolution of the Universe. Instead, if $\mpl/3<\varphi <  \lambda^{-1/4}\mpl$, the values of the field $\varphi$ slowly decrease and the Universe then inflates. Inflation takes place while the inflaton is displaced. The evolution of the scale factor implies the required exponential expansion during the initial stages of inflation. Then it slows down since $(-m\mpl/4\sqrt{3\pi})t^2$ grows with time. The slow-roll parameters in this case are identical, that is
\be
\epsilon=\eta=\dfrac{\mpl^2}{4\pi\varphi^2}\,.
\ee
Accordingly, inflation ends when $|\varphi|\approx \mpl/4\pi$. The initial value of the inflaton field is obtained by demanding sufficient inflation, i.e. that $N\gsim 70$. This is fulfilled by $\varphi_i\gsim 3\mpl$, since the number of e-foldings is given by
\be
N\approx 2\pi\left(\dfrac{\varphi_0}{\mpl} \right)^2-\dfrac{1}{2}\,.
\ee
The observations of the density perturbations by the COBE satellite \cite{COBE} provide with constraints on the inflaton mass,
\be
m\approx 10^{-6}\mpl\,,
\ee
and on the self-coupling \cite{LL}, which is very small $\lambda\approx 10^{-13}$.

\begin{center}

\begin{figure}[t]

\begin{center}
\resizebox{0.89\textwidth}{!}{
\begin{minipage}[]{0.5\textwidth}
\caption{Chaotic inflation. If $\mpl/3<\varphi <  \lambda^{-1/4}\mpl$, the values of the field $\varphi$ slowly decrease and the Universe then inflates. Inflation takes place while the inflaton is displaced, as in figure \cite{Linde}.}
\end{minipage}

\hspace{0.01\linewidth}
\begin{minipage}[]{0.5\textwidth}

	\includegraphics[width=8.5cm]{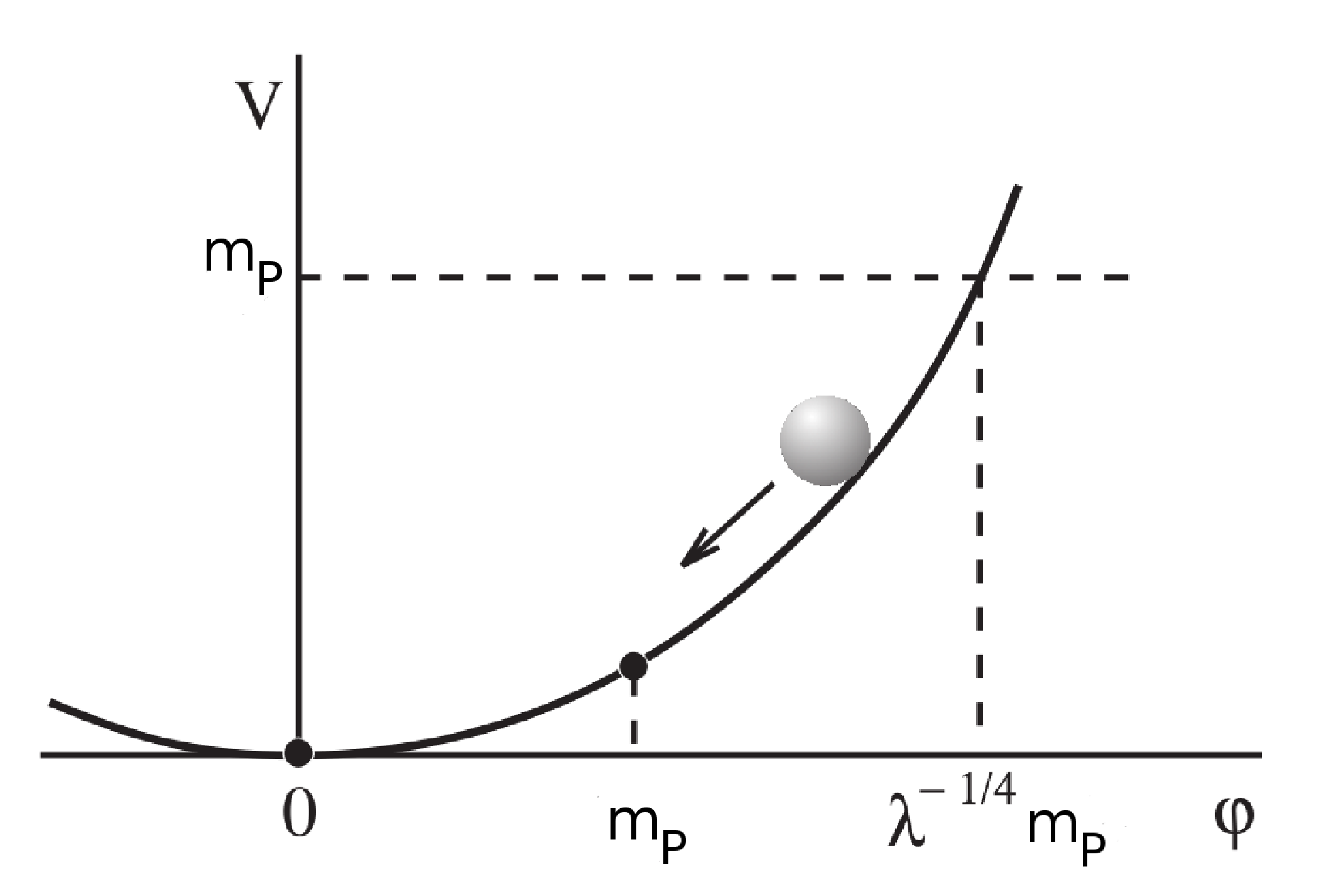}

 \end{minipage} 
}
\end{center}

\label{fig:chaotic}

\end{figure}

\end{center}

\subsubsection{Hybrid inflation}

Instead of considering only a single inflaton field, one might want to define multiple scalar fields \cite{Lindehyb}. In this case, the potential can be written as
\be
V=\dfrac{\lambda}{4}\left( \chi^2-\dfrac{M^2}{\lambda}\right)^2+\dfrac{1}{2}g^2\varphi^2\chi^2+\dfrac{1}{2}m^2\varphi^2\,.
\ee
If $\varphi^2$ is large, this can be approximated by the single-field potential
\be
V=\dfrac{M^4}{4\lambda}+\dfrac{1}{2}m^2\varphi^2\,,
\label{approx}
\ee
since $\varphi$ tends to roll down toward the potential minimum at $\chi=0$. Inflation does not end for the approximated potential (\ref{approx}), however the mass of $\chi$ becomes negative for $\varphi<\varphi_c\equiv M/g$. This implies that the field rolls down to one of the true minima at $\varphi=0$ and $\chi=\pm M/\sqrt{\lambda}$, as illustrated in Fig.(1.3). Inflation ends after the symmetry breaking for $\varphi<\varphi_c$, due to the rolling of the field $\chi$. If the initial value of the inflaton is $\varphi_i$, the number of e-foldings is
\be
N\approx \dfrac{2\pi M^4}{\lambda m^2\mpl^2}\ln{\dfrac{\varphi_i}{\varphi_c}}\,,
\ee
that follows from the approximated potential.

\subsection{Reheating}
Right after the end of inflation, which cooled the Universe via the expansion, a reheating period occurs. The Universe is thermalised, as the potential energy of the inflaton is transferred to radiation. In the original idea, the "old reheating" \cite{Dolgov}, the inflaton decays perturbatively. However, this is not an efficient mechanism for baryogenesis at the GUT scale. It was later found that a consistent scenario should include a nonperturbative stage called \emph{preheating}, where an explosive production of particles occurs in the early stages of reheating \cite{Lindereh,Branden}.

\subsubsection{Old reheating}

\begin{center}

\begin{figure}[t]

\begin{center}
\resizebox{14cm}{!}{
\begin{minipage}[]{0.5\textwidth}
\caption{Hybrid inflation.  The inflaton rolls down to one of the true minima at $\varphi=0$ and $\chi=\pm M/\sqrt{\lambda}$. Inflation ends after the symmetry breaking for $\varphi<\varphi_c$, due to the rolling of the field $\chi$.}
\end{minipage}

\hspace{0.01\linewidth}
\begin{minipage}[]{0.4\textwidth}

	\includegraphics[width=6cm]{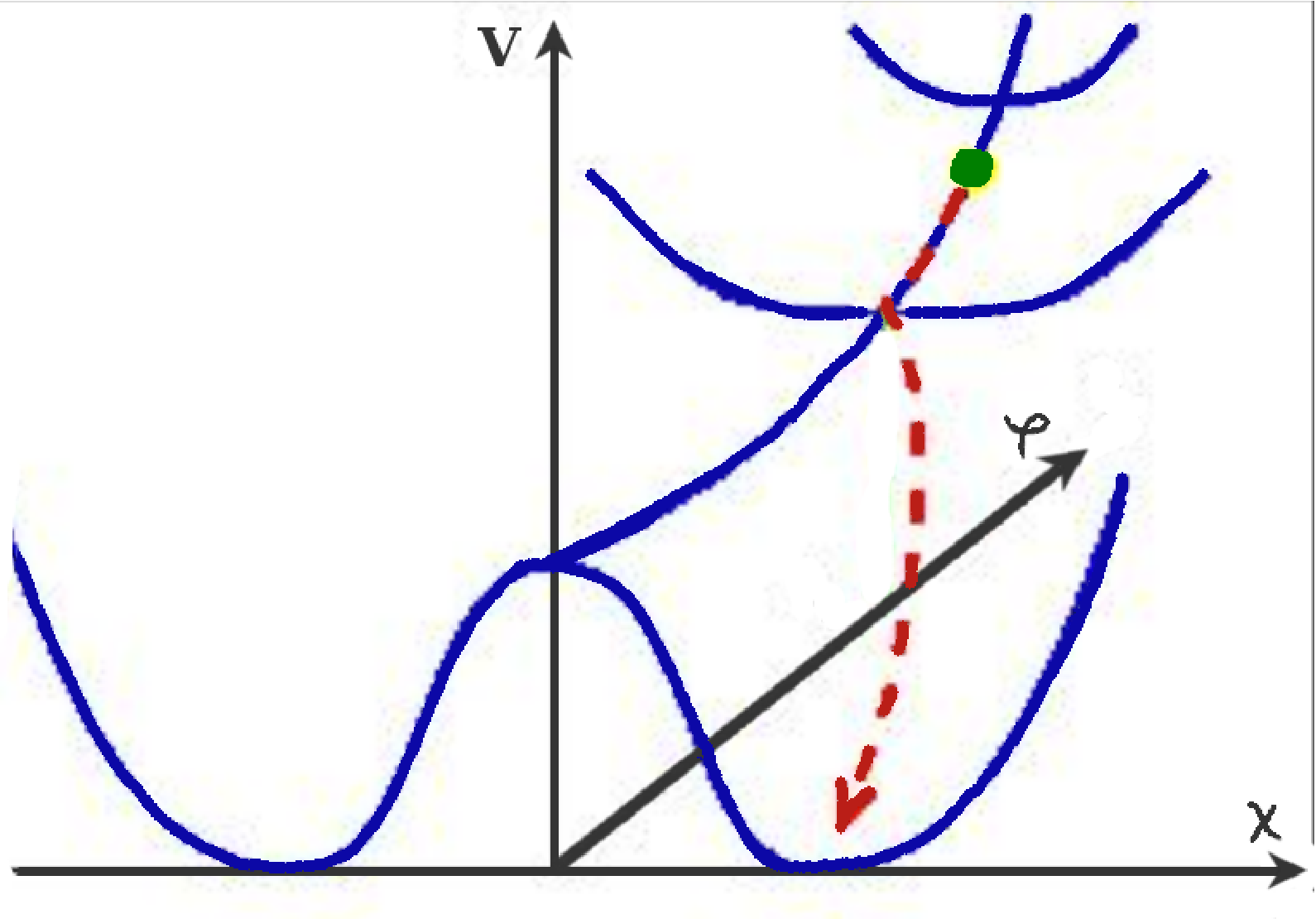}

 \end{minipage} 
}
\end{center}
\label{fig:hybrid}

\end{figure}

\end{center}

	At the end of inflation, the inflaton field reaches the minimum of the potential and starts to oscillate coherently about this minimum. This is due to the friction term proportional to the Hubble rate in the Friedman equation (\ref{friedinfl1}). We will now sketch some properties of the reheating scenario, since a general background is enough to deal with the content of this thesis.
Let us consider a simple quadratic potential
\be
V(\varphi)=\dfrac{1}{2}m^2\varphi^2\,.
\ee
The inflaton is then described by sinusoidal oscillations, with decreasing amplitude $A(t)$,
\be
\varphi=A(t)\sin{mt}\,,\qquad A(t)=\dfrac{\mpl}{\sqrt{3\pi}mt}\,,
\ee
which give a decreasing energy density that is modified by the inflaton decay,
\be
\rho=\dfrac{1}{2}\dot{\varphi}^2+V(\varphi)\approx\dfrac{1}{2}m^2A^2(t)\propto a^{-3}\,.
\ee
In the old reheating model, there is a single inflaton field $\varphi$ which is coupled to a scalar field $\chi$ and to a fermion $\psi$,
\be
\mathcal{L}_{int}=-\sigma\varphi\chi^2-h\varphi\bar{\psi}\psi\,,
\ee
with the coupling constants $\sigma$ and $h$. Here we have assumed that $m_\chi$ and $m_\psi$ are negligible with respect to the inflation mass $m$ for simplicity.

	Quantum corrections are also taken into account, through the decay processes of $\varphi$. Let us call $\Gamma=\Gamma(\varphi\rightarrow\chi\chi)+\Gamma(\varphi\rightarrow\bar{\psi}\psi)$ the total decay rate of the inflaton. The partial decay rates into scalar and fermion pairs result as \cite{Dolgov},
\be
\Gamma(\varphi\rightarrow\chi\chi)=\dfrac{\sigma^2}{8\pi m}\,,\qquad \Gamma(\varphi\rightarrow\bar{\psi}\psi)=\dfrac{h^2m}{8\pi}\,,
\ee
which remain valid as long as $\Gamma\ll m$, namely $\sigma^2\ll m^2$ and $h^2\ll 1$. By adding a phenomenological decay term $\Gamma\dot{\varphi}$ to the evolution equation,
\be
\ddot{\varphi}+3\Hub\dot{\varphi}+\Gamma\dot{\varphi}+m^2\varphi=0\,,
\ee
we now obtain the solution
\be
A(t)=\dfrac{\mpl}{\sqrt{3\pi}mt}e^{-\Gamma t/2}\,.
\ee
For small couplings $\sigma$ and $h$, $\Gamma<3\Hub$ at the beginning of inflation. Thus particle production may become important, in comparison to the total energy density, under certain conditions. This happens when $3\Hub\lsim\Gamma$, since the Hubble rate decreases as $1/t$. An estimation of the energy density of the Universe can be made by setting $\Gamma^2=(3\Hub)^2=24\pi\rho/\mpl^2$\,:
\be
\rho=\dfrac{\Gamma^2\mpl^2}{24\pi}\,.
\ee
By assuming that this is all transferred to light particles, which are instantaneously thermalized at temperature $T_R$, we obtain
\be
\rho=\dfrac{g_*\pi^2T^4_R}{30}=\dfrac{\Gamma^2\mpl^2}{24\pi}\,.
\ee
$g_*$ is the effective number of degrees of freedom at $T=T_R$, and it is larger than 100. The estimation of the reheating temperature holds as
\be
T_R\lsim 0.1\sqrt{\Gamma \mpl}\,,
\ee
which by virtue of the relation $\Gamma\ll m \lsim 10^{-6}\mpl$ can be rewritten as
\be
T_R\ll 10^{-4}\mpl\,.
\ee
	This is a much smaller value than the temperature at the GUT scale $T_{GUT}\approx 10^{-3}\mpl\approx 10^{16}\G$. This means that the old reheating scenario does not provide a good GUT scale baryogenesis.

\subsubsection{Preheating}
	We have seen that the old reheating model cannot describe correctly the phenomenology of particle creation in the early Universe. The theory of preheating \cite{KLS,Greene}, on the other hand, is free of this problem. It postulates that before the perturbative decay, the inflaton might have started to decay by means of a parametric resonance. This is a much more explosive process, which is non-perturbative.

	Preheating has a very complex dynamics, that is still not completely understood. Since it might affect significantly both Dark Matter production and the density porturbations, it has been studied extensively in recent years \cite{Kofman,Jokinen,preheating}. By including also this mechanism, the reheating process consists of three different stages:
i) Preheating, where particles are produced non perturbatively by parametric resonance, ii) Perturbative decay of the inflaton, iii) Thermalization of the particles.

	We now consider this scenario in the hybrid inflation model, following Ref. \cite{Tsujikawa}. The potential is given by
\be
V(\varphi,\chi)=\dfrac{1}{2}m^2\varphi^2+\dfrac{1}{2}g^2\varphi^2\chi^2\,,
\ee
and we assume that spacetime and inflaton $\varphi$ give a classical background, upon which the scalar field $\chi$ is quantum. By expanding $\chi$ in plane waves
\be
\chi=\dfrac{1}{(2\pi)^{3/2}}\int\left(a_k\chi_k(t)e^{-i\mbox{k}\cdot\mbox{x}}+a^{\dagger}_k\chi^*_k(t)e^{i\mbox{k}\cdot\mbox{x}} \right)d^3\mbox{k}\,,
\ee
and using the Friedman-Robertson-Walker metric, for each Fourier component $\chi_k(t)$ the equation of motion is the following\,,
\be
\ddot{\chi}_k+3\Hub\dot{\chi}_k+\left(\dfrac{k^2}{a^2}+g^2\varphi^2 \right)\chi_k=0\,.
\ee
By rescaling $\chi_k$ and thus introducing the scalar field $X_k\equiv a^{3/2}\chi_k$, the above equation is then recasted as
\be
\ddot{X}_k+\omega^2_kX_k=0\,,
\label{Xdyn}
\ee
where the frequency of the mode $k$ is
\be
\omega^2_k\equiv \dfrac{k^2}{a^2}+g^2\varphi^2-\dfrac{3}{4}\left(\dfrac{2\ddot{a}}{a}+\Hub^2 \right)\,.
\label{Xfreq}
\ee
During reheating, the term in brackets in (\ref{Xfreq}) becomes negligible, so it can be safely ignored. We then obtain from (\ref{Xdyn}) the Mathieu equation
\be
\dfrac{d^2X_k}{dz^2}+(A_k-2q\cos 2z)X_k=0\,,
\label{Mathieu}
\ee
where $z=mt$. The amplitude of the resonance is
\be
A_k=2q+\dfrac{k^2}{m^2a^2}\,,\label{amp}
\ee
with
\be
q=\dfrac{g^2A^2(t)}{4m^2}\,.
\ee
	It is clear that the strength of the resonance in (\ref{Mathieu}) depends on $A_k$ and $q$. Regions of stability and instability are determined by the ratio $A_k/q$. In the unstable region there is production of $X_k\propto\exp(\mu_kz)$ and of particles with momentum $k$ ($\mu_k$ is the Floquet index). For $q$ smaller than unity, only a few modes grow and we have a narrow resonance. On the contrary, if $q$ is much larger than the unity, resonance occurs for a broad range of the momentum $k$-space. Since the growth rate of the produced particles is proportional to $q$, this is a much more efficient mechanism than the narrow resonance. It is called "broad resonance" \cite{KLS}.

	From the above we see that the initial amplitude of the inflaton and the coupling $g$ are fundamental to determine whether the resonance will be broad or narrow. Since the inflaton mass should be $m\approx 10^{-6}\mpl$ to satisfy the COBE normalizations, $q$ is large if $g\gsim 10^{-4}$ with $A(t_i)\approx 0.2\mpl$. Then it can be shown that the broadest resonance is given by $A_k=2q$. Eq.(\ref{amp}) also implies that particles with low momenta are mostly produced. However, particles with high momenta can be created if $g$ and $A_k$ are large. This follows from the expression of the maximum comoving momentum \cite{Lindereh},
\be
k\lsim\sqrt{\dfrac{gmA_k}{2}}\,,
\ee
that is obtained by using the nonadiabaticity condition $d\omega_k/dt\gg\omega_k^2$. For initially large $q$, there is a \emph{stochastic} resonance of each mode \cite{Lindereh}. In fact, the frequency $\omega_k$ decreases by cosmic expansion and changes within each oscillation of the inflaton. Therefore there is no correlation between the phases of the fields $\chi$ and $\varphi$. This is important. In the first stage of preheating, the $\chi$ fields cross many instability bands, without spending enough time on each band to provide with efficient resonance. This is in contrast to what happens in the case of Minkowski spacetime.
	However, as the cosmic expansion slows down, $q$ becomes smaller and the fields stay in each resonance band for a longer time. Particle production ends when the variables decrease below the lower boundary of the first resonance band, by effect of the expansion of the Universe.

	Interestingly, the resonance can be terminated by one more mechanism. There is a backreaction effect of the $\chi$ particles which are produced, that modifies the equation of motion of the inflaton as follows,
\bea
\ddot{\varphi}+3H\dot{\varphi}+(m^2+g^2\langle\chi^2\rangle)\varphi=0\,,
\eea
with the following expectation value of $\chi^2$,
\bea
\langle\chi^2\rangle \equiv \frac{1}{2\pi^2} \int k^2|\chi_k|^2~dk\,.
\eea 
	An initial value $q~\gsim~3000$, which corresponds to $g~\gsim~3.0 \times 10^{-4}$, gives a growth of the variance $\langle\chi^2\rangle$ of the order $m^2/g^2$. In this case the backreaction becomes important \cite{Tsujikawa}. This affects the oscillations of the inflaton field, which become incoherent, and the resonance is stopped.

\section{The matter content of the Universe}

	In this chapter we have discussed models which aim to provide for a theoretical explanation of cosmological observations. It has been shown that a conservative way to explain the observed acceleration in the expansion of the Universe is assuming a certain form of the energy momentum tensor $T_{\mu\nu}$ in the Einstein equations (\ref{Einstein}).

	This corresponds to introducing exotic types of matter, such as Dark Matter and dark energy. In this final section we review the amount of each species as a fraction of the overall matter content, as it is obtained in agreement with recent experimental data.

	As we have seen in Section \ref{sect:friedman}, a key quantity to consider is the density parameter $\Omega$. It is defined as the ratio of the actual (observed) density $\rho$ to the critical density $\rho_c$ of the Friedman Universe. The critical density is the watershed between an expanding and a contracting Universe, since it corresponds to a static, flat cosmology with spatial curvature $k=0$. With this assumption, the first Friedman equation (\ref{Fried}) gives for the critical density today
\be
\rho_c(t_0)=\dfrac{3\Hub_0^2}{8\pi G}=\dfrac{3\Hub_0^2\mpl^2}{8\pi}=1.88\times 10^{-29}h^2\dfrac{\rm g}{\rm cm^3}\,,
\ee
where $G$ is the Newton's gravitational constant, $\Hub_0$ is the actual value of the Hubble parameter and the Planck mass is here expressed in grams as $\mpl=2.176\times 10^{-5}\rm g$.
	The dimensionless parameter $h$ is defined in term of the Hubble scale as
\be
\Hub=100h \rm km/sec/Mps\,,
\ee
and its value has been recently measured as \cite{WMAP5}
\be
h=0.73\pm 0.005\,.
\ee
	The density parameter of the species $i$ is accordingly defined in function of the critical density,
\be
\Omega_i\equiv \dfrac{\rho_i}{\rho_c}=\dfrac{8\pi G\rho_i}{3\Hub^2}\,.
\ee
	It is interesting to note that the above $\rho_c(t_0)$ corresponds roughly to 10 protons per $\rm m^3$. But actually the dominant matter is not baryonic, thus there are on an average only 0.5 protons per cubic meter! A general expression for $\Omega$ where the density parameter equals exactly unity is commonly used. According to the $\Lambda$CDM model, the important components of $\Omega$ are due to baryons (i.e. ordinary matter), Cold Dark Matter and dark energy. The WMAP satellite has measured the spatial geometry of the Universe to be nearly flat, thus we have really $k=0$ and the total energy density is (almost) equal to the critical one:
\be
 \Omega_{tot}=\Omega_B+\Omega_{DM}+\Omega_{\Lambda}\,.
\ee
	The total cosmological energy density is close to be critical, $\Omega_{tot}=1\pm 0.02$. This value has been obtained experimentally from the position of the first peak of the angular spectrum of the CMB radiation and the large scale structure (LSS) of the Universe. Let us now consider the different matter species separately.

	The ordinary, or baryonic matter, makes a very small contribution, $\Omega_B=0.044\pm0.004$, as found e.g. from the eights of the peaks in angular fluctuations of the CMB and from the production of light elements in the BBN.

	Dark matter makes instead a more relevant contribution, $\Omega_{DM}=0.22\pm0.04$. This value is mostly made of two components: particles which are relativistic (called Hot Dark Matter), and those which constitute an almost inert relic in the intergalactic spaces, called Cold Dark Matter (CDM).
There is also a third kind of particles, called Warm Dark Matter (WDM). WDM candidates usually are non-annihilating but rather weakly-interacting
particles. They barely escape cosmological constraints like the free-streaming and self-damping bounds. In this case, structure formation occurs bottom-up from above their free-streaming scale, and top-down below their free streaming scale. They can be associated with any kind of particles, as long as they are just at the limit of the region allowed by self-damping and free-streaming\footnote{We shall consider free-streaming for gravitinos in Chapter \ref{chapt:Infl}.}.

 Dark matter is mysterious, since it interacts only gravitationally and as such it cannot be detected directly. However, its effects on galactic rotation curves, gravitational lensing, equilibrium of hot gas in rich galactic clusters, cluster evolution and LSS provide with the actual value of $\Omega_{DM}$. In this thesis, in particular, we focus on the possibility of producing Dark Matter gravitinos in the early Universe.

	Finally, the most mysterious and controversial component of the matter content: dark energy. Nobody has a clear idea about what it really is, either a purely repulsive form of energy or some peculiar scalar particles (moduli, quintessence fields...). It drives the accelerated cosmological expansion and is uniformly distributed in the Universe. The dimming of supernovae with high redshift, LSS, the CMB spectrum and the age of the Universe give the value $\Omega\approx 0.74$.
Since this is overwhelmingly the dominant component of the total matter content, we say that the actual Universe is dominated by a cosmological constant, in contrast with previous periods of radiation and matter domination.

\chapter{About supergravity and gravitino cosmology}\label{chapt:sugra}

	In this chapter we consider supergravity (SUGRA), namely the theory that corresponds to local supersymmetry. The gauge field is the gravitino, which obtains a mass through the super-Higgs mechanism \cite{Ferrara,Nilles,Wess}. Our discussion is based on the review by Lyth and Riotto \cite{LR} and on Takeo Moroi's thesis \cite{M}, to which we refer for deeper analyses.

\section{Basics of supersymmetry and supergravity}

	The Standard Model of Particle Physics (SM) is an established and successful framework, which has been tested experimentally with great accuracy. Nevertheless, it is common conviction that it has to be somehow extended to more fundamental theories. There are not really inconsistencies within the SM, however the theory does not seem completely natural (remember the famous "hierarchy problem").

	To this aim, supersymmetry (SUSY) is certainly the most promising possibility. In very simple words, it interchanges SM (or "ordinary") particles with new particles, which have higher masses and are called "superpartners". SUSY provides with a natural solution of the hierarchy problem and it gives a remarkably good unification of the gauge constants. Moreover, via conservation of R parity, it introduces automatically a candidate for Dark Matter.

	We still lack experimental evidence of the superpartners, thus SUSY must be broken to account for the mass difference between the two sets of particles. In the case of global supersymmetry (the transformations are not space-time dependent), the SUSY breaking cannot be phenomenologically satisfactory. For example, there is no experimental candidate for the massless spin-1/2 particle called goldstino, that is created by the global SUSY breaking.

	As a solution to this drawback, one might consider local transformations. The gauge particle is the spin-3/2 gravitino, which after the breaking of supersymmetry absorbs the goldstino modes and becomes massive. Since SUSY requires the gravitino to be coupled to a spin-2 ordinary particle, we find this in the graviton. Gravity is then automatically embedded, and supergravity (SUGRA) is the theory which arises accordingly. This provides with new interesting insights, from both theoretical and experimental perspectives.

\begin{center}

\begin{figure}[t]

\begin{center}
\resizebox{14cm}{!}{
\begin{minipage}[]{0.5\textwidth}

\includegraphics[width=8cm]{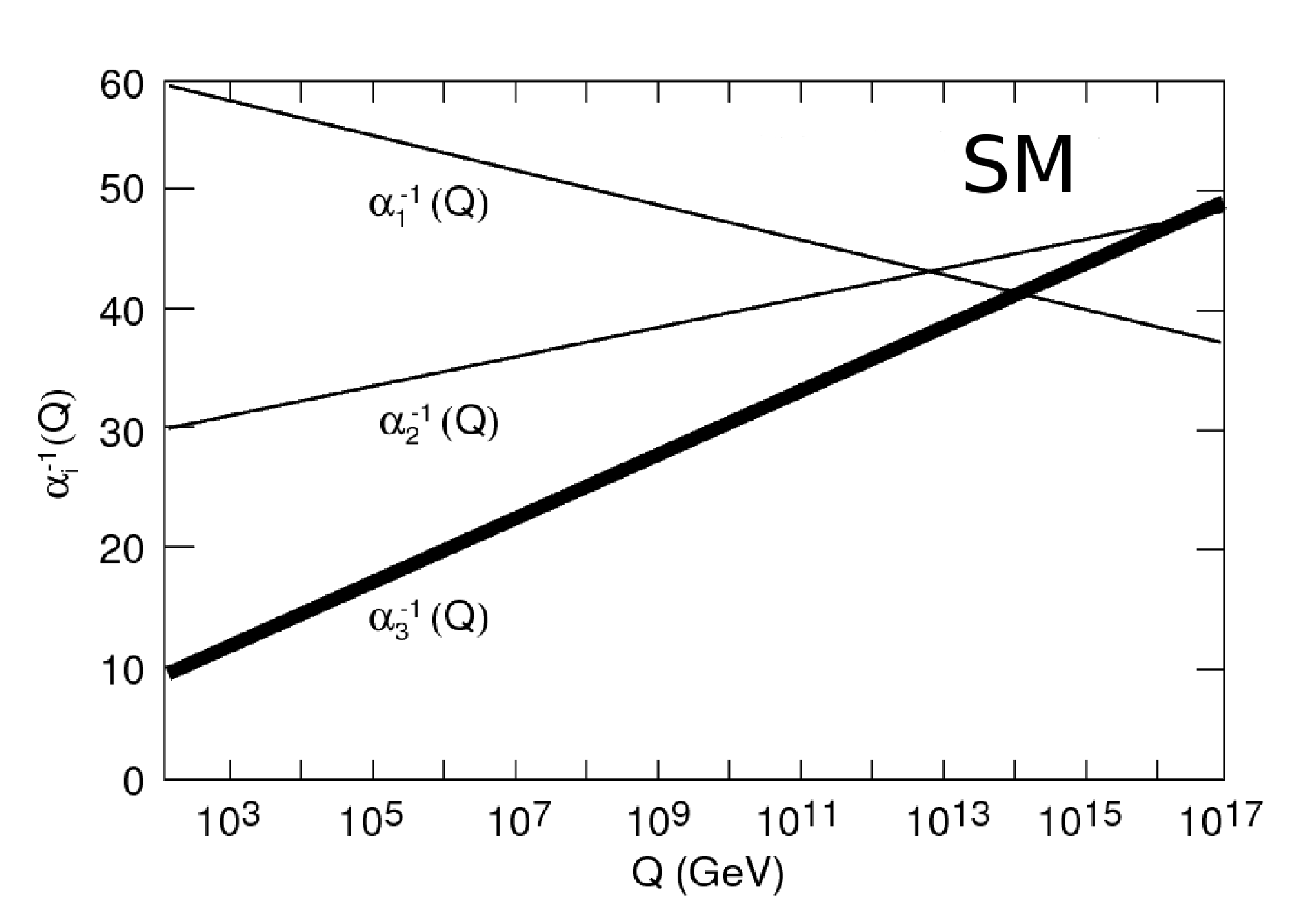}

\end{minipage}

\begin{minipage}[]{0.5\textwidth}

\includegraphics[width=8cm]{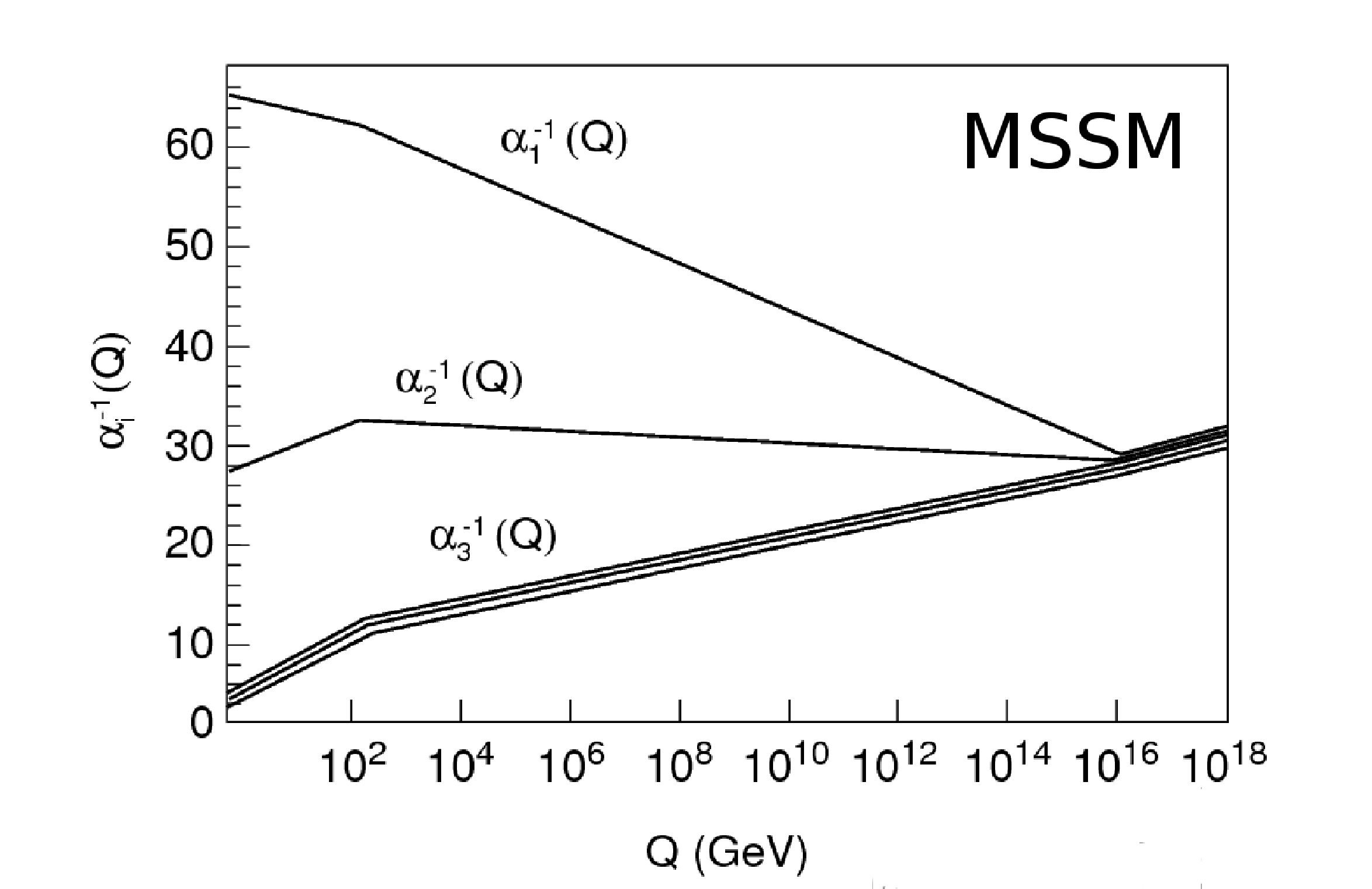}

 \end{minipage}
}
\end{center}

\caption{Unification of the couplings in the SM (left) and in the MSSM (right).}
\end{figure}

\end{center}

\subsection{The supersymmetry algebra}

	SUSY theories of type N=1 (see for instance \cite{Nilles,WZ,Dimopoulos}), which have only one fermionic generator, take part to the lowest class of supersymmetric models. Yet they are relevant for inflation. In the following, we describe their basics with the conventions of Wess and Bagger \cite{Wess}.

	Let us consider a supersymmetry algebra given by 
\be
\{ Q_{\alpha}, \bar{Q}_{\dot\beta}\}
=2 \sigma^{\mu}_{\alpha\dot\beta} P_{\mu}\,, 
\ee
where $Q_{\alpha}$ and $\bar{Q}_{\dot\beta}$ are the supersymmetric 
generators and the bars stand for conjugation. The two-component Weyl spinors are labelled by $\alpha$ and $\beta$, which run from 1 to 2. The quantities with dotted and those with undotted indices transform respectively under the $(0,\frac{1}{2})$ 
representation and the $(\frac{1}{2},0)$ conjugate representation of the Lorentz group.
$\sigma^\mu$ is a matrix four vector, $\sigma^\mu=(
-{\bf 1},\vec{\sigma})$ and $P_\mu$ is the four-momentum operator.

	The supersymmetry algebra that contains fields of spin less than or equal to one have two irreducible representations, the chiral and vector superfields. More precisely, chiral fields contain a Weyl spinor and a complex scalar, vector fields contain a Weyl spinor and a massless vector.
A chiral superfield can be expanded in terms of the Grassmann variable $\theta$ in superspace,
\be
\phi(x,\theta)= \phi(x) + \sqrt{2}
\theta \psi(x) + \theta^2F(x)\,,
\ee
where $x$ is a point in spacetime, $\phi(x)$ is the complex scalar, $\psi$ the fermion and $F$ is an auxiliary field. Following a common habit, here we maintain the same symbol to represent a superfield and its scalar component.
	Under a supersymmetry transformation with anticommuting parameter $\epsilon$, the component fields transform as
\bea
\delta \phi &= &\sqrt{2} \epsilon \psi, 
\label{atransform} \\
\delta \psi &= &\sqrt{2} \epsilon F + \sqrt{2} i
\sigma^{\mu} \bar \epsilon \partial_{\mu} \phi, \label{psitransform} \\
\delta F &= &-\sqrt{2}i \partial_{\mu} \psi \sigma^{\mu} \bar
\epsilon \label{ftransform}\,.
\eea
On the other hand, the vector superfields in the Wess-Zumino gauge, for the simplest case of an abelian group U(1), can be written as
\be
\mathbf{V}=-\theta \sigma^{\mu} \bar \theta
A_{\mu} + i \theta^2 \bar \theta \bar \lambda
-i \bar \theta^2 \theta \lambda + \frac{1}{2}
\theta^2 \bar \theta^2 D\,,
\ee
where $\bar\lambda$ is the complex conjugate of the two-component Weyl spinor $\lambda$.
	Here $A_{\mu}$ is the gauge field, $\lambda$ is the gaugino and $D$ is an auxiliary field.
The analog of the gauge invariant field strength is a chiral field,
\be
W_{\alpha} = -i \lambda_{\alpha}
+ \theta_{\alpha}D -\dfrac{i}{2}
(\sigma^{\mu} \bar \sigma^{\nu} \theta)_{\alpha} F_{\mu \nu}
+ \theta^2 \sigma^{\mu}_{\alpha \dot \beta} \partial_{\mu}
\bar \lambda^{\dot \beta}\,,
\ee
where $F_{\mu\nu}=\partial_\mu A_\nu-\partial_\nu A_\mu$ and $\bar{\sigma}^\mu=(-{\bf 1},-\vec{\sigma})$. 
Under the supersymmetry transformations, the gaugino $\lambda$ transforms as
\be
\delta\lambda 
= i\epsilon D + \epsilon \sigma^{\mu} \bar{\sigma}^\nu F_{\mu\nu}\,.
\label{lamtran}
\ee
	Global supersymmetry corresponds to invariance with respect to SUSY transformations with parameter $\epsilon$ independent of spacetime coordinates. In local supersymmetry (i.e. in supergravity), $\epsilon$ is instead $x_\mu$-dependent, thus we write $\epsilon(x)$. In the latter case, one has to introduce one more supermultiplet that contains the graviton and the gravitino.

	Global SUSY can be seen, with a rough estimate, as the limit of supergravity when the Planck mass goes to infinity. However, this is a good approximation only if the vacuum expectation values (\emph{vevs}) of all the relevant scalars, which have not been integrated out, are much less than $\mpl$.

	Moreover, the overall picture is not completely clear. In the true vacuum, global SUSY would predict a large positive potential $V$, while the observed value is almost zero. Supergravity instead predicts $V\approx 0$, through subtraction of an unknown large value. This generates the cosmological constant problem. The second exception deals with the limit $\mpl\rightarrow\infty$, that does not make any sense during inflation. The Planck mass is indeed a fundamental parameter during that epoch. The naive limit of an infinite Planck mass should be replaced by less immediate mechanisms. This generates a problem for model building, since supergravity theories imply in general an inflaton potential that is not sufficiently flat for inflation.

	By keeping this in mind, we now discuss the Lagrangian of a globally supersymmetric theory. The fundamental concepts of superpotential, scalar potential and $F$ and $D$ terms will be now introduced.

	We begin with the most general renormalizable Lagrangian written in superspace,
\be
{\cal L} = \sum_n\int d^4 \theta\;
\phi_n^{\dagger} e^{M} \phi_n+ 
\dfrac{1}{4 k} \int d^2 \theta\; W_{\alpha}^2
+ \int d^2 \theta\; W(\phi_n) + {\mathrm{h.c.}}
\ee
with the superpotential $W(\phi_n(x,\theta))$ and with the matrix $M=T^a\mathbf{V}^a$. A vector superfield $\mathbf{V}^a$ corresponds to each generator $T^a$ of the gauge group $G$, in the representation that is determined by the $\phi_n$. In the adjoint representation, ${\mathrm{Tr}} (T^a T^b)=k\delta^{ab}$. Consider now for simplicity the gauge theory of a single U(1), with coupling constant $g$ and charges $q_n$. The covariant derivative of a chiral superfield,
\be
D_\mu\phi_n= \partial_\mu\phi_n - i\frac{g}{2} A_\mu q_n\phi_n\,,
\ee
allows to rewrite the Lagrangian in terms of the component fields as follows: 
\bea
{\cal L} &=& \sum_n\left ( D_\mu \phi_n^* D^\mu \phi_n
+ i D_\mu \bar{\psi}_n\bar{\sigma}^{\mu} 
\psi_n+ \vert F_n\vert^2 \right )\nonumber\\
\label{componentla}
&-&\frac{1}{4} F^2_{\mu \nu} 
-i \lambda \sigma^\mu \partial_\mu 
\bar{\lambda} +\frac{1}{2}D^2 + \frac{g}{2} D\sum_n
q_n\phi_n^*\phi_n\nonumber\\
&-&\left[ i \sum_n\frac{g}{\sqrt{2}} \bar{\psi}_n\bar{\lambda} \phi _n
- \sum_{nm} \dfrac{1}{2}\frac{\partial^2 W}{\partial \phi_n
\partial \phi_m} \psi_n\psi_m \right.
\nonumber\\
&+& \left.   
\sum_nF_n\left(\frac{\partial W}{\partial \phi_n}\right)\right]
+{\mathrm{c.c.}}
\eea
Clearly, $D$ and $F_n$ do not propagate, since they do not have kinetic terms in the above equation. This implies that they are auxiliary fields and their equations of motion are just simple constraints,
\bea
F_n&=&- \left(\frac{\partial W }{\partial \phi_n}\right)^*\,, \label{fi1} \\
D &=& -\frac{g}{2} \sum_nq_n\vert\phi_n\vert^2 \,.
\label{d1}
\eea
The above equations (\ref{fi1}) and (\ref{d1}) can then be used to extract the scalar potential from the Lagrangian Eq.(\ref{componentla}), and to separate it into the $F$ and $D$ term as $V=V_F + V_D$, where
\bea
V_F &\equiv& \sum_n|F_n|^2, \\
V_D &\equiv& \half D^2\,.
\eea
In any renormalizable theory, $W$ is at most cubic in the fields (thus the potential is quartic), though in several cases higher order terms are admitted. We will see below that the above splitting is extremely important in any generic SUSY theory, including supergravity.

Terms like $(1/2)m\phi_1^2$ in $W$ are usually forbidden in theories which arise from the Lagrangian (\ref{componentla})\footnote{With the only exception of the $\mu$ term $\mu H_uH_d$, that gives mass to the Higgs doublet.}. This means that the masses in the observable sector come only from SUSY breaking and from the spontaneous breaking of an internal symmetry.

The famous Fayet-Iliopoulos term \cite{FI}
\be
-2\xi \int d^4 \theta ~V\,,
\label{dterm}
\ee
corresponds to adding a contribution $-\xi$ to the $D$ field. Eq.(\ref{d1}) thus becomes
\be
D= - \frac{g}{2}\sum_n q_n|\phi_n|^2-\xi\,.
\ee
The $D$ term is accordingly modified as
\be
V_D = \half\left(\frac{g}{2}\sum_n q_n |\phi_n|^2+\xi\right)^2\,.
\ee
If $\xi$ and the charges are redefined so that
\be
V_D = \half g^2 \left( \sum_n q_n |\phi_n|^2 + \xi \right)^2 \,,
\ee
we obtain
\be
D =  - g \left( \sum_n q_n |\phi_n|^2 + \xi \right) \,.
\label{ddef}
\ee
The additional term $\xi$ is usually related to internal symmetries, and it can appear in the theory from the very beginning. For instance, weakly coupled string theories allow anomalous U(1) symmetries which provide with terms such as (\ref{dterm}) in the Lagrangian.

\subsection{Global supersymmetry breaking}

	If supersymmetric models aim to be phenomenologically realistic, supersymmetry must be broken. In fact, we did not observe so far any superpartner of the particles we could detect at colliders. This means that the ordinary particles and their superpartners are not degenerate in mass.
The breaking of global supersymmetry may occur either explicitly (i.e., by adding soft breaking terms in the Lagrangian) or spontaneously.

\subsubsection{Soft SUSY breaking}

	Spontaneous breaking of supersymmetry can be done in several ways, and is still object of discussion. The main idea behind the explicit, or "soft" breaking is instead very simple: since in the effective theory at low energy (below $1\T$) SUSY is broken explicitly, we just add to the effective Lagrangian density some SUSY-breaking terms \cite{Martin}. We do not worry about their origin. However, the choice is somehow constrained, since the naturalness of the theory admits only certain soft breaking parameters.

	The couplings must be "soft", namely they have positive mass dimension to ensure a natural hierarchy between the electroweak and the Planck scales. These can typically be the scalar and gaugino masses or the cubic terms in the scalar field potential ($A^{ijk}\phi_i\phi_j\phi_k$, with couplings $A^{ijk}$). Soft quartic terms vanish in a flat direction of unbroken SUSY. In fact, phenomenology requires the couplings and the scalar and gaugino masses to be smaller than $1\T$ \cite{HK}.

	The squark and slepton masses come almost entirely from the supersymmetry breaking (with the exception of the stop). They should be $\gsim 100 \G$, since they have never been observed experimentally. Such considerations give for the masses of the scalars $\widetilde{m}$ an expected range between $100\G$ and $1\T$. The possible soft SUSY breaking terms in the Lagrangian are \cite{Martin}
\be
\mathcal{L}_{soft}=-\left (
\half M_a\, \lambda^a\lambda^a + \dfrac{1}{6} A^{ijk} \phi_i\phi_j\phi_k + \half M^{ij} \phi_i\phi_j + t^i \phi_i \right )
+ \mathrm{c.c.} 
- (m^2)_j^i \phi^{j*} \phi_i \,.
\label{soft}
\ee
	We recognise here the gaugino masses $M_a$ for each gauge group, the scalar (squared) mass terms $M^{ij}$ and $(m^2)_j^i$, and the tadpole coupling $t^i$. The latter is associated with a gauge singlet, and as such it is absent from the MSSM. In Ref.\cite{soft} it is shown that the above soft breaking terms do not generate any quadratic divergences from quantum corrections to the scalar masses, to all orders in perturbation theory.

	The terms in Eq.(\ref{soft}) break supersymmetry because they contain only scalars and gauginos, not their respective superpartners. Thus they give masses to all the scalars and gauginos of the theory, even though the gauge bosons and the matter fermions are massless or very light. Gauge symmetry allows both $M_a$ and $(m^2)_j^i$. The other couplings instead are allowed by gauge invariance only if there is a corresponding term in the superpotential.

	The soft breaking terms are put in the Lagrangian by hand. As already remarked, we do not know how they are generated unless we invoke the spontaneous supersymmetry breaking.
Generally speaking, the effective theory, which is valid only below the breaking scale, describes uniquely the \emph{observable sector}, which contains the fields of the Standard Model and their supersymmetric partners.

	In the full theory the spontaneous SUSY breaking occurs in a \emph{hidden sector}, where the fields do not obey to SM gauge interactions.
The effective theory in the observable (or \emph{visible}) sector is obtained when the hidden sector is integrated out. This happens when the scalar field potential, at fixed values of the fields which remain in the action, is minimised with respect to the hidden sector fields. This gives a well-defined theory if their masses are much larger than those of the fields that are not integrated out\footnote{In fact, in this case their motion about the minimum of the scalar potential is negligible.} (or if the two sets are coupled only gravitationally), as it is discussed in \cite{LR} and in \cite{Nilles}. 

 How the SUSY breaking is mediated from the hidden to the observable sector defines different classes of theories. If this happens through gravity, the theory is called "gravity-mediated". On the other hand, if the mediation occurs through gauge interactions, we talk about "gauge mediation".

\subsubsection{Fayet-Iliopoulos supersymmetry breaking: the $D$-term}

	We now investigate the  case in which supersymmetry is spontaneously broken. This means that there is still invariance (or covariance) of the Lagrangian under SUSY transformations, but there is no invariance of the vacuum. More into detail, the generators $Q_\alpha$ do not annihilate the vacuum. They produce instead a chiral fermion $\psi_\alpha$ or a gaugino $\lambda_\alpha$. This happens if the \emph{vev} for $\{Q_\alpha,\psi_\beta\}$ or $\{Q_\alpha,\lambda_\beta\}$ is non-zero. The transformation rules of the chiral fermion (\ref{psitransform}) and of the gaugino (\ref{lamtran}) define these quantities. SUSY is spontaneously broken if at least one of the auxiliary fields $F_n$ or $D$ has a non-vanishing \emph{vev}. In the true vacuum, the scale of global supersymmetry breaking is defined as
\be
\mu^4= \sum_n |F_n|^2 + \half D^2=V.
\ee
	Spontaneous supersymmetry breaking can occur either at tree-level or dynamically (namely, due to quantum corrections). In any case, the mass spectrum is modified, since the symmetry between scalar and fermion masses is lost \cite{Martin,Dine}. In the simplest case, $D=0$ and there is just one $F_n$. This is the \OR model, that we will discuss in the next subsection. Here we focus on the alternative $D\neq 0$: the $D$ term drives the breaking mechanism. This corresponds to the "Fayet-Iliopoulos model".

Consider a non-zero $D$-term in the SUSY Lagrangian, which we write as \cite{Martin}
\be
\mathcal{L}_{FI}=-kD\,.
\label{FI}
\ee
The dimension of the constant $k$ is $[mass]^2$. Since this term is linear, we expect $D$ to get a non vanishing \emph{vev}. By recalling Eq.(\ref{ddef}), we can write the relevant part of the scalar potential as
\be
V = k D -\half D^2 - g D \sum_n q_n |\phi_n|^2\,.
\ee
This gives the equation of motion
\be
D=k - g \sum_n q_n |\phi_n|^2\,,
\label{Deom}
\ee
therefore if the scalar fields $\phi_n$ (which are charged under the U(1) group previously introduced) all have non-zero masses $m_n$, the scalar potential becomes
\be
V  = \sum_n |m_n|^2 |\phi_n|^2 +\half (k -g \sum_n q_n |\phi_n|^2)^2\neq 0\,.
\ee
	Since the minimum occurs for non vanishing $D$, there is spontaneous supersymmetry breaking. The scalars have squared masses $(|m_n|^2-gq_nk)$, whilst for their fermion partners we have $|m_n|^2$.

	It seems that the above mechanism provides spontaneous SUSY breaking at the tree level in a rather simple fashion. Anyway, there is a fundamental problem with the Fayet-Iliopoulos term. Up to now we have considered only an Abelian group. For non-Abelian Yang Mills theories, a term like (\ref{FI}) would break gauge invariance. It follows that only $D$-terms in U(1) can be used, which is bad for phenomenology. In the MSSM, in fact, the squarks and sleptons do not have mass terms in the superpotential, so they would get non-zero \emph{vevs} in order to make the equation of motion (\ref{Deom}) vanish. This would break colour symmetry, but not supersymmetry. Accordingly, a F-I term must be subdominant with respect to other mechanisms of SUSY breaking, if not totally absent. 

	Therefore we can conclude this discussion by saying that, even though the $D$-term breaking has not been completely ruled out nowadays, it does not seem to be efficient in giving the right masses to the MSSM particle spectrum (and in particular, to the gauginos). The $F$-term, or \OR model, gives instead a more transparent and less problematic phenomenology.

\subsubsection{\OR supersymmetry breaking: the $F$-term\label{section:OR}}

	If only the $F$-term is involved, the corresponding models of tree-level SSSB are called of the \OR type \cite{OR}. The SUSY breaking is due to a non-zero \emph{vev} of the $F$-term. The superpotential is chosen so that the equations $F_i=0$ have no simultaneous solutions and the scalar potential $V=\sum_i{|F_i|^2}$ is positive at its minimum.
More into details, let us consider the set of three chiral supermultiplets ($S,X,Y$) and the superpotential
\be
W = - \mu^2 S + \lambda S X^2 + m XY\,,
\ee
which we will discuss further in Chapter \ref{chapt:Infl}. The key element in $W$ is the term $- \mu^2 S$, which is required in renormalizable theories with tree-level SUSY breaking, if the superpotential contains only polynomial terms \cite{GR}. In fact, since $\lambda$ and $m$ can be assumed to be positive with no loss of generality, the term $-\mu^2 S$ implies the equations
\be
F_S=\dfrac{\partial W}{\partial S}=-\mu^2+\lambda X^2=0\,, \qquad F_Y=\dfrac{\partial W}{\partial Y} =mX=0\,,
\ee
which are incompatible. The equation for $X$,
\be
F_X=\dfrac{\partial W}{\partial X} = 2 \lambda SX + m Y =0\,,
\ee
is nothing more than a constraint on the fields. Therefore supersymmetry is broken, and the scalar potential is the following:
\be
V(S,X,Y) = |-\mu^2+\lambda X^2|^2+m^2|X|^2+|2 \lambda S X + m Y|^2\,.
\ee
The absolute minimum of $V$ is at $X=Y=0$, with $S$ left undetermined. This means that at the minimum
\be
F_S=-\mu^2 \Rightarrow V=\mu^4\,.
\ee
	Thus $\mu$ is the SUSY breaking scale at the tree level, and $S$ is a \emph{flat direction} of the scalar potential, namely a subset in the scalar field space along which the potential vanishes (by rescaling). In the MSSM, around 300 such fields have been catalogued \cite{Gherghetta}. Flat directions are interesting also in the context of cosmology, since they might explain the origin of the inflaton \cite{Anupam}.

	Let us now consider the particle spectrum of the \OR model. If $V$ is expanded around $S=0$, there are 6 real scalars with squared masses
\be
0,\; 0,\;m^2,\;m^2,\;m^2-2\lambda\mu^2,\;m^2+2\lambda\mu^2\,,
\ee
where the 0 eigenvalues correspond to the complex scalar $\phi_S$, which is the scalar component of the chiral superfield $S$:
\be
S(x,\theta)=\phi_S(x)+\sqrt{2}\theta\psi_S(x)+\theta^2F_S(x)\,.
\ee
The fermionic sector contains three Weyl fermions with squared masses
\be
0,\;m^2,\;m^2\,,
\ee
thus the fermionic partner $\psi_S$ of the massless scalar $\phi_S$ is massless as well. It is anyway important to remark that $\psi_S$ is exactly massless, since it is the goldstino. Instead, the scalar is massless because of the flat direction. However, quantum corrections remove ("lift") the flat direction, and generate the one-loop Coleman-Weinberg potential which provides the scalar $\phi_S$ with a mass
\be
m_{\phi_S}^2 = \dfrac{1}{4\pi^2} \lambda^2 m^2 \left[
\ln (1 - r^2) - 1 + \frac{1}{2} \left ( r + \dfrac{1}{r} \right )
\ln \left ( \frac{1 + r}{1-r} \right )
\right ],
\ee
where $r=2\lambda\mu^2/m^2$ (this mechanism will be considered explicitly in Chapter \ref{chapt:Infl}). We see from the above equations that the two sectors are not degenerate, as it follows from the spontaneous breaking of supersymmetry. It is straightforward to check that the general supertrace formula, written as
\be
\sum (-1)^F m_\alpha^2 =0,
\label{supertrace}
\ee
where $(-1)^F$ is $1$ for bosons and $-1$ for fermions, is verified in this model. This is a general result of renormalizable theories at tree level.
This simple model seems already satisfactory, but there is a free parameter, namely the SUSY breaking scale $\mu$, which needs to be fixed by hand. Crucially, $\mu$ must be much smaller than $\mpl$ to be consistent with the phenomenology of the weak interaction. The scales of the soft terms are generated naturally only in theories with dynamical supersymmetry breaking, where they appear by virtue of dimensional transmutation \cite{GR,dynsssb}. Such a mechanism will not be addressed in this work.

\subsection{The general supergravity Lagrangian}

	So far we have considered only local supersymmetry. The SUSY algebra and soft, $D$-term and $F$-term breaking have been defined and discussed. However, the gravitino, which is the main topic of this thesis, does not exist until one introduces supergravity. This is what we are going to do in this section.

	As it is well known, gravity is a non-renormalizable field theory. The same holds for supergravity, which by construction is induced by gravity. As such, it is regarded as an effective theory, that is valid only below a certain ultraviolet cutoff $\Lambda_{UV}$. In the case at hand, this scale is the (reduced) Planck mass $\Mpl=2.44\times 10^{18}\G$.

	The supersymmetry transformations, which are defined by Eqs.(\ref{atransform}), (\ref{psitransform}) and (\ref{ftransform}), still remain valid in supergravity. The difference is found in the Lagrangian, since in addition to the superpotential $W$, we now define two more functions: the \K potential $K$ and the gauge kinetic function $f$. $W$ and $f$ are holomorphic in the complex scalar fields, but $K$ is not. In this context, the only physically meaningful object is the so-called \K function
\be
G(\phi,\phi^*)=\frac{K(\phi,\phi^*)}{\Mpl^2}+\ln{\frac{|W(\phi)|^2}{\Mpl^6}}\,.
\ee
	We stress that within supergravity, $G$ and $f$ are completely arbitrary. The choices which are made in the literature are indeed based on purely phenomenological grounds.

	Before writing down the Lagrangian, let us consider the above three fundamental functions, namely the \K potential $K$, the superpotential $W$ and the gauge kinetic function $f$. The \K potential is a real-valued function $K(\phi,\phi^*)$ in the chiral scalar fields, with mass dimension two. Since it is not a holomorphic function, it is not strongly constrained by symmetries. It determines the kinetic terms of the scalar fields as follows,
\be
\mathcal{L}_{kin}=(\partial_\mu\phi^{*i})g_{i^*j}(\partial^\mu\phi^{*i}),
\ee
where we have introduced the \K metric $g$, whose components are defined as
 \be
    g_{ij^*} = \frac{\partial^2 K}{\partial\phi^i\partial\phi^{*j}} \equiv \partial_i \partial_{j*} K,
  \ee
  along with its inverse $g^{ij^*}$, namely $g^{ij^*}g_{j^*k}=\delta^i_k $. The \K connection is given by
  \be
    \Gamma^k_{ij}=g^{kl^*}\frac{\partial g_{jl^*}}{\partial\phi^i}\, .
  \ee
In the case of canonical \K potential, the scalar fields are canonically normalized at the origin. This corresponds to $g_{ij^*}=\delta_{ij^*}$, and $K$ can be written in the following generic form,
  \be
    K=K_h (h,h^*) + \alpha_{ij}(h,h^*)\phi^i\phi^{*i} + 
    \left[ Z_{ij}(h,h^*)\phi^i\phi^j + h.c.\right]\, .
  \ee
  The hidden part of $K$ that is independent of observable fields has been labelled with $K_h$. The above definition, along with the \K metric, defines an interesting topology which is based on the concept of \K manifold. The above metric and connection define indeed the covariant derivatives of the superpotential,
\bea
    &&D_iW= W_i + \Mpl^{-2} K_iW\,,\\
    &&D_iD_jW = W_{ij} + \Mpl^{-2}
    (K_{ij}W+K_iD_jW+K_jD_iW)-\Gamma^k_{ij}D_kW+\mathcal{O}(\Mpl^{-3})\, ,\nonumber
\\
  \eea
  with the compact notation $K_i=\partial_i K$, $W_i =\partial_i W$, $K_{ij}  =\partial_i\partial_j K$, and $W_{ij} =\partial_i\partial_j W$.

The next object we consider is the superpotential $W$. As already discussed, it is an analytic function in the chiral scalar fields, and it has mass dimension three:
  \be
    W=W_h(h)+\frac{1}{2}\mu_{i j }(h)\phi^i \phi^j
    +\frac{1}{6} y_{ijk}(h) \phi^i \phi^j \phi^k\,.
  \ee
In contrast with $K$, the superpotential is holomorphic, thus it must be consistent with gauge invariance. Here we have left the superpotential of the hidden sector $W_h(h)$ implicit. The simplest option for $W$ is a direct sum of the superpotentials of the hidden and observable sectors,
  \be
    W = W_h(h) + W_{obs}(\phi)\,,
  \ee
even though in theories with dynamical SUSY breaking, a mixing of the two sectors through a hidden field $\mu_{i j }(h)$ is generally required \cite{GR}.

	The gauge kinetic function $f$ determines the kinetic terms of the gauge fields and of the corresponding gauginos. Since it is holomorphic, internal symmetries constrain its form exactly as it happens with the superpotential.
$f$ is dimensionless in the scalar fields $\phi$, and it multiplies the kinetic term for the vector supermultiplet. This means that the derivatives of $f_{ab}(\phi)$ with respect to the scalars~$\phi^i$,
  \bea
    \partial_{i} f_{ab} \equiv \frac{\partial f_{ab}} { \partial\phi^i}\, ,
  \eea
have negative mass dimension. This is relevant in theories of spontaneous SUSY breaking, as in certain models $  \partial_{i} f_{ab} = \mathcal{O} (\Mpl^{ -1 })$. Accordingly, terms in the supergravity Lagrangian (\ref{sugral}) which are proportional to $(\Mpl^{-1}\partial_i f_{ab})$ are $\mathcal{O}(\Mpl^{-2})$, thus subdominant.

	The general supergravity Lagrangian has a number of symmetries. It is invariant under the following transformations, where $F(\phi)$ is an arbitrary holomorphic function,
\bea
     && K(\phi,\phi^*)\rightarrow K(\phi,\phi^*) + F(\phi)+F^*(\phi^*)\,,\label{Ktra}\\
      && W\rightarrow e^{-F(\phi)/{\Mpl^2}}W\,,
\eea
provided the following $F$-dependent Weyl rotations of the spinor fields:
\bea
  \chi_L^i &\rightarrow & e^{\frac{i}{2}\textrm{Im}F/{\Mpl^2}\,\gamma_5}\chi_L^i  \,,\\
  \lambda^a &\rightarrow &e^{-\frac{i}{2}\textrm{Im}F/{\Mpl^2}\,\gamma_5}\lambda^a  \,,\\
  \psi_\mu &\rightarrow &e^{-\frac{i}{2}\textrm{Im}F/{\Mpl^2}\,\gamma_5}\psi_\mu\,.
\eea
Eq.(\ref{Ktra}) leaves the metric $g_{ij^*}$ and the \K connection $\Gamma^k_{ij}$ unchanged. The isometries of the \K manifold are described by the real Killing potentials, which we call $D_{a}(\phi,\phi^*)$. These are found by solving the corresponding Killing equation in the Killing vectors,
    \bea
     && X^{i}_a  = -i g^{ij*} \partial_{j*} D_a \, ,\\
    &&  X^{*j}_a = i  g^{ij*} \partial_{i} D_a \, , 
    \eea
which obey the field transformation
\bea
&&\phi^i \rightarrow \phi^{i\prime} = \phi^i + X^{ai}(\phi) \epsilon\,,
\\
&&\phi^{i*} \rightarrow \phi^{i*\prime} = \phi^{i*} + X^{*ai}(\phi^*) \epsilon\,,
\eea
with the infinitesimal parameter $\epsilon$. The Killing equations
\bea
&&\nabla_i X^a_j + \nabla_j X^a_i = 0\,,
\label{kill1}
\\
&&\nabla_i X^a_{j^*} + \nabla_{j^*} X^a_i = 0\,,
\label{kill2}
\eea
follow from the explicit form of the Lie derivative $\mathcal{L}_X$ of the (K\"ahler) metric $g_{ij^*}$:
\bea
\mathcal{L}_X g_{\tilde{i}\tilde{j}} &\equiv&
X^{a\tilde{k}} \frac{\partial}{\partial \phi^{\tilde{k}}}
g_{\tilde{i}\tilde{j}} +
g_{\tilde{i}\tilde{k}} \frac{\partial}{\partial \phi^{\tilde{j}}}
X^{a\tilde{k}} +
g_{\tilde{j}\tilde{k}} \frac{\partial}{\partial \phi^{\tilde{i}}}
X^{a\tilde{k}}
\nonumber \\
&=&
\nabla_{\tilde{i}} X^a_{\tilde{j}} + \nabla_{\tilde{j}} X^a_{\tilde{i}}= 0\,.
\eea
In the above equations, $X^a_{\tilde{i}}\equiv g_{\tilde{i}\tilde{j}} X^{a\tilde{j}}$ and the index~$\tilde{i}$ represents both $i$ and
$i^*$. The covariant derivative $\nabla$ of a generic "vector" $A_i$ has the generic form
\be
\nabla_i A_j\equiv \frac{\partial}{\partial\phi^i}A_j-\Gamma^k_{ij}A_k\,.
\ee
Eq.(\ref{kill1}) is identically satisfied, while (\ref{kill2}) allows to express the Killing vectors $X^{ai}(\phi)$ and
$X^{*ai}(\phi^*)$ as derivatives of the Killing potential $D^a$:
\bea
&&X^{ai}(\phi) = -i g^{ij^*} \frac{\partial}{\partial \phi^{*j}}
D^a\,,
 \\
&&X^{*ai}(\phi^*) = i g^{ij^*} \frac{\partial}{\partial \phi^i}
D^a\,.
\eea
This is interesting, since the above can be used to define an analytic function $F^a$ of the scalar fields $\phi$ as follows,
\be
F^a \equiv -i g^{ij^*} \frac{\partial D^a}{\partial \phi^{*j}}
\frac{\partial K}{\partial \phi^{i}} + i D^a.
\ee
This object appears in the covariant derivatives of the fields, which we will address later. For example, with a minimal K\"ahler potential $K^{min}=\phi^i\phi^{*i}$ and the generators of a gauged Lie group $T^a_{ij}$, the Killing vectors and the Killing potential are the following:
\bea
X^{ai} (\phi) &=& -i T^a_{ij} \phi^j\,,
\\
X^{*ai} (\phi^*) &=& i \phi^{*j} T^a_{ji}\,,
\\
D^a &=& \phi^{*i} T^a_{ij} \phi^j\,.
\label{Dminimal}
\eea
	By using the \K potential, the kinetic function $f$ and the superpotential $W$, the general form of the supergravity Lagrangian can be now obtained. 
As a formal remark, we point out that the following is an all-order \textit{on-shell} Lagrangian. This means that the auxiliary field of the SUGRA multiplet is eliminated via the field equations of the vielbein. In fact, the kinematical null-torsion constraint $T^a=0$, that is used in this standard approach, allows to express the spin connection $\omega_\mu^{ab}$ in function of the vielbein $e^a_\mu$ and of the gravitino $\psi_\mu$. Therefore the spin connection is no longer an independent degree of freedom \cite{Ferrara,Fre,Tesi}.

In the following equation we impose $\Mpl=1$, $f^R\equiv{\mathrm{Re}}f$ and $f^I\equiv{\mathrm{Im}}f$ and ($a,b,c,...$) are indices of the adjoint representation of the gauge group with coupling constant $g$ and structure constant $f^{abc}$. The indices are raised and lowered with $f^R_{ab}$ and its inverse. The Lagrangian is very complicated, and it contains the scalar fields $\phi$, chiral (matter) fermions $\chi$, gauge bosons $A^a_\mu$ and gauginos $\lambda$, the vielbein ${e_\mu}^a$ and the gravitino $\psi_\mu$ \cite{Ferrara,Wess,M}.

\newpage
\bea
\label{sugral}
{\cal L}_{\mathrm{SUGRA}} &=&
- \frac{1}{2} eR
+ e g_{ij^*} D_\mu \phi^i  D^\mu \phi^{*j}
%
%
- \frac{1}{2} e g^2 D_a D^a + i e g_{ij^*}
\bar{\chi}^j \bar{\sigma}^\mu D_\mu \chi^i
\nonumber \\ &&
+e \epsilon^{\mu\nu\rho\sigma}
\bar{\psi}_\mu \bar{\sigma}_\nu D_\rho \psi_\sigma
-\frac{1}{4} e f^R_{ab} F_{\mu\nu}^a F^{\mu\nu(b)}
+\frac{1}{8} e \epsilon^{\mu\nu\rho\sigma} f^I_{ab}
F_{\mu\nu}^a F_{\rho\sigma}^b
\nonumber \\ &&
+\frac{i}{2} e \left(
\lambda_a \sigma^\mu D_\mu \bar{\lambda}^a
+ \bar{\lambda}_a \bar{\sigma}^\mu
D_\mu \lambda^a
\right)
-\frac{1}{2} f^I_{ab} D_\mu
\left( e \lambda^a \sigma^\mu \bar{\lambda}^b
\right)
\nonumber \\ &&
+\sqrt{2} e g g_{ij^*} X^{*j}_a \chi^i \lambda^a
+\sqrt{2} e g g_{ij^*} X^{i}_a
\bar{\chi}^j \bar{\lambda}^a
\nonumber \\ &&
-\frac{i}{4} \sqrt{2} e g \partial_{i} f_{ab}
D^a \chi^i \lambda^b
+\frac{i}{4} \sqrt{2} e g \partial_{i^*} f^*_{ab}
D^a \bar{\chi}^i \bar{\lambda}^b
\nonumber \\ &&
-\frac{1}{4} \sqrt{2} e \partial_{i} f_{ab}
\chi^i \sigma^{\mu\nu} \lambda^a F_{\mu\nu}^b
-\frac{1}{4} \sqrt{2} e \partial_{i^*} f^*_{ab}
\bar{\chi}^i \bar{\sigma}^{\mu\nu} \bar{\lambda}^a
F_{\mu\nu}^b
\nonumber \\ &&
+\frac{1}{2} e g D_a \psi_\mu \sigma^\mu \bar{\lambda}^a
-\frac{1}{2} e g D_a
\bar{\psi}_\mu \bar{\sigma}^\mu \lambda^a
\nonumber \\ &&
-\frac{1}{2} \sqrt{2} e g_{ij^*} D_{\nu} \phi^{*j}
\chi^i \sigma^\mu \bar{\sigma}^\nu \psi_\mu
-\frac{1}{2} \sqrt{2} e g_{ij^*} D_{\nu} \phi^{i}
\bar{\chi}^j \bar{\sigma}^\mu \sigma^\nu \bar{\psi}_\mu
\nonumber \\ &&
-\frac{i}{4} e \left(
\psi_\mu \sigma^{\nu\rho} \sigma^\mu \bar{\lambda}_a
+ \bar{\psi}_\mu \bar{\sigma}^{\nu\rho}
\bar{\sigma}^\mu \lambda_a \right) \left(
F_{\nu\rho}^a + \hat{F}_{\nu\rho}^a \right)
\nonumber \\ &&
+ \frac{1}{4} e g_{ij^*} \left(
i \epsilon^{\mu\nu\rho\sigma} \psi_\mu \sigma_\nu \bar{\psi}_\rho
+ \psi_\mu \sigma^\sigma \bar{\psi}^\mu \right)
\chi^i \sigma_\sigma \bar{\chi}^i
\nonumber \\ &&
-\frac{1}{8} e \left( g_{ij^*} g_{kl^*} - 2 R_{ij^*kl^*} \right)
\chi^i \chi^k \bar{\chi}^j \bar{\chi}^l
\nonumber \\ &&
+ \frac{1}{16} e \left[ 2 g_{ij^*} f^R_{ab}
+ f^{R({cd})-1} \partial_i f_{bc} \partial_{j^*} f^*_{ad} \right]
\bar{\chi}^j \bar{\sigma}^\mu \chi^i
\bar{\lambda}^a \bar{\sigma}_\mu \lambda^b
\nonumber \\ &&
+\frac{1}{8} e \nabla_i \partial_j f_{ab}
\chi^i \chi^j \lambda^a \lambda^b
+\frac{1}{8} e \nabla_{i^*} \partial_{j^*} f^*_{ab}
\bar{\chi}^i \bar{\chi}^j \bar{\lambda}^a
\bar{\lambda}^b
\nonumber \\ &&
+\frac{1}{16} e f^{R({cd})-1} \partial_{i} f_{ac} \partial_{j} f_{bd}
\chi^i \lambda^a \chi^j \lambda^b
%
%
+\frac{1}{16} e f^{R({cd})-1} \partial_{i^*}
f^*_{ac} \partial_{j^*} f^*_{bd}
\bar{\chi}^i \bar{\lambda}^a \bar{\chi}^j
\bar{\lambda}^b
\nonumber \\ &&
-\frac{1}{16} e g^{ij^*} \partial_i f_{ab} \partial_{j^*} f^*_{cd}
\lambda^a \lambda^b \bar{\lambda}^c \bar{\lambda}^d
%
%
+\frac{3}{16} e \lambda_a \sigma^\mu \bar{\lambda}^a
\lambda_b \sigma_\mu \bar{\lambda}^b
\nonumber \\ &&
+\frac{i}{4} \sqrt{2} e \partial_i f_{ab}
\left( \chi^i \sigma^{\mu\nu} \lambda^a
\psi_\mu \sigma_\nu \bar{\lambda}^b
-\frac{1}{4} \bar{\psi}_\mu \bar{\sigma}^\mu \chi^i
\lambda^a \lambda^b \right)
\nonumber \\ &&
+\frac{i}{4} \sqrt{2} e \partial_{i^*} f^*_{ab}
\left( \bar{\chi}^i \bar{\sigma}^{\mu\nu} \bar{\lambda}^a
\bar{\psi}_\mu \bar{\sigma}_\nu \lambda^b
-\frac{1}{4} \psi_\mu \sigma^\mu \bar{\chi}^i
\bar{\lambda}^a \bar{\lambda}^b \right)
\nonumber \\ &&
-e e^{K/2} \left\{
W^* \psi_\mu \sigma^{\mu\nu} \psi_\nu
+ W \bar{\psi}_\mu \bar{\sigma}^{\mu\nu} \bar{\psi}_\nu \right\}
\nonumber \\ &&
+ \frac{i}{2} \sqrt{2} e e^{K/2} \left\{
D_i W \chi^i \sigma^{\mu} \bar{\psi}_\mu
+ D_{i^*} W^* \bar{\chi}^i \bar{\sigma}^{\mu} \psi_\mu \right\}
\nonumber \\ &&
-\frac{1}{2} e e^{K/2} \left\{
D_i D_j W \chi^i \chi^j
+ D_{i^*} D_{j^*} W^* \bar{\chi}^i \bar{\chi}^j \right\}
\nonumber \\ &&
+ \frac{1}{4} e e^{K/2} g^{ij^*} \left\{
D_{j^*} W^* \partial_i f_{ab}
\lambda^a \lambda^b
+ D_i W \partial_{j^*} f^*_{ab}
\bar{\lambda}^a \bar{\lambda}^b \right\}
\nonumber \\ &&
-e e^K \left[
g^{ij^*} ( D_i W )( D_{j^*} W^*) - 3 W^* W \right].
\eea
The covariant derivatives are defined as
\bea
D_\mu \phi^i &\equiv&
\partial_\mu \phi^i
- g A_\mu^a X^i_a\,,
\\
D_\mu \chi^i &\equiv&
\partial_\mu \chi^i + \frac{1}{2} {\omega_{\mu}}^{ab}\sigma_{ab} \chi^i
+\Gamma^i_{jk} D_\mu \phi^j \chi^k
- g A_\mu^a \frac{\partial X^i_a}{\partial \phi^j}\chi^j
\nonumber \\ &&
- \frac{1}{4} \left( K_j D_\mu \phi^j -
K_{j^*} D_\mu \phi^{*j} \right) \chi^i
-\frac{i}{2} g A_\mu^a {\mathrm{Im}} F_a \chi^i\,,
\\
D_\mu \lambda^a &\equiv&
\partial_\mu \lambda^a
+ \frac{1}{2} {\omega_{\mu}}^{ab}\sigma_{ab} \lambda^a
- g f^{abc} A_\mu^b \lambda^c
\nonumber \\ &&
+ \frac{1}{4} \left( K_j D_\mu \phi^j -
K_{j^*} D_\mu \phi^{*j} \right) \lambda^a
+\frac{i}{2} g A_\mu^b {\mathrm{Im}} F_b \lambda^a\,,
\\
D_\mu \psi_\nu &\equiv&
\partial_\mu \psi_\nu
+ \frac{1}{2} {\omega_{\mu}}^{ab}\sigma_{ab} \psi_\nu
\nonumber \\ &&
+ \frac{1}{4} \left( K_j D_\mu \phi^j -
K_{j^*} D_\mu \phi^{*j} \right) \psi_\nu
+\frac{i}{2} g A_\mu^b {\mathrm{Im}} F_b \psi_\nu\,.
\eea
$F_{\mu\nu}^a$ is the field strength (or curl) of the gauge field $A_\mu^a$, and $\hat{F}_{\mu\nu}^a$ is defined as
\be
\hat{F}_{\mu\nu}^a \equiv
F_{\mu\nu}^a - \frac{i}{2} \left(
\psi_\mu \sigma_\nu \bar{\lambda}^a
+ \bar{\psi}_\mu \bar{\sigma}_\nu \lambda^a
+ \psi_\nu \sigma_\mu \bar{\lambda}^a
+ \bar{\psi}_\nu \bar{\sigma}_\mu \lambda^a \right)\,.
\ee

	The supergravity Lagrangian (\ref{sugral}) is invariant under local supersymmetry
transformations. These are parametrised by an anticommuting Majorana spinor $\xi$ of mass dimension $-1/2$, that will be written here as $\xi(x)$, with abuse of notation. For the bosons (the vielbein $e_\mu^a$, the scalars $\phi^i$ and the gauge fields $A_\mu^a$), we have
\bea
\delta {e_\mu}^a &=& - i \left( \xi \sigma^a \bar{\psi}_\mu
+ \bar{\xi} \bar{\sigma}^a \psi_\mu \right)\,,
\\
\delta \phi^i &=& \sqrt{2} \xi \chi^i\,,
\\
\delta A_\mu^a &=& i \left( \xi \sigma_\mu \bar{\lambda}^a
+ \bar{\xi} \bar{\sigma}_\mu \lambda^a \right)\,.
\eea
For the matter fermions $\chi^i$, the transformation rule is the following,
\bea
\delta \chi^i &=&
i \sqrt{2} \sigma^\mu
\left( D_\mu \phi^i - \frac{1}{\sqrt{2}} \psi_\mu \chi^i \right)
- \Gamma^i_{jk} \delta \phi^j \chi^k
+ \frac{1}{4} \left( K_j \delta\phi^j - K_{j^*} \delta\phi^{j^*} \right)
\chi^i -
\nonumber \\ &&
- \sqrt{2} F^i  \xi+ \frac{1}{2\sqrt{2}} \xi g^{ij^*} \partial_{j^*} f^*_{(ab)}
\bar{\lambda}^a \bar{\lambda}^b\,,
\eea
and for the gauginos $\lambda^a$ one finds
\bea
\delta \lambda^a &=&
\hat{F}_{\mu\nu}^a \sigma^{\mu\nu} \xi
- \frac{1}{4} \left( K_j \delta\phi^j - K_{j^*} \delta\phi^{j^*} \right)
\lambda^a
- i g D^a \xi +
\nonumber \\ &&
+ \frac{1}{2\sqrt{2}} \xi f^{R(ab)-1} \partial_i f_{(bc)}
\chi^i \lambda^c
- \frac{1}{2\sqrt{2}} \xi f^{R(ab)-1} \partial_{i^*} f_{(bc)}^*
\bar{\chi}^i \bar{\lambda}^c\,.
\eea
Finally, for the gravitino $\psi_\mu$ we obtain
\bea
\delta \psi_\mu &=&
2 D_\mu \xi
- \frac{i}{2} \sigma_{\mu\nu} \xi
g_{ij^*} \chi^i \sigma^\nu \bar{\chi}^j
+ \frac{i}{2} \left( {e_\mu}^a e_{\nu a} + \sigma_{\mu\nu} \right) \xi
\lambda_a \sigma^\nu \bar{\lambda}^a -
\nonumber \\ &&
- \frac{1}{4} \left( K_j \delta\phi^j - K_{j^*} \delta\phi^{j^*} \right)
\psi_\mu
+ i e^{K/2} W \sigma_\mu \bar{\xi}\,.
\eea
The auxiliary field $F^i$, that plays a crucial role in global SUSY, here appears by virtue of the following definition:
\be
F^i \equiv e^{K/2} g^{ij^*} D_{j^*} W^*\,.
\ee

	Let us now focus on the particle content of the Lagrangian, that consists of three distinct sectors. Matter fermions are described in terms of left-handed four-spinors
  \be
     \chi_L^i = 
     \begin{pmatrix}
         \chi_{\alpha}\\ 0 
     \end{pmatrix}\,,
   \ee
   with the two-component Weyl spinor $\chi_{\alpha}$. The scalar superpartners are denoted as $\phi^i$, and the multiplet is in the fundamental representation of the gauge group.
The gauge bosons $A^a_\mu$ and the gauginos
  \be
    \lambda^a = i \begin{pmatrix}
      -  l^{a}_{\hphantom{a\,}\alpha}  \\
      \hphantom{-}  \bar{l}^{a\,\dot\alpha} 
     \end{pmatrix}\,,
  \ee
namely Majorana fields which are their superpartners, constitute the gauge supermultiplet, in the adjoint representation of the gauge group $\mathcal{G}$. The indices $a,b,...$ run accordingly. Since SUGRA is a Yang-Mills theory\footnote{N=1 supergravity is indeed the gauge theory of the Poincar\' e supergroup ISO(4,1).}, $F^a_{\mu\nu}$ are the associated field strengths of the $A^a_\mu$. Moreover, the auxiliary fields $D_{a}$ are the generalizations of the $D$ terms in the vector supermultiplets in global supersymmetry.

	The gravity multiplet is a bit different. The graviton (which we indicate as the vielbein $e_\mu^m$), enters the SUGRA Lagrangian via the determinant of the vielbein $e=\det e_\mu^m$. It also appears in the curvature (Ricci) scalar $R$. By recalling the fundamental relation
\be
g_{\mu\nu}=\eta_{mn}e_\mu^me_\nu^n,
\ee
we distinguish between the flat spacetime indices ($m,n,\dots$) and the Lorentz indices ($\mu, \nu,\dots$). We will consider the gravitino into details in the next section. It is a spin-$3/2$ field, which is written in terms of the Majorana vector-spinor,
  \begin{equation}
    \psi_\mu = i\begin{pmatrix} 
      - \psi_{\mu\,\alpha}  
      \\ \hphantom{-} \bar{\psi}_{\mu}^{\,\dot{\alpha}} 
     \end{pmatrix}\,.
  \end{equation}
	The gravitino is massless in the limit of unbroken local SUSY. As we will see in the next section, the SUSY breaking provides it with a (\emph{vev}-dependent) mass, through the super-Higgs mechanism (see for instance Ref.\cite{M}).

\subsection{Spontaneous breaking of local SUSY}

	In contrast to global SUSY, supergravity can be broken only spontaneously, not explicitly. However, the condition for the breaking is still a non-vanishing \emph{vev} of at least one of the auxiliary fields $D$ and $F^n$. The similarity with the spontaneous gauge symmetry breaking of the Standard Model is however less evident in this case. The \emph{vevs} of the $D$ and $F^n$ can indeed get contributions not only from scalar fields, but also from fermion condensation (for instance, from gaugino condensation).

	The scalar potential at the tree level is given by
\be
V=V_D+V_F\,,
\ee
similarly to the case of global SUSY. However, here $V_F$ is not only determined by the $F$-terms, because the \K potential makes the difference,
\be
V_F = e^{K/\Mpl^2} \left[ \left( W_n+\dfrac{1}{\Mpl^2}WK_n \right) g^{m^* n}\left( W_m+\dfrac{1}{\Mpl^2}WK_m\right)^* - 3\dfrac{|W|^2}{\Mpl^2}  \right]\,.
\ee
	The ultraviolet cutoff $\Lambda_{UV}$ can be now recalled and combined with the reduced Planck mass $\Mpl=2.43\times 10^{18}\G$. Taking $\Mpl$ to infinity and keeping the cutoff fixed, we have
\be
V_F=W_ng^{m^*n}(W_m)^*\,.
\ee
Namely, we have obtained non-renormalizable global supersymmetry. Renormalizable global SUSY is recovered if also $\Lambda_{UV}$ goes to infinity. If instead $\Lambda_{UV}\rightarrow\infty$ and $\Mpl$ stays fixed, the theory corresponds to minimal supergravity, which contains only the graviton $e^a_\mu$ and the gravitino $\psi_\mu$. In such a model, the total Lagrangian is the sum of the Hilbert-Einstein and Rarita-Schwinger Lagrangians (the latter is basically the kinetic term of the gravitino field):
\be
\mathcal{L}=\mathcal{L}_{HE}+\mathcal{L}_{RS}=-\frac{\Mpl^2}{2}eR+
e \epsilon^{\mu\nu\rho\sigma}
\bar{\psi}_\mu \bar{\sigma}_\nu D_\rho \psi_\sigma\,.
\ee
$\mathcal{L}_{RS}$ ensures invariance of the action under \emph{local} SUSY transformations, and it is for this reason that the gravitino is the gauge field of local supersymmetry. Usually we consider $\Lambda_{UV}=\Mpl$, as in the Lagrangian (\ref{sugral}). Global supersymmetry is recovered when the cutoff (namely, the reduced Plank mass) goes to infinity.

	We now study into detail the spontaneous breaking of supergravity, following \cite{Nilles} and \cite{M}. As already discussed in the previous sections, supersymmetry is broken by \emph{vevs} acquired by fields in the hidden sector, which are gauge singlets. This generates soft breaking terms of order
\be
m_{soft}\approx \frac{\left\langle F\right\rangle}{\Mpl}\,,
\ee
which vanish in the limit of exact supersymmetry. That is, when the \emph{vev} of the auxiliary field approaches zero, $\left\langle F\right\rangle\rightarrow 0$.

	It is very important to remark that the different ways of breaking SUSY (through gravity, gauge interactions,...) do not change the structure of these parameters. What is different in each model is their \emph{size}, which determines the mass spectrum of the various particle multiplets. Thus the Lagrangian (\ref{sugral}) is completely general. Accordingly, this occurs also for the interaction terms and the Feynman rules which are used throughout this thesis.

	After that global SUSY is spontaneously broken, a massless Nanbu-Goldstone fermion called the goldstino is generated. In local supersymmetry, this component is absorbed by the gravitino, which obtains a mass (super-Higgs mechanism). The gravitino mass is therefore directly related to the SUSY breaking scale. Let us refer once again to the general supergravity Lagrangian Eq.(\ref{sugral}). In the following, we will consider only the minimal (or canonical) \K potential and the kinetic function at the lowest order, namely
\be
K=\sum_i \phi^i\phi^{*i}+\dots,\qquad f_{(ab)}=\delta_{ab}+\dots
\label{minimkah}
\ee
which is a good enough approximation for the purposes of this thesis. The dots in the above equation label higher order terms in the inverse reduced Planck mass $\Mpl$, which describe subdominant interactions. We remind that $K$ and $f$ generate respectively the kinetic terms of the chiral and of the vector multiplets. The choice (\ref{minimkah}) of a canonical \K potential means that $g_{ij^*}=\delta_{ij^*}+\dots$, namely it defines a flat \K manifold. Extracting from Eq.(\ref{sugral}) the quadratic fermion terms without derivatives \cite{M}\,,
\bea
{\cal L}_{\mathrm{F}}^{(2)} &=&
- e^{G/2}
\left( \psi_\mu \sigma^{\mu\nu} \psi_\nu
+ \bar{\psi}_\mu \bar{\sigma}^{\mu\nu} \bar{\psi}_\nu \right)
+\dfrac{1}{2} g D^a \psi_\mu \sigma^{\mu} \bar{\lambda}^a
-\dfrac{1}{2} g D^a \bar{\psi}_\mu \bar{\sigma}^{\mu} \lambda^a-
\nonumber \\ &&
- e^{G/2}
\left( \dfrac{i}{\sqrt{2}} G_i \chi^i \sigma^\mu \bar{\psi}_\mu
+ \dfrac{i}{\sqrt{2}} G_{i^*} \bar{\chi}^i \bar{\sigma}^\mu \psi_\mu \right)
-e^{G/2}
\left\{ \dfrac{1}{2} \left( G_{ij} + G_i G_j \right) \chi^i \chi^j+\right. \nonumber \\
&&
\left.
+ \dfrac{1}{2} \left( G_{i^*j^*} + G_{i^*} G_{j^*} \right)
\bar{\chi}^i \bar{\chi}^j \right\}
+ \sqrt{2} g \left(
-i D^a_i \chi^i \lambda^a
+i D^a_{i^*} \bar{\chi}^i \bar{\lambda}^a \right),
\eea
we notice mixing terms of the gravitino with either the matter fermion $\chi^i$ or the gaugino $\lambda^a$, provided that the \emph{vevs} of $e^{K/2}D_iW$ or $D^{a}$ are nonzero. It is possible to eliminate such terms by a shift of the gravitino field,
\be
\psi_\mu\longmapsto\psi_\mu+\dfrac{1}{3}\bar{\eta}\bar{\sigma}_\mu\,,
\ee
where the spinor $\eta$ is defined such that
\be
\bar{\eta} \equiv  \frac{i}{\sqrt{2}} G_{i^*} \bar{\chi}^i + \frac{1}{2} e^{-G/2} g D^a \bar{\lambda}^a\,.
\ee
The new form of the Lagrangian is
\bea
{\cal L}_{\mathrm{F}}^{(2)} &=&
-e^{G/2}
\left( \psi_\mu + \frac{1}{3} \bar{\eta} \bar{\sigma}_\mu \right)
\sigma^{\mu\nu}
\left( \psi_\nu - \frac{1}{3} \sigma_\nu \bar{\eta} \right)
%
%
-e^{G/2}
\left( \frac{1}{2} G_{i^*j^*} + \frac{1}{6} G_{i^*} G_{j^*} \right)
\bar{\chi}^i \bar{\chi}^j
\nonumber \\ &&
- \frac{1}{6} e^{-G/2}
g^2 D^a D^b \bar{\lambda}^a \bar{\lambda}^b
%
%
+ i \left(
\sqrt{2} g D^a_{i^*} - \frac{\sqrt{2}}{3} g G_{i^*}  D^a \right)
\bar{\chi}^i \bar{\lambda}^a +h.c.\nonumber\\
\eea
By restoring the explicit dependence on the Planck mass, we accordingly find the gravitino mass as
\be
m_{3/2}=\Mpl e^{G/2} \equiv\dfrac{e^{K/2\Mpl^2}}{\Mpl^2}  |W|.
\label{mg}
\ee
It can be shown that the spin 1/2 fermion $\eta$ is a massless eigenstate of the fermion mass matrix in the basis $(\chi^i,\lambda^a)$, thus it is the goldstino. We have thus verified that this particle is absorbed (or "eaten") by the gravitino field $\psi_\mu$.

	We conclude this section with a discussion on the supertrace formula in supergravity, that was here introduced in the context of global SUSY, Eq.(\ref{supertrace}):
\be
{\mathrm{Str}}{\bf M}^2 \equiv
\sum_{J=0}^{3/2} (-1)^J {\mathrm{tr}}{\bf M}_J^2\,.
\ee
	Here each ${\bf M}_J^2$ is the squared mass matrix of the corresponding particle multiplet, namely the set of scalar, vector and spinor fields. The mass matrix of the scalar fields can be obtained from the second derivatives of the scalar potential $V$, in the case of a vanishing cosmological constant. The trace is then
\bea
{\mathrm{tr}}{\bf M}^2_0 &=&
 \left( - n_\phi -1 \right) g^2 D^a D^a
- \frac {1}{2} e^{-G} \left( g^2 D^a D^a \right)^2
+ 2 e^G \left( G_{ij} G_{i^*j^*} + n_\phi \right)
\nonumber \\ &&
+ 2 g^2 \left( D^a_{i} D^a_{i^*} + D^a D^a_{ii^*} \right)\,,
\label{trM0}
\eea
where $n_\phi \equiv \sum_i g_{ii^*}$ is the number of chiral multiplets \cite{M}. The mass terms for the vector bosons are derived from the covariant derivatives of the scalar fields. The trace of the mass-squared matrix is
\be
{\mathrm{tr}}{\bf M}^2_1 = 6 g^2 \phi^{*i} T^a_{ij} T^a_{jk} \phi^k = 6 g^2 D^a_{i} D^a_{i^*}\,,
\label{trM1}
\ee
where the Killing potential $D^a$ for the minimal \K potential is obtained from Eq.(\ref{Dminimal}). Finally, the super-Higgs mechanism provides with the fermion mass matrix, which gives
\be
{\mathrm{tr}}{\bf M}_{1/2}^2 =
2 e^G \left( G_{ij} G_{i^*j^*} - 1 \right)
- 2 g^2 D^a D^a
- \frac{1}{2} e^{-G} \left( g^2 D^a D^a \right)^2
+ 8 g^2 D^a_{i} D^a_{i^*}\,.
\label{trM1/2}
\ee
The only spin 3/2 particle is the gravitino, therefore one obtains
\be
{\mathrm{tr}}{\bf M}_{3/2}^2 = 4\Mpl^2 e^{G} = 4 \mg^2\,.
\label{trM3/2}
\ee
The supertrace formula in supergravity can be written by adding Eqs.(\ref{trM0}), (\ref{trM1}), (\ref{trM1/2}) and (\ref{trM3/2}):
\be
{\mathrm{Str}}{\bf M}^2 = 2 \left( n_\phi - 1 \right) \mg^2
- \left( n_\phi - 1 \right) g^2 D^a D^a
+ 2 g^2 D^a D^a_{ii^*}\,.
\ee
	Interestingly, when SUSY breaking occurs via $F$-term condensation ($D^a=0$), the above formula is totally dependent on the gravitino mass. Accordingly, for unbroken exact supersymmetry ($\mg\rightarrow0$) it gives zero, as is it generally required. The fact that in this case the scalar particles are heavier than their fermionic superpartners is also relevant for phenomenology.


\section{Gravitinos}

	According to the formalism which was developed long ago by Dirac and other greats, charged fermions correspond to Dirac spinors. The Dirac Lagrangian and the Dirac equation constitute the basis for the understanding of the Standard Model of particle physics.

With the advent of supersymmetric theories, which are progressively setting a new standard for particle physics, one has to consider also the Majorana fermions. They appear as supersymmetric partners of the SM bosons (either scalars or vectors), and as such they are represented by neutral (charge auto-conjugated) spinors. In particular, the gravitino is the SUSY partner of the spin 2 graviton. Accordingly, we begin this section by recalling some basic properties, which are used to introduce the gravitino field and perform the calculations in Chapter \ref{chapt:WW}. We remand to the Appendix for the notations which are used in this thesis.

\subsection{Majorana fermions}\label{sect:majo}

	Let us first introduce the charge conjugation operator $C$ for a scalar field $\phi$, namely
\be
\phi^c(x)=C\phi(x)=\eta_c\phi^*(x)\,,
\ee
where the phase factor is unimodular ($|\eta_c|=1$) and takes the values $\eta_c=\pm 1$. In the chiral representation,
$C$ is represented in matrix form as follows,
\be
C=-i\gamma^2\gamma^0=\left(\begin{tabular}{cc} 
$ \epsilon_{\beta\alpha}$ & 0 \\  0  & $\epsilon^{\dot{\beta}\dot{\alpha}} $\\ 
\end{tabular}\right)\,,
\ee
with the Pauli matrix $\sigma^2$ and the antisymmetric tensor in two components
\be
\epsilon^{\alpha\beta}=-\epsilon_{\alpha\beta}=i\sigma^2=\left(\begin{tabular}{cc} 
 0  & 1 \\  -1  & 0 \\ 
\end{tabular}\right)\,.
\ee
$C$ interchanges particles and antiparticles. The action on a spinor $\psi$ holds as
\be
\psi^c=C\bar{\psi}^T\,.
\ee
$\psi^c$ is called the charge conjugated spinor of $\psi$. In four-component notation, this corresponds to
\be
\psi=\left(\begin{tabular}{c} 
$ \xi_\alpha $ \\ $\bar{\eta}^{\dot{\alpha}}$\\ 
\end{tabular}\right)
\Rightarrow
\bar{\psi}^T=\left(\begin{tabular}{c}
$\eta^\alpha $\\ $ \bar{\xi}_{\dot{\alpha}} $\\ 
\end{tabular}\right)
\,.
\ee
One then obtains
\be
\psi^c=\left(\begin{tabular}{c} 
$ \eta_\beta $ \\ $\bar{\xi}^{\dot{\beta}}$\\ 
\end{tabular}\right)\,.
\ee
A four-component Majorana spinor $\psi_M$ satisfies the property \cite{Majorana}
\be
\psi_M=\psi^c=C\bar{\psi}^T\,,
\ee
therefore $\eta=\xi$. In this specific case, we obtain the following:
\be
\psi_M=\left(\begin{tabular}{c} 
$ \xi_\alpha $ \\ $ \bar{\xi}^{\dot{\alpha}} $\\ 
\end{tabular}\right)
 =
 \left(\begin{tabular}{c} 
$ \psi_L $ \\ $ i\sigma^2\psi_L^* $\\ 
\end{tabular}\right)\,. 
 \ee
The \emph{Majorana flip identities} are very useful in the calculations of scattering amplitudes. We include them below. If $\psi$ and $\lambda$ are two anticommuting Majorana spinors, they obey the following \cite{HK}:
\bea
&&\bar{\psi}\lambda=\bar{\lambda}\psi\,,\\
&&\bar{\psi}\gamma_5\lambda=\bar{\lambda}\gamma_5\psi\,,\\
&&\bar{\psi}\gamma_\mu\lambda=-\bar{\lambda}\gamma_\mu\psi\,,\\
&&\bar{\psi}\gamma_\mu\gamma_5\lambda=\bar{\lambda}\gamma_\mu\gamma_5\psi\,,\\
&&\bar{\psi}\sigma_{\mu\nu}\lambda=-\bar{\lambda}\sigma_{\mu\nu}\psi\,.\\
\eea
The above equations imply the useful formula
\be
\bar{\psi}\gamma_\mu P_L\lambda=-\bar{\lambda}\gamma_\mu P_R\psi\,.
\ee

\subsection{The free massive gravitino field}\label{gravspinsum}

	The Lagrangian for the free gravitino field\footnote{This subsection refers mainly to Moroi's PhD thesis \cite{M} and to the references quoted there.} can be read off the general SUGRA Lagrangian (\ref{sugral}).
 We remark that it is written in a generic curved space, thus the Lorentz indeces ($\mu,\nu,\dots$) in Eq.(\ref{sugral}) are \emph{not} flat. In the following, we will instead work in Minkowski space, consistently with quantization, and use \emph{flat} Lorentz indices. This is described in the Appendix, and corresponds to a $0^{th}$ order locally supersymmetric system.

 	From Eq.(\ref{sugral}), by using the formula (\ref{mg}) for the gravitino mass, one finds 
\be
{\cal L}_{\psi_\mu} = -\frac{1}{2} \epsilon^{\mu\nu\rho\sigma}
\bar{\psi}_\mu \gamma_5 \gamma_\nu \partial_\rho \psi_\sigma
- \frac{1}{4} \mg
\bar{\psi}_\mu \left[ \gamma^\mu , \gamma^\nu \right] \psi_\nu\,,
\ee
modulo a total divergence. By using the Majorana condition for the gravitino, $\psi_\mu = C\bar{\psi}_\mu^T$, the above can be rewritten as
\be
{\cal L}_{\psi_\mu} = \frac{1}{2} \epsilon^{\mu\nu\rho\sigma}
\psi^T_\mu C^{\dagger} \gamma_5 \gamma_\nu \partial_\rho \psi_\sigma
+ \frac{1}{4} \mg
\psi^T_\mu C^{\dagger} \left[ \gamma^\mu , \gamma^\nu \right] \psi_\nu\,.
\ee
By variation with respect to $\psi_\mu$, this provides with the following field equation \cite{RS}
\be
\epsilon^{\mu\nu\rho\sigma}
\gamma_5 \gamma_\nu \partial_\rho \psi_\sigma + \frac{1}{2} \mg \left[ \gamma^\mu , \gamma^\nu \right] \psi_\nu=0\,,
\label{rs}
\ee
that is called the Rarita-Schwinger equation. Operating $\partial_\mu$ on Eq.(\ref{rs}) yields
\be
\slashed{\partial} \gamma^\mu \psi_\mu - \gamma^\mu \slashed{\partial} \psi_\mu = 0\,.
\label{cond1}
\ee
Operating $\gamma_\lambda\gamma_\mu$, we obtain one more equation
\be
2 i \left( \partial_\lambda \gamma^\mu \psi_\mu - \slashed{\partial} \psi_\lambda \right)
+ \mg \left( \gamma_\lambda \gamma^\mu \psi_\mu + 2 \psi_\lambda \right) = 0\,.
\label{cond2}
\ee
Multiplication of (\ref{cond2}) by $\gamma^\lambda$ provides with a third equation,
\be
i \left( \slashed{\partial} \gamma^\mu \psi_\mu - \gamma^\mu \slashed{\partial} \psi_\mu \right) + 3\mg \gamma^\mu \psi_\mu = 0\,,
\ee
which, by taking into account (\ref{cond1}), gives the first constraint on the massive ($\mg\neq0$) gravitino field:
\be
\gamma^\mu \psi_\mu=0\,.
\label{constraint1}
\ee
The second condition follows from substitution of (\ref{constraint1}) into (\ref{cond1}),
\be
\partial^\mu\psi_\mu=0\,.
\label{constraint2}
\ee
By using (\ref{constraint1}) in (\ref{cond2}), one finally obtains the Dirac equation for each vector component of the gravitino field:
\be
\left( i \slashed{\partial} - \mg \right) \psi_\mu = 0\,.
\label{constraint3}
\ee
Equations (\ref{constraint1}), (\ref{constraint2}) and (\ref{constraint3}) can be indeed solved by constructing a particular spinor, that is composite of a wave function $u$ for a spin 1/2 Dirac field, and of a polarization vector $\epsilon_{\mu}$ for a spin 1 field. Working in momentum space, one uses the normalization condition
\be
\bar{u}({\bf p},s) u({\bf p},s^{\prime}) = 2 \mg \delta_{ss^{\prime}}\,,
\label{normspi}
\ee
and the polarization vectors
\bea
&&\epsilon_\mu({\bf p},1) = \frac{1}{\sqrt{2}}
(0, \cos\theta\cos\phi - i \sin\phi, \cos\theta\sin\phi + i \cos\phi,
-\sin\theta )\,,
\label{pol_+} \\
&&\epsilon_\mu({\bf p},0) = \frac{1}{\mg}
(|{\bf p}|, -E\sin\theta\cos\phi,
-E\sin\theta\sin\phi, -E\cos\theta)\,,
\\
&&\epsilon_\mu({\bf p},-1) = \frac{-1}{\sqrt{2}}
(0, \cos\theta\cos\phi + i \sin\phi,
\cos\theta\sin\phi - i \cos\phi, -\sin\theta )\,,\nonumber\\
\label{pol_-}
\eea
for the momentum vector $p^\mu = (E,|{\bf p}|\sin\theta\cos\phi,|{\bf p}|\sin\theta\sin\phi,|{\bf p}|\cos\theta)$ of a massive particle. Namely, $p^2=p_\mu p^\mu = \mg^2$. These are normalized as
\be
\epsilon^*_\mu({\bf p},r)\epsilon^\mu({\bf p},r^\prime)=-\delta_{rr^{\prime}}.
\label{normpol}
\ee
The polarization vectors (\ref{pol_+}) -- (\ref{pol_-}) obey the constraint
\be
p^\mu \epsilon_\mu({\bf p},r) = p^\mu \epsilon_\mu^*({\bf p},r) = 0\,.
\ee
It is straightforward to prove that the wave function for the gravitino, as it has been constructed above, satisfies the following equations:
\bea
\gamma^\mu \widetilde{\psi}_\mu ({\bf p},\lambda) &=& 0\,,
\\
p^\mu \widetilde{\psi}_\mu ({\bf p},\lambda) &=& 0\,,
\\
\left( \slashed{p} - \mg \right) \widetilde{\psi}_\mu ({\bf p},\lambda)
&=& 0\,.
\eea
Here $\widetilde{\psi}_\mu ({\bf p},\lambda)$ is the gravitino wave function in momentum space. The above constraints imply that in coordinate space, $\psi_\mu\simeq e^{-ipx}\widetilde{\psi}_\mu$ satisfies Eqs.(\ref{constraint1}), (\ref{constraint2}) and (\ref{constraint3}). The normalization of $\widetilde{\psi}_\mu ({\bf p},\lambda)$ follows from Eqs.(\ref{normspi}) and (\ref{normpol}):
\be
\bar{\widetilde{\psi}}_\mu ({\bf p},\sigma)
\widetilde{\psi}^\mu ({\bf p},\sigma^{\prime}) =
-2 \mg \delta_{\sigma\sigma^{\prime}}.
\ee
The formula for the polarization tensor of a gravitino with mass $\mg$ and momentum $p_\mu$ \cite{BBB} can be accordingly written as
\bea
&\Pi_{\mu\nu}(p)&=\sum_{s}\widetilde{\psi}_\mu(p,s)\bar{\widetilde{\psi}}_\nu(p,s)=\nonumber\\
&&-(\slashed{p}+\mg)\left[\left(\eta_{\mu\nu}-\dfrac{p_\mu
p_\nu}{\mg^2}\right)-\dfrac{1}{3}\left(\eta_{\mu\theta}-\dfrac{p_\mu
p_\theta}{\mg^2}\right)\left(\eta_{\nu\xi}-\dfrac{p_\nu
p_\xi}{\mg^2}\right)\gamma^\theta\gamma^\xi\right]\,,\nonumber\\
\label{gravproj} 
\eea
where the sum is performed over the gravitino helicities $\pm1/2$ and $\pm3/2$. $\Pi_{\mu\nu}(p)$ satisfies the following constraints:
\bea
&& \gamma^\mu \Pi_{\mu\nu}(p)
= \Pi_{\mu\nu}(p) \gamma^\nu = 0\,,
\label{spinsumeq1}\\ &&
p^\mu \Pi_{\mu\nu}(p) = \Pi_{\mu\nu}(p) p^\nu = 0\,,
\label{spinsumeq2}\\ &&
\left( \slashed{p} - \mg \right) \Pi_{\mu\nu}(p) =
\Pi_{\mu\nu}(p) \left( \slashed{p} - \mg \right) = 0\,.
\label{spinsumeq3}\eea
In the high energy limit, i.e. when the centre of mass energy is much larger than the gravitino mass $\mg$, the spin sum is greatly simplified and acquires an interesting structure \cite{BBB}:
\be
\Pi_{\mu\nu}(p)\simeq-\slashed{p}\eta_{\mu\nu}+\dfrac{2}{3}\slashed{p}\dfrac{p_\mu p_\nu}{\mg^2}\,.
\ee
	One can clearly see here the sum over the $\pm 3/2$ helicities of the gravitino in the first term, while the $\pm 1/2$ helicities of the goldstino (the longitudinal degrees of freedom of the gravitino) are represented by the second term. Such a splitted form is very important to understand the results of Chapter \ref{chapt:WW}.

	The solution of Eqs.(\ref{spinsumeq1}), (\ref{spinsumeq2}), (\ref{spinsumeq3}) can be expanded in plane waves as follows,
\be
\psi_\mu (x) = \int \frac{d^3{\bf p}}{(2\pi )^3 2p_0} \sum_\lambda
\left\{ e^{i{\bf px}} \widetilde{\psi}_\mu ({\bf p},\lambda)
a_{{\bf p}\lambda} (t)
+ e^{-i{\bf px}} \widetilde{\psi}_\mu^{C} ({\bf p},\lambda)
a_{{\bf p}\lambda}^\dagger(t) \right\}\,,
\ee
where we have taken into account the fact that $\widetilde{\psi}_\mu$ is a Majorana spinor. This can be proven also for $\psi_\mu$, namely $\psi_\mu = \psi_\mu^C$. Clearly, $p_0 \equiv \sqrt{|{\bf p}|^2+\mg^2}$. In the above equation, the time dependence of the coefficient $a_{{\bf p}\lambda}(t)$ is fixed by the Dirac equation (\ref{constraint3}).

\subsection{Interactions of the gravitino\label{sect:feynsugra}}

	The quantization of the gravitino field is a rather standard procedure, which is discussed in the literature \cite{Wess}, \cite{M}. Here we do not consider the subject, rather we go straight to the interactions of the gravitino, and write down the Feynman rules which are used in Chapter \ref{chapt:WW}.
	The starting point is the general SUGRA Lagrangian, from which we can extract the interactions of the gravitino with the matter multiplets. It is clear from Eq.(\ref{sugral}) that this looks like a very complicated task.

	However, the energy scales of the processes which will be considered in the following allow to neglect most of the terms in the Lagrangian (\ref{sugral}).
	First of all, at centre of mass energies $\sqrt{s}$ which are much lower than the Planck scale $\Mpl$, some operators are suppressed by at least a factor $\sqrt{s}/\Mpl$. Secondly, the gravitinos in the scattering processes that are studied in this thesis appear only as external lines (on-shell particles). Thus, by invoking the constraint (\ref{constraint1}), we can safely ignore the interaction terms which contain $\gamma^\mu\psi_\mu$ or $\bar{\psi_\mu}\gamma^\mu$. Thus the Lagrangian we use to derive the Feynman rules is simply the following \cite{Ferrara,M}:
\be
{\cal L}_{\psi J} =
-\frac{1}{\sqrt{2}\Mpl} \left(D_{\nu} \phi^{*i}
\bar{\psi}_\mu \gamma^\nu \gamma^\mu \chi^i_R
+ D_{\nu} \phi^{i}
\bar{\chi}^i_L \gamma^\mu \gamma^\nu \psi_\mu\right)
%
%
-\frac{i}{8\Mpl} \bar{\psi}_\mu
\left[ \gamma^\nu , \gamma^\rho \right] \gamma^\mu
\lambda^a F_{\nu\rho}^a\,.
\ee
The Feynman rules which follow are reported in Figures (\ref{fig:gvg}), (\ref{fig:4part}), (\ref{fig:hgf}) in the Appendix.


\subsection{Cosmology: the gravitino problem}

	So far we have considered the theory of supergravity and how the gravitino emerges in this framework. We conclude this chapter by discussing the role of this particle in cosmology, which constitutes the motivation of this thesis. The gravitino is relevant for several aspects. Not only it has a primary role on theoretical grounds, since it is the gauge field of local supersymmetry. Either stable or unstable gravitinos, when embedded in cosmological backgrounds, can generate enormous problems to theories which are in agreement with observations. This is known in the literature as the \emph{gravitino problem}.

	The production of gravitinos in the early universe occurs through a variety of ways, via thermal or non-thermal mechanisms. In the first case, in the reheating era at the end of inflation (see Chapter \ref{chapt:intro}) a thermal bath is generated via the decay of the inflaton field. The particles in the plasma scatter off each other, and this process generates gravitinos as the final states of $2\rightarrow2$ hard scatterings \cite{Linde,BBB,PP,Wein1,Krauss,Moroi,Pradler}. It has been shown that an important amount of gravitinos was produced right through these reactions \cite{Khlopov,L}. This is the subject of paper \cite{WW}, where we have found an unexpected effect when the gauge bosons which scatter in the primordial bath are massive.

	Non-thermal production is perhaps a more complicated topic. It can follow from several sources: from the decay of the next to lightest SUSY particle (NLSP) \cite{Feng,Ellis,Kohri,Covi}, of the moduli fields \cite{Randall,Endo,Nakamura,DineKitano}, of the inflaton \cite{Kallosh,Maroto,Kawasaki,Asaka} and of the SUSY breaking field \cite{DineFischler,Coughlan,Banks,Ibe,IK}. The latter mechanism is discussed in the article \cite{Infl}, where it is shown that the reheating temperature can be related by the parameters of the model considered, if a right amount of gravitino Dark Matter is required.

	The gravitino mass $\mg$ can range in principle from the $\e$ scale up to the $\T$ scale and beyond \cite{Martin}. It is strongly model dependent, since it is proportional to the SUSY breaking scale and therefore to the condensation value of the $F$ field, as we have discussed in this chapter.

	In general, gauge mediation predicts the gravitino to be the lightest supersymmetric particle, or LSP \cite{GR}. Then, if R-parity is conserved, the gravitino is stable and it can be a very attractive candidate for Dark Matter \cite{gravitinoDM}. It can however generate problems: since too many gravitinos might overclose the universe, in the standard Big Bang cosmology it is set an upper limit of $\mg\lsim 1 \k$ \cite{PP}.
	This constraint can be anyway relaxed if we assume inflation, since it dilutes the initial abundance of gravitinos\footnote{Historically, the first bound was calculated to be between $\mathcal{O}(1\,\M)$ and $\mathcal{O}(100\,\G)$ \cite{L}.}. However, the problem persists also in this case. The thermal scatterings reproduce the particle after reheating. The number density of the secondary gravitinos is proportional to the reheating temperature, thus $T_R$ should be constrained in order to avoid particle overproduction.

	In other classes of theories, for instance in gravity mediation, the gravitino is unstable and if $\mg$ is smaller than $10 \T$, it has a lifetime $\tau_{\g}$ which is usually longer than 1 sec. This means that it decays after the beginning of the Big Bang Nucleosynthesis (BBN) and it generates hadronic and electromagnetic showers which can be very energetic \cite{M}. This implies disintegration of the primordial light elements \cite{Wein1}. Since cosmological observations have so far verified the BBN predictions to a very high precision, one must impose constraints on the scenario, in order to preserve the agreement between theory and observations \cite{BBN,Kaz,Jedamzik}.
It is therefore evident that the gravitino generates several problems in cosmology. In the literature lots of effort had been devoted to an extensive study of these issues (see for example, \cite{PP,Wein1,Krauss}).

	In particular, BBN constraints on both unstable and stable gravitinos have been recently derived in Ref.\cite{Kaz}, which contains an updated analysis of the gravitino problem in this context. We refer to this paper to complete the above generic discussion.
	For unstable gravitinos, let us consider the decay into a neutralino LSP. It is found that the upper bound strongly depends on $\mg$. In order to preserve the light elements $^3 \rm He$, $^4 \rm He$, $^6\rm Li$ and $D$, the upper limit on the reheating temperature has a mild dependence on the mass spectrum of the MSSM particles. However, it strongly depends on the gravitino mass:
\be
10^6 \G\lsim T_R \lsim 10^{10} \G \qquad \mathrm{if} \qquad 300\G \lsim \mg \lsim 30 \T\,.
\ee
	The above range for $T_R$ is interesting, since it provides with a very stringent constraint on theories of thermal leptogenesis, which actually require a reheating temperature that is at least of the order $10^9\G$ \cite{Fujii,Allahverdi,Delepine,GNRRS,Buchmuller,Panotopoulos,Giudice,BE}.

	In the case of stable LSP gravitinos, the bounds depend on which particle is the NLSP (the authors have considered the bino, the stau and the sneutrino). Very stringent constraints have been found if the bino or the stau are the NLSP. Namely, $\mg\gsim 10\G$ is excluded if the mass of the NLSP is lighter than $1\T$. If the sneutrino is the lightest supersymmetric particle, the BBN constraints are sensibly weaker since the sneutrino decays mainly into gravitino and neutrino (i.e. weakly interacting particles). In any case, the constraints are generally very restrictive. The decay of the NLSP alone cannot provide with the correct amount of gravitino Dark Matter. Therefore the largest contribution should be produced by other mechanisms, for instance through thermal scatterings or by decay of the scalar condensate.

	Both of these possibilities are investigated in the research papers \cite{WW} and \cite{Infl}, which are discussed respectively in Chapters \ref{chapt:WW} and \ref{chapt:Infl} of this thesis.

\chapter{Gauge boson scattering and gravitinos}\label{chapt:WW}

	In this chapter, the phenomenology of the Standard Model and of the MSSM will be discussed. Topics like gauge invariance and the MSSM mass spectrum are closely related to the results of paper \cite{WW}, that concerns the scattering of two massive W bosons
\be
W^a+W^b\longrightarrow \w^c+\g\,,\qquad W^++W^-\longrightarrow \widetilde{\chi}^i_0+\g\,,
\label{wbosons}
\ee
with a gravitino and a gaugino in the final state, both in the gauge and matter eigenstates. 
In the next section we analyse the massless limit of the above process, namely the gluon scattering in supergravity.

\section{Scattering of gauge bosons in supergravity}

\subsection{The massless limit: gluon scattering}

	As a starting point, we consider in analogy with Ref.\cite{BBB} the production of gravitinos through a process in QCD. Namely, we study the scattering of two gluons $g^a_{\alpha}$ and $g^b_{\beta}$ with a gravitino $\widetilde G$ and a gaugino (gluino) $\tilde{g}^c$ in the final state. This is process [A] in \cite{BBB},
\be
g^a+g^b\longrightarrow\gl^{c}+\g\,,
\label{qcdpro}
\ee
where $a,b,c$ are indices of the $SU(3)_C$ algebra.

	The reaction (\ref{qcdpro}) corresponds to the four Feynman diagrams in Fig.(\ref{gluondiag}). Three of them contain the channels $s,t,u$, while the forth is a contact diagram, with a point-like supergravity interaction \cite{Ferrara}.
\begin{center}
\begin{figure}[ht]
\resizebox{0.99\textwidth}{!}{
 \begin{picture}(614,121) (4,2)
    \SetWidth{0.5}
    \SetColor{Black}
    \Text(16.06,117.75)[]{\Large{\Black{$g^a$}}}
    \Text(14.35,8.41)[]{\Large{\Black{$g^b$}}}
    \Text(146.04,116.99)[]{\Large{\Black{$\widetilde{G}$}}}
    \Text(144.51,8.41)[]{\Large{\Black{$\tilde{g}^c$}}}
    \Text(80.29,81.05)[]{\Large{\Black{$g^c$}}}
    \Text(476.36,117.75)[]{\Large{\Black{$g^a$}}}
    \Text(320.85,117.75)[]{\Large{\Black{$g^a$}}}
    \Text(170.98,117.75)[]{\Large{\Black{$g^a$}}}
    \Text(300.73,117.46)[]{\Large{\Black{$\widetilde{G}$}}}
    \Text(451.13,88.7)[]{\Large{\Black{$\tilde{g}^c$}}}
    \Text(170.04,8.41)[]{\Large{\Black{$g^b$}}}
    \Text(477.13,8.41)[]{\Large{\Black{$g^b$}}}
    \Text(320.67,8.41)[]{\Large{\Black{$g^b$}}}
    \Text(395.31,64.23)[]{\Large{\Black{$\tilde{g}^b$}}}
    \Text(157.51,64.23)[]{\Large{\Black{$+$}}}
    \Text(308.91,64.99)[]{\Large{\Black{$+$}}}
    \Text(468.72,65.76)[]{\Large{\Black{$+$}}}
    \Text(451.13,38.23)[]{\Large{\Black{$\widetilde{G}$}}}
    \SetWidth{0.5}
    \Vertex(536.77,64.99){2.16}
    \Gluon(103.22,64.23)(137.63,18.35){5.73}{6.86}
    \Vertex(103.22,64.99){2.16}
    \Gluon(57.35,63.46)(22.94,18.35){5.73}{6.86}
    \Gluon(57.35,64.23)(103.22,64.23){5.73}{5.14}
    \Line(101.99,67.42)(134.1,111.77)\Line(104.46,65.63)(136.58,109.97)
    \Vertex(58.11,64.23){2.16}
    \Gluon(22.94,110.11)(56.58,64.23){5.73}{6.86}
    \Vertex(376.2,97.11){2.16}
    \Line(376.2,29.06)(376.2,97.11)
    \Line(376.2,97.11)(439.66,89.46)
    \Vertex(376.2,29.06){2.16}
    \Gluon(325.73,109.34)(376.2,97.11){5.73}{6}
    \Gluon(326.5,17.59)(376.2,29.06){5.73}{6.74}
    \Line(375.91,30.56)(439.37,42.79)\Line(376.49,27.55)(439.95,39.79)
    \Vertex(227.09,97.11){2.16}
    \Vertex(227.09,29.06){2.16}
    \Line(103.22,64.23)(137.63,18.35)
    \Text(299.73,9.94)[]{\Large{\Black{$\tilde{g}^c$}}}
    \Text(246.21,63.46)[]{\Large{\Black{$\tilde{g}^a$}}}
    \Line(227.09,29.06)(290.56,17.59)
    \Gluon(176.63,17.59)(227.09,29.06){5.73}{6.86}
    \Gluon(175.86,109.34)(227.09,97.11){5.73}{6.77}
    \Gluon(227.09,97.11)(227.09,28.29){5.73}{6.86}
    \Line(226.79,98.61)(291.02,111.61)\Line(227.4,95.61)(291.63,108.61)
    \Gluon(482.48,110.87)(536.77,64.99){5.73}{6.8}
    \Gluon(483.24,17.59)(536.77,64.99){5.73}{7.03}
    \Gluon(536.77,64.99)(596.41,18.35){5.73}{7.98}
    \Line(535.85,66.22)(596.26,111.33)\Line(537.68,63.77)(598.09,108.88)
    \Line(536.77,64.23)(596.41,18.35)
    \Text(607.11,116.99)[]{\Large{\Black{$\widetilde{G}$}}}
    \Text(607.88,10.7)[]{\Large{\Black{$\tilde{g}^c$}}}
    \Gluon(227.09,29.06)(290.56,17.59){5.73}{6.86}
    \Gluon(376.2,97.11)(439.66,89.46){5.73}{6.86}
    \Gluon(376.2,97.11)(376.2,29.06){5.73}{6.86}
    \Line(227.09,97.11)(227.09,28.29)
\end{picture}
}
\label{gluondiag}
\caption{The four diagrams which contribute to
$g^a+g^b\longrightarrow\gl^{c}+\g$}
\end{figure}
\end{center}
	The matrix element $\mathcal{M}$ of the process (\ref{qcdpro}) can be written as the sum of the four subamplitudes
\be
\mathcal{M}=M_s+M_t+M_u+M_x\,,
\ee
where $s$, $t$ and $u$ are the Mandelstam variables \cite{P}. By using the Feynman rules in Appendix \ref{app:feyn}, it is easy to write the four amplitudes as follows:
\bea
M_s&=&\dfrac{g_sf_{abc}}{4\Mpl}\left[
\eta^{\alpha\beta}(k-k^{\prime})^\sigma+\eta^{\beta\sigma}(2k^{\prime}+k)^\alpha
-\eta^{\alpha\sigma}(2k+k^{\prime})^\beta\right]\times\nonumber\\
&&\times\dfrac{1}{s}\left\{\bar{\psi}_{\mu}^s(p)\left[\slashed{k}+\slashed{k}^{\prime},\gamma_\sigma\right]
\gamma^\mu v^c_{s^\prime}(p^\prime)\right\}\epsilon^a_\alpha(k)\epsilon^b_\beta(k^{\prime})\,,
\label{Ms}
\\
M_t&=&\dfrac{g_sf_{abc}}{4\Mpl}
\left\{\bar{\psi}_{\mu}^s(p)
\left[\slashed{k},\gamma^\alpha\right]\gamma^\mu(\slashed{k}^{\prime}-\slashed{p}^{\prime}+\mgl)
\gamma^\beta v^c_{s^\prime}(p^\prime)\right\}\dfrac{\epsilon^a_\alpha(k)\epsilon^b_\beta(k^{\prime})}{t-\mgl^2}\,,
\label{Mt}
\\
M_u&=&-\dfrac{g_sf_{abc}}{4\Mpl}
\left\{\bar{\psi}_{\mu}^s(p)
[\slashed{k}^{\prime},\gamma^\beta]\gamma^\mu(\slashed{k}-\slashed{p}^{\prime}+m_{\tilde{g}})
\gamma^\alpha v^c_{s^\prime}(p^\prime)\right\}\dfrac{\epsilon^a_\alpha(k)\epsilon^b_\beta(k^{\prime})}{u-\mgl^2}\,,
\label{Mu}
\\
M_x&=&-\dfrac{g_sf_{abc}}{4\Mpl}\left\{\bar{\psi}_{\mu}^s(p)
[\gamma^\alpha,\gamma^\beta]\gamma^\mu
v^c_{s^\prime}(p^\prime)\right\}\epsilon^a_\alpha(k)\epsilon^b_\beta(k^{\prime})\, .
\label{Mx}
\eea
	In the above equations, $\Mpl=(8\pi G_N)^{-1/2}= 2.43\times 10^{18}$ GeV is the reduced Planck mass, and $g_s$ and $f_{abc}$ are, respectively, the strong coupling constant and the structure constants of the group $\qcd$. The spinors are $\bar{\psi}_{\mu}^s(p)$, namely the wave function of a gravitino with four-momentum $p$ and helicity state $s$, which is represented by a four-component Majorana spinor, with spin 3/2 and Lorentz index $\mu$. $v^c_{s^\prime}(p^\prime)$ is the wave function of an antiwino with four-momentum $p^{\prime}$, helicity state $s^{\prime}$ and $\ew$ index $c$ in the final state. It is represented by a spin 1/2 Majorana spinor (see Section \ref{sect:majo}).

	In the following, $\bar{\psi}_{\mu}^s(p)$ and $v^c_{s^\prime}(p^\prime)$ are replaced respectively by $\bar{\psi}_{\mu}$ and $v^{c}$ for a clearer notation.
The wave functions of the gluons, $\epsilon^a_\alpha(k)$ and $\epsilon^b_\beta(k^{\prime})$, are the polarization vectors of two bosons with, respectively, four-momenta $k$ and $k^{\prime}$, $\rm SU(3)_C$ indices $a$ and $b$ and Lorentz indices $\alpha$ and $\beta$.
	The kinematics of the process defines the Mandelstam variables as follows,
\be 
\left\{
\begin{tabular}{l}
$s=(k+k^{\prime})^2$\,,\\
$t=(p^{\prime}-k^{\prime})^2$\,,\\
$u=(p^{\prime}-k)^2$\,.
\end{tabular}
\right. \label{mandelstam}
\ee
The conservation of energy-momentum is written as \cite{P}:
\be
s+t+u=\sum_i m^2_i=\mgl^2+\mg^2\,.
\ee

\subsubsection{Ward identities and connection to unitarity}

	Before calculating the squared amplitude relative to the process (\ref{qcdpro}), let us recall the generalized Ward identities for Yang-Mills theories \cite{P,CL}. In the non-abelian case, they are called Slavnov-Taylor identities \cite{ST}. We basically follow the book by Cheng and Lee \cite{CL}, where a detailed calculation can be found. This section will provide only with the comments which are relevant to our results.

	The Ward identities relate different Green's functions, namely these reflect the symmetries of the theory such as gauge invariance. For instance, in electromagnetism these imply one single condition, namely that particles with the same bare (electric) charges have the same renormalized charges. In non-abelian gauge theories the number of these conditions is much larger. Moreover, the Ward identities guarantee that all the unphysical singularities eventually disappear in the physical amplitude.

	Let us consider an SU(2) theory with fermions $f$ in a doublet representation. The requirement that the scattering matrix $S$ must be unitary,
\be
SS^{\dagger}=S^{\dagger}S=1 \qquad\Longleftrightarrow\qquad \sum_lS_{al}S^*_{bl}=\delta_{ab}\,,
\ee
by virtue of the relation
\be
S_{ab}=\delta_{ab}+i(2\pi)^4\delta^4(p_a-p_b)T_{ab}\,,
\ee
can be written in terms of the scattering amplitude $T_{ab}$ as follows:
\be
\mathrm{Im}\:T_{ab}=\half\sum_l T_{al}T^*_{bl}(2\pi)^4\delta^4(p_a-p_l).
\label{unitarity}
\ee
	The unitarity of the $S$-matrix thus implies certain conditions which any scattering amplitude has to satisfy. In particular, the imaginary part of $T^{ab}$ connects the initial and final states as a consequence of energy conservation.

	Suppose now to study a fermion-antifermion scattering $f\bar{f}\rightarrow f\bar{f}$ in this basic $SU(2)$ theory, with off-shell gauge bosons in the intermediate states \cite{FF}. In order to recall the property (\ref{unitarity}), one can use the Cutkosky rule \cite{Cut} to calculate the imaginary part of the scattering amplitude.  One then replaces the boson propagators by their imaginary parts, which are multiplied by the on-shell scattering amplitudes $T(f\bar{f}\rightarrow AA)$ and $T^*(AA\rightarrow f\bar{f})$. This is equivalent to making a cut in the original diagram, as in Fig.(3.2).
\begin{center}
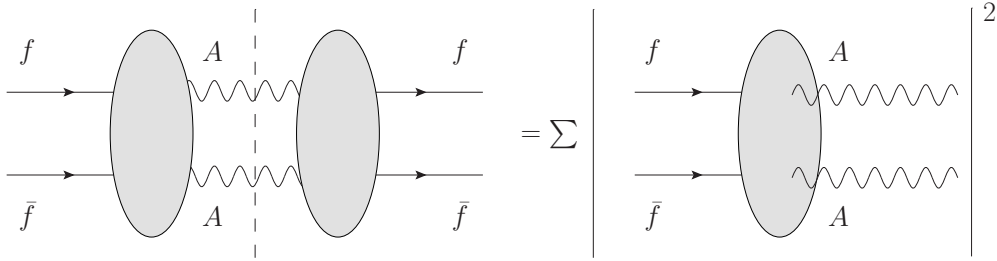
\begin{figure}[t]
\resizebox{0.9\textwidth}{!}{
  \begin{picture}(614,177) (-7,-19)
    \SetWidth{0.5}
    \SetColor{Black}
    \ArrowLine(252.64,81.69)(328.43,81.69)
    \ArrowLine(252.64,31.16)(328.43,31.16)
    \DashLine(189.48,132.22)(189.48,-19.37){8.42}
    \Text(315.8,106.95)[]{\Large{\Black{$f$}}}
    \Text(315.8,5.89)[]{\Large{\Black{$\bar{f}$}}}
    \ArrowLine(37.9,81.69)(113.69,81.69)
    \ArrowLine(37.9,31.16)(113.69,31.16)
    \Text(164.22,106.95)[]{\Large{\Black{$A$}}}
    \Text(164.22,5.89)[]{\Large{\Black{$A$}}}
    \Photon(129.69,82.53)(230.75,82.53){6.32}{6.5}
    \Photon(129.69,30.32)(230.75,30.32){6.32}{6.5}
    \GOval(126.32,56.42)(63.16,25.26)(0){0.882}
    \GOval(240.01,56.42)(63.16,25.26)(0){0.882}
    \Text(50.53,106.95)[]{\Large{\Black{$f$}}}
    \Text(50.53,5.89)[]{\Large{\Black{$\bar{f}$}}}
    \Text(370.54,56.42)[]{\Large{\Black{$=\sum$}}}
    \Line(395.8,132.22)(395.8,-19.37)
    \ArrowLine(421.07,81.69)(496.86,81.69)
    \ArrowLine(421.07,31.16)(496.86,31.16)
    \Text(547.39,106.95)[]{\Large{\Black{$A$}}}
    \Text(547.39,5.89)[]{\Large{\Black{$A$}}}
    \GOval(509.49,56.42)(63.16,25.26)(0){0.882}
    \Photon(517.07,80)(618.13,80){6.32}{6.5}
    \Photon(517.07,29.47)(618.13,29.47){6.32}{6.5}
    \Line(626.55,133.06)(626.55,-18.53)
    \Text(640.02,131.37)[]{\Large{\Black{$2$}}}
    \Text(433.7,106.95)[]{\Large{\Black{$f$}}}
    \Text(433.7,5.89)[]{\Large{\Black{$\bar{f}$}}}
  \end{picture}
      }
\label{fig:cut}
\caption{The Cutkosky rule applied to fermion-antifermion scattering.}
\end{figure}
\end{center}
	It can be shown \cite{CL} that at the lowest order the unitarity condition for this scattering reduces to
\be
\int{d\rho \left(T^{ab}_{\mu\nu}T^{ab*}_{\alpha\beta}\eta^{\mu\alpha}\eta^{\nu\beta}-2S^{ab}S^{ab*} \right)}=
\int{d\rho T^{ab}_{\mu\nu}T^{ab*}_{\alpha\beta}P^{\mu\alpha}(k_1)P^{\nu\beta}(k_2)}\,,
\label{unicond}
\ee
where $P^{\mu\alpha}(k_1)$ and $P^{\nu\beta}(k_2)$ are the spin sums of the gauge bosons:
\bea
P^{\mu\alpha}(k_1)&=&\sum_{r}\epsilon^{(r)\mu}(k_1)\epsilon^{(r)\alpha*}(k_1)\,,\\
P^{\nu\beta}(k_2)&=&\sum_{r}\epsilon^{(r)\nu}(k_2)\epsilon^{(r)\beta*}(k_2)\,.
\eea
	In the unitarity condition Eq.(\ref{unicond}), $d\rho$ is a volume element in the phase space of the two bosons. $T^{ab}_{\mu\nu}$ corresponds to $f\bar{f}\rightarrow A^a_\mu A^b_\nu$ and $S^{ab}$ is the amplitude of the process $f\bar{f}\rightarrow c^ac^{b\dagger}$, where $c$ are the ghost fields. In the left hand side of (\ref{unicond}), these give an explicit contribution to the scattering amplitude. In the right hand side, they appear instead as the longitudinal degrees of freedom of the gauge bosons in the above spin sums. 
		
		In other words, the spin sum of the massless gauge boson in non abelian theories cannot just be $\eta_{\mu\nu}$ (as in QED), but it must contain also the unphysical ghosts in the form of longitudinal polarizations of the gauge bosons.

	This is our starting point. In the case of gluon scattering in supergravity, we encounter exactly the same form of equation (\ref{unicond}) and verify the unitarity condition. In contrast, in Section \ref{sect:ww} we show that in the case of massive W bosons, the longitudinal degrees of freedom in the r.h.s. of Eq.(\ref{unicond}) are missing, by virtue of the SUGRA vertex in the $s$ channel.

\subsubsection{Squared amplitude of the gluon scattering}

	Let us now return to the gluon scattering in supergravity, by keeping in mind the discussion in the previous section. The sum of the four contributions Eqs.(\ref{Ms}), (\ref{Mt}), (\ref{Mu}) and (\ref{Mx}) is the matrix element $\mathcal{M}$, that can be written as follows:
\bea
&&\mathcal{M}=M^{\alpha\beta}_{ab}\epsilon^a_\alpha\epsilon^b_\beta=\dfrac{g_s f_{abc}}{4M_P}\Bigg(\left[
\eta^{\alpha\beta}(k-k^{\prime})^\sigma+\eta^{\beta\sigma}(2k^{\prime}+k)^\alpha
-\eta^{\alpha\sigma}(2k+k^{\prime})^\beta\right]\nonumber\\
&&\times\dfrac{1}{s}\left\{\bar{\psi}_{\mu}\left[\slashed{k}+\slashed{k}^{\prime},\gamma_\sigma\right]\gamma^\mu v^{c}\right\}+\dfrac{1}{t-\mgl^2}\left\{\bar{\psi}_{\mu}
\left[\slashed{k},\gamma^\alpha\right]\gamma^\mu(\slashed{k}^{\prime}-\slashed{p}^{\prime}+\mgl)
\gamma^\beta v^{c}\right\}-\nonumber\\
&&
-\dfrac{1}
{u-\mgl^2}\left\{\bar{\psi}_{\mu}
[\slashed{k}^{\prime},\gamma^\beta]\gamma^\mu(\slashed{k}-\slashed{p}^{\prime}+\mgl)
\gamma^\alpha v^{c}\right\}-\bar{\psi}_{\mu}[\gamma^\alpha,\gamma^\beta]\gamma^\mu v^{c}\Bigg)\epsilon^a_\alpha\epsilon^b_\beta\,.
\nonumber\\
\label{matrixg}
\eea
	The Ward identities for the above amplitude assume the following form \cite{CL},
\be \left\{
\begin{tabular}{l}
$k_\alpha M_{ab}^{\alpha\beta}=S_{ab}k^{\prime\beta}$\\
$k^{\prime}_\beta M_{ab}^{\alpha\beta}=T_{ab}k^\alpha$\\
\end{tabular}
\right. ,
\label{ward}
\ee
where $k_\alpha$ and $k^{\prime}_{\beta}$ are the four momenta of the two on-shell gauge bosons. $S_{ab}$ and $T_{ab}$ are the Lorentz scalars which have been defined in the previous section. According to the above discussion, here we call them ghost amplitudes. By using equations (\ref{ward}), we can write the squared amplitude, summed over
the polarizations of the initial and final states, in a very meaningful way. A straightforward calculation gives:
\bea
\sum_{spin}|\mathcal{M}|^2&=&\sum_{spin}|M_{ab}^{\alpha\beta}\epsilon^a_\alpha(k)\epsilon^b_\beta(k^{\prime})|^2=\nonumber\\
&=&\sum_{spin}|M_{ab}^{\alpha\beta}|^2-\sum_{spin}(S^{ab}T^*_{ab})-\sum_{spin}(T^{ab}S^*_{ab})\,.
\label{squared}
\eea
By using Eq.\ref{matrixg}), the scalar amplitude can be rewritten as
\be
T_{ab}=\dfrac{g_sf_{abc}}{4\Mpl}\left\{\dfrac{\bar{\psi}_{\mu}
\left[\slashed{k},\slashed{k}^{\prime}\right]\gamma^\mu
v^c}{s}\right\}=S_{ab}\,,
\label{ghost}
\ee
therefore (\ref{squared}) can be recast in a very simple form:
\be
\sum_{spin}|\mathcal{M}|^2=\sum_{spin}|M_{ab}^{\alpha\beta}|^2-2\sum_{spin}|S_{ab}|^2\,.
\ee
The above formula follows from the polarization sum of a massless vector boson with momentum $k$ and helicity state $r$ \cite{CL},
\be
\sum_{r}\epsilon^{a(r)}_\alpha(k)\epsilon^{l(r)*}_\nu(k)=-\left(
\eta_{\alpha\nu}-\dfrac{k_\alpha\eta_\nu+k_\nu\eta_\alpha}{k\eta}\right)\delta^{al}\equiv T+L\,,
\label{glproj}
\ee
where the two different contributions of the transverse and longitudinal components are labeled by $T$ and $L$. Clearly, $\sum_{spin}|M_{ab}^{\alpha\beta}|^2$ comes from $T\equiv-\eta_{\alpha\nu}$, while $L$ gives $-2\sum_{spin}|S_{ab}|^2$. The polarization sums of the fermions result as follows. For an anti-gaugino (anti-gluino) \cite{HK} with momentum $p^\prime$, group index $c$, helicity state $s^\prime$ and mass $m_{\tilde{g}}$, we have
\be
\sum_{s^\prime}v^c_{s^\prime}(p^\prime)\bar{v}^n_{s^\prime}(p^\prime)=(\slashed{p}^{\prime}-m_{\tilde{g}})\delta^{cn}\,.
\ee
For a gravitino with momentum $p$, helicity state $s$ and mass $m_{\widetilde G}$, the general expression for the full projector (\ref{gravproj}) has been given in Section \ref{gravspinsum} as
\bea
&&\Pi_{\mu\nu}(p)=\sum_{s}\psi^s_\mu(p)\overline{\psi}^s_\nu(p)=\nonumber\\
&&-(\slashed{p}+\mg)\left[\left(\eta_{\mu\nu}-\dfrac{p_\mu
p_\nu}{\mg^2}\right)-\dfrac{1}{3}\left(\eta_{\mu\theta}-\dfrac{p_\mu
p_\theta}{\mg^2}\right)\left(\eta_{\nu\xi}-\dfrac{p_\nu
p_\xi}{\mg^2}\right)\gamma^\theta\gamma^\xi\right]\,.\\
\label{gravprojtot}
\eea
	One can simplify the above object by assuming that the centre of mass energy is much larger than the gravitino mass.
 In this case, the full gravitino projector (\ref{gravproj}) reduces to \cite{BBB}:
\be
\Pi_{\mu\nu}(p)\simeq-\slashed{p}\eta_{\mu\nu}+\dfrac{2}{3}\slashed{p}\dfrac{p_\mu p_\nu}{\mg^2}\,.
\label{gravprojred}
\ee
 	This approximation is consistent with a large class of cosmological scenarios with a reheating temperature $T_R$ that is well above the electroweak threshold. This corresponds to Refs.\cite{BBB} and \cite{Pradler}. Even though in models with low $T_R$ this is not always true, for the gluon scattering at hand we make the above assumption\footnote{It can be shown that the full gravitino projector (\ref{gravprojtot}) does not give a qualitatively different squared amplitude.}. The reason for using equation (\ref{gravprojred}) is that the contributions from the gravitino polarizations $\pm 3/2$ and $\pm 1/2$ are clearly distinct.

	The kinematics is derived by taking into account Eq.(\ref{mandelstam}),
\be
\left\{
\begin{tabular}{lll}
$k^2=k^{\prime2}=0\,,$&$2pp^{\prime}=(s-\mgl^2-\mg^2)\,,$&$2kp=(\mg^2-t)$\,,\\
\addlinespace[0.1cm]
$p^2=\mg^2\,,$& $2k^{\prime}p^{\prime}=(\mgl^2-t)$\,,&\\
\addlinespace[0.1cm]
$p^{\prime2}=\mgl^2\,,$& $2kp^{\prime}=(\mgl^2-u)$\,, &\\
\addlinespace[0.1cm]
$2kk^{\prime}=s\,,$ & $2k^{\prime}p=(\mg^2-u)$\,. &\\
\end{tabular} \right.
\ee
The final expression for the squared matrix element of the scattering
\be
g^a+g^b\longrightarrow \gl^{c}+\g,
\nonumber
\ee
summed over the spins of the initial and final states, and averaged over the initial states, can be accordingly written as:
\be
\sum_{spin}|\overline{\mathcal{M}}|^2=\dfrac{1}{4}\sum_{spin}|\mathcal{M}|^2=\dfrac{g_s^2|f_{abc}|^2}{\Mpl^2}\left(1+\dfrac{\mgl^2}{3\mg^2}\right)
\left(s+2t+2\dfrac{t^2}{s}\right).
\label{qcdres}
\ee
	This agrees with the result obtained in \cite{BBB}. In the next section we perform the same analysis for on-shell \emph{massive} bosons and see that things change non trivially, despite what is generally expected.

\subsection{Scattering of massive gauge bosons}\label{sect:ww}

	Here we calculate the cross section of the WW scattering into gravitino and gaugino, and compare it to the results of the previous section. Since the supergravity vertices are universal, one expects that the scattering amplitude of
\be
W^a(k)+W^b(k^{\prime})\longrightarrow \w^c(p^{\prime})+\g(p)\,,
\ee
where $a,b,c$ are now indices of the $\ew$ algebra and $\w$ is a wino, should not be formally different from Eq.(\ref{matrixg}). The Feynman diagrams of the above process, whose structure reflects this universality, are drawn in Fig.(\ref{fig:ww}).
\begin{figure}[t]{
\resizebox{\textwidth}{!}{
 \begin{picture}(614,121) (4,2)
\SetWidth{0.5}
    \SetColor{Black}
    \Text(16.28,120.16)[]{\large{\Black{$W^a$}}}
    \Text(18.61,9.3)[]{\large{\Black{$W^b$}}}
    \Text(148.07,119.39)[]{\Large{\Black{$\widetilde{G}$}}}
    \Text(146.52,9.3)[]{\large{\Black{$\widetilde{W}^c$}}}
    \Text(81.4,82.95)[]{\large{\Black{$W^c$}}}
    \Text(174.43,8.53)[]{\large{\Black{$W^b$}}}
    \Text(159.7,66.67)[]{\Large{\Black{$+$}}}
    \Text(606.23,115.51)[]{\Large{\Black{$\widetilde{G}$}}}
    \Text(607.78,14.73)[]{\large{\Black{$\widetilde{W}^c$}}}
    \SetWidth{0.5}
    \Vertex(104.66,66.67){2.19}
    \Vertex(58.92,65.89){2.19}
    \Photon(58.14,66.67)(24.03,112.41){5.81}{8}
    \Photon(58.14,65.89)(104.66,65.89){5.81}{8}
    \Photon(58.92,65.89)(24.03,20.16){5.81}{9}
    \Line(104.66,67.45)(139.54,19.38)
    \Vertex(544.21,66.67){2.19}
    \Photon(544.21,66.67)(594.6,20.93){5.81}{10}
    \Line(544.21,67.45)(595.38,20.93)
    \Line(543.93,67.8)(591.99,112.77)\Line(546.05,65.54)(594.11,110.5)
    \Photon(544.21,67.45)(500.8,112.41){5.81}{9}
    \Photon(544.99,67.45)(500.03,20.93){5.81}{9}
    \Text(492.72,118.61)[]{\large{\Black{$W^a$}}}
    \Text(492.5,13.95)[]{\large{\Black{$W^b$}}}
    \Text(480.64,66.67)[]{\Large{\Black{$+$}}}
    \Text(319.4,66.67)[]{\Large{\Black{$+$}}}
    \Vertex(393.04,99.23){2.19}
    \Line(393.04,28.68)(393.04,100.78)
    \Vertex(393.04,28.68){2.19}
    \Line(393.04,99.23)(455.84,91.48)
    \Text(333.35,119.39)[]{\large{\Black{$W^a$}}}
    \Text(467.47,90.7)[]{\large{\Black{$\widetilde{W}^c$}}}
    \Text(337.23,9.3)[]{\large{\Black{$W^b$}}}
    \Text(410.87,65.89)[]{\large{\Black{$\widetilde{W}^b$}}}
    \Text(467.47,39.54)[]{\large{\Black{$\widetilde{G}$}}}
    \Line(392.71,30.2)(455.5,44.15)\Line(393.38,27.17)(456.17,41.12)
    \Photon(342.65,113.96)(392.27,100.01){5.81}{8}
    \Photon(393.04,31.01)(345.75,19.38){5.81}{8}
    \Photon(393.04,100.78)(455.84,92.25){5.81}{8}
    \Text(170.78,119.39)[]{\large{\Black{$W^a$}}}
    \Vertex(234.12,99.23){2.19}
    \Line(233.35,30.23)(297.69,18.61)
    \Vertex(233.35,30.23){2.19}
    \Photon(234.12,99.23)(233.35,30.23){5.81}{8}
    \Text(306.99,119.84)[]{\Large{\Black{$\widetilde{G}$}}}
    \Text(307.99,6.85)[]{\large{\Black{$\widetilde{W}^c$}}}
    \Text(252.73,65.12)[]{\large{\Black{$\widetilde{W}^a$}}}
    \Line(235.37,101.53)(298.16,113.93)\Line(235.97,98.48)(298.77,110.89)
    \Photon(234.12,101.56)(180.63,112.41){5.81}{8}
    \Photon(232.57,30.23)(180.63,18.61){5.81}{8}
    \Line(234.12,100.01)(234.12,31.01)
    \Photon(233.35,30.23)(298.46,18.61){5.81}{8}
    \Line(104.18,69.13)(136.74,114.09)\Line(106.69,67.31)(139.25,112.27)
    \Photon(104.66,66.67)(139.54,19.38){5.81}{8}
    \Photon(393.04,100.78)(393.04,28.68){5.81}{8} 
    \end{picture}
}
\caption{The four diagrams which contribute to $W^a+W^b\longrightarrow \w^c+\g$.}
\label{fig:ww}
}
\end{figure}
	The four corresponding amplitudes are practically the same as in Eqs.(\ref{Ms}), (\ref{Mt}), (\ref{Mu}), (\ref{Mx}):
\bea
M_s&=&\dfrac{g\epsilon_{abc}}{4M_P}\left[
\eta^{\alpha\beta}(k-k^{\prime})^\sigma+\eta^{\beta\sigma}(2k^{\prime}+k)^\alpha
-\eta^{\alpha\sigma}(2k+k^{\prime})^\beta\right]\times
\nonumber\\
&&\times\dfrac{1}{s-m_W^2}\left\{\bar{\psi}_{\mu}^s(p)\left[\slashed{k}+\slashed{k}^{\prime},\gamma_\sigma\right]
\gamma^\mu v^c_{s^\prime}(p^\prime)\right\} \epsilon^a_\alpha(k)\epsilon^b_\beta(k^{\prime})\,,
\nonumber\\
M_t&=&\dfrac{g\epsilon_{abc}}{4M_P}\dfrac{1}
{t-m_{\w}^2} \left\{\bar{\psi}_{\mu}^s(p)
\left[\slashed{k},\gamma^\alpha\right]\gamma^\mu(\slashed{k}^{\prime}-\slashed{p}^{\prime}+m_{\w})
\gamma^\beta v^c_{s^\prime}(p^\prime)\right\} \epsilon^a_\alpha(k)\epsilon^b_\beta(k^{\prime})\,,
\nonumber\\
M_u&=&-\dfrac{g\epsilon_{abc}}{4M_P}\dfrac{1}
{u-m_{\w}^2}\left\{ \bar{\psi}_{\mu}^s(p)
[\slashed{k}^{\prime},\gamma^\beta]\gamma^\mu(\slashed{k}-\slashed{p}^{\prime}+m_{\w})
\gamma^\alpha v^c_{s^\prime}(p^\prime)\right\} \epsilon^a_\alpha(k)\epsilon^b_\beta(k^{\prime})\,,
\nonumber\\
M_x&=&-\dfrac{g\epsilon_{abc}}{4M_P}\left\{\bar{\psi}_{\mu}^s(p)
[\gamma^\alpha,\gamma^\beta]\gamma^\mu
v^c_{s^\prime}(p^\prime)\right\}\epsilon^a_\alpha(k)\epsilon^b_\beta(k^{\prime})\,.
\label{wwamp}
\eea
	These can be rearranged in a single matrix element, which takes the form
\bea
&&\mathcal{M}=M^{\alpha\beta}_{ab}\epsilon^a_\alpha\epsilon^b_\beta=\dfrac{g\epsilon_{abc}}{4\Mpl}\Bigg(\left[
\eta^{\alpha\beta}(k-k^{\prime})^\sigma+\eta^{\beta\sigma}(2k^{\prime}+k)^\alpha
-\eta^{\alpha\sigma}(2k+k^{\prime})^\beta\right]\times\nonumber\\
&&\times\dfrac{1}{s-m_W^2}\left\{\bar{\psi}_{\mu}\left[\slashed{k}+\slashed{k}^{\prime},\gamma_\sigma\right]\gamma^\mu v^{c}\right\}+\dfrac{1}{t-\mw^2}\left\{\bar{\psi}_{\mu}
\left[\slashed{k},\gamma^\alpha\right]\gamma^\mu(\slashed{k}^{\prime}-\slashed{p}^{\prime}+\mw)
\gamma^\beta v^{c}\right\}\nonumber\\
&&
-\dfrac{1}
{u-\mw^2}\left\{\bar{\psi}_{\mu}
[\slashed{k}^{\prime},\gamma^\beta]\gamma^\mu(\slashed{k}-\slashed{p}^{\prime}+\mw)
\gamma^\alpha v^{c}\right\}-\bar{\psi}_{\mu}[\gamma^\alpha,\gamma^\beta]\gamma^\mu v^{c}\Bigg)\epsilon^a_\alpha\epsilon^b_\beta\,.
\nonumber\\
\label{matrixw}
\eea
The explicit dependence of the polarization vector of the W on the momentum has been omitted. As it was done before, the above amplitude is recast by using the Ward identities Eq.(\ref{ward}),
\be \left\{
\begin{tabular}{l}
$k_\alpha M_{ab}^{\alpha\beta}=S_{ab}k^{\prime\beta}$\\
$k^{\prime}_\beta M_{ab}^{\alpha\beta}=T_{ab}k^\alpha$\\
\end{tabular}
\right. \,.
\ee
	In this case, $S_{ab}$ and $T_{ab}$ are no longer due to the ghosts, but to the physical degrees of freedom which are carried by the Goldstone bosons. Accordingly, we will regard them as Goldstone amplitudes. These have the following expression,
\be
S_{ab}=\dfrac{g\epsilon_{abc}}{4\Mpl}
\left\{\dfrac{\bar{\psi}_{\mu}\left[\slashed{k},\slashed{k}^{\prime}\right]\gamma^\mu
v^c}{s-m_W^2}\right\}=T_{ab}\,.
\ee
As expected, the above differs from (\ref{ghost}) only trivially, by the W boson mass $m_W$ in the denominator. The spin-summed square of the matrix element in Eq.(\ref{matrixw}) can be now written as:
\be
\sum_{spin}|\mathcal{M}|^2=\sum_{spin}|M^{\alpha\beta}_{ab}|^2-\sum_{spin}|S_{ab}|^2\,.
\label{totalw}
\ee
	By using the full gravitino projector (\ref{gravproj}) and averaging over the initial states, we get
\be
\sum_{spin}|\overline{\mathcal{M}}|^2=4\frac{g^2|\epsilon^{abc}|^2}{9\Mpl^2}
\left[\left(1+\dfrac{\mw^2}{3\mg^2}\right)\left(s+2t+2\frac{t^2}{s}\right)-\dfrac{t(s+t)}{3\mg^2}+\mathcal{O} \left(\dfrac{m_i^2}{\mg^2}\right)f(s,t)\right]\,.
\label{totalm0}
\ee
$m_i$ can be either $m_W$ or $m_{\w}$, and $f$ is a function of dimension $[mass]^2$. In the previous section we have obtained for SUSY QCD the following,
\be
\sum_{spin}|\overline{\mathcal{M}}|^2=\dfrac{1}{4}\sum_{spin}|\mathcal{M}|^2=\dfrac{g_s^2|f^{abc}|^2}{M_P^2}\left(1+\dfrac{m_{\tilde{g}}^2}{3{m^2_{\g}}}\right)
\left(s+2t+2\frac{t^2}{s}\right)\,.
\ee
Therefore the relevant difference with respect to Eq.(\ref{totalm0}) is found in the term
\be
\dfrac{-t(s+t)}{3\Mpl^2\mg^2}\,,
\label{aiuto}
\ee
	that contributes to the differential and to the total cross section of the process as follows:
\be
\left(\dfrac{d\sigma}{dt}\right)_{cm}\approx\frac{g^2|\epsilon^{abc}|^2}{64\pi
\Mpl^2}\dfrac{(1-\cos^2{\theta})}{{3m^2_{\g}}}\Longrightarrow \sigma_{tot}\approx \dfrac{s}{\Mpl^2\mg^2}
\,.\label{wcs}
\ee
	Clearly, the above result implies violation of unitarity of the scattering matrix for sufficiently high centre of mass energies, since the cross section grows with energy. Fortunately, the structure of (\ref{aiuto})\footnote{Namely, the fact that the Planck mass appears in the denominator.} makes the unitarity breaking energy coincide precisely with the supersymmetry breaking scale (see Chapter \ref{chapt:sugra}).
	Since supergravity is the low energy limit of some GUT theory yet to be discovered, in principle the above result does not affect observations. Unitarity is guaranteed up to the scale of validity of this effective theory. For instance, since the SUSY breaking energy in supergravity corresponds to \cite{Nilles}
\be
\mu_{\mathrm{SUSY}}=\sqrt{\left\langle F\right\rangle}=(\sqrt{3}\mg\Mpl)^{1/2}\,,
\ee
we find that unitarity is broken e.g. at $\sqrt{s}\approx\mathcal{O}(10^{14}\,\G)$ if $\mg\approx\mathcal{O}(1 \,\M)$.

	Let us now summarise our results. After calculating the gluon scattering in supergravity (\ref{qcdpro}), which provided an expected result, we have considered its extension to massive gauge bosons. We have found that the cross section for the WW scattering contains an extra term which breaks unitarity at the SUSY breaking scale. Now we would like to understand this unexpected result.

	To begin with, we observe that Eq.(\ref{wcs}) has some similarity with the case of WW scattering in the Standard Model \cite{P}. The Goldstone equivalence theorem, that will be discussed in the next section, states that the longitudinal polarizations of the massive on-shell W are crucial. We expect that these will be relevant for the case at hand as well.

	Consider the squared amplitude written as (\ref{totalw}). The main difference with respect to the gluon scattering (\ref{qcdpro}),
\be
\sum_{spin}|\mathcal{M}|^2=\sum_{spin}|M_{ab}^{\alpha\beta}|^2-\sum_{spin}(T^{ab}S^*_{ab})-\sum_{spin}(S^{ab}T^*_{ab})=\sum_{spin}|M_{ab}^{\alpha\beta}|^2-2\sum_{spin}|S^{ab}|^2\,,
\label{total}
\ee
is found in the different contribution of the scalar amplitudes. It can be shown that $M_{ab}^{\alpha\beta}$ is formally the same\footnote{Modulo constant factors such as the couplings and the structure constants.} as Eq.(\ref{matrixw}) and that the masses of the W bosons in the kinematics do not give any contribution which would cancel (\ref{aiuto}). The sum over the polarizations of a massless gauge boson with four-momentum $k$ is Eq.(\ref{glproj}):
\be
\sum_{r}\epsilon^{a(r)}_\alpha(k)\epsilon^{l(r)*}_\nu(k)=-\left(
\eta_{\alpha\nu}-\dfrac{k_\alpha\eta_\nu+k_\nu\eta_\alpha}{k\eta}\right)\delta^{al}\equiv T+L\,.
\ee
On the other hand, one obtains Eq.(\ref{totalw}) from the projector of a W boson with momentum $k$, mass $m_W$ and helicity $r$:
\be
\sum_{r}\epsilon^{a(r)}_\alpha(k)\epsilon^{l(r)*}_\nu(k)=-\left(
\eta_{\alpha\nu}-\dfrac{k_\alpha k_\nu}{m_W^2}\right)\delta^{al}\equiv T+L\,.
\label{wproj}
\ee
	We have written (\ref{totalw}) and (\ref{total}) in such a way that the two distinct contributions of the above projectors stay separated. Thus $\sum_{spin}|M_{ab}^{\alpha\beta}|^2$ comes from the transverse polarizations $\eta_{\alpha\nu}$ (i.e. from T). The longitudinal degrees of freedom\footnote{It is anyway important to point out that the gluons do not have any proper longitudinal polarizations, since these are massless. Such degrees of freedom are the Faddeev-Popov ghosts.} give instead $\sum_{spin}|S_{ab}|^2$. 
As a cross check, by substituting the spin sums (\ref{wproj}) directly into the square of (\ref{matrixw}), we obtain the same result. In Table (\ref{table:TL}) we compare the divergences generated by the distinct terms T and L, for both massless and massive W. These lead respectively to Eqs.(\ref{qcdres}) and (\ref{totalm0}).

	It is now evident that (\ref{aiuto}) is somehow related to the longitudinal degrees of freedom of the W, namely to its mass. We try to find a theoretical explanation for Eq.(\ref{totalm0}) in the next subsection.
 \begin{center}  
\begin{figure}
\begin{center}
\begin{tabular}[t]{crr}
  \hline
  \addlinespace[0.1cm]
  Transverse (T)  and
 Longitudinal (L)  & $m_W=0$ & $m_W\neq 0$  \\
 \addlinespace[0.1cm]
  \hline  
  \addlinespace[0.1cm]
   LL & $-2\frac{t(s+t)}{3{m^2_{\g}}}$  & $-\frac{t(s+t)}{3{m^2_{\g}}}$\\
  \addlinespace[0.1cm]
  \hline  
  \addlinespace[0.1cm]
  TL  & $2\frac{t(s+t)}{3{m^2_{\g}}}$  & $\frac{t(s+t)}{3{m^2_{\g}}}$\\
  \addlinespace[0.1cm]
  \hline  
  \addlinespace[0.1cm]
  LT  & $2\frac{t(s+t)}{3{m^2_{\g}}}$  & $\frac{t(s+t)}{3{m^2_{\g}}}$\\  
  \addlinespace[0.1cm]
  \hline  
  \addlinespace[0.1cm]
   TT & $-2\frac{t(s+t)}{3{m^2_{\g}}}$  & $-2\frac{t(s+t)}{3{m^2_{\g}}}$\\
   \addlinespace[0.1cm]
  \hline 
   \hline 
   \addlinespace[0.1cm]
Total contribution  & $0$  & $-\frac{t(s+t)}{3{m^2_{\g}}}$\\
  \addlinespace[0.1cm]
  \hline
\end{tabular}
\end{center}
   \label{table:TL} 
    \caption{The contributions of the spin sums (\ref{glproj}) and (\ref{wproj}) to $\sum_{spin}|\mathcal{M}|^2$.}
\end{figure}
\end{center}

\subsubsection{On the mass of gauge bosons and unitarity violation at high energies}\label{sect:theo}

	We have seen so far that the gluon scattering into gravitinos provides with an expected result (\ref{qcdres}), in which the squared amplitude factorizes a mass splitting that vanishes in the limit of exact supersymmetry. On the other hand, if the gluons are replaced by W bosons in the broken phase, a new term that breaks unitarity at a certain energy scale appears, Eq.(\ref{aiuto}).

	Since the interactions of the gravitino with the gluon and the W are exactly the same (modulo constant factors), we concentrate on the longitudinal degrees of freedom of the W. These are carried by the Goldstone bosons after the spontaneous breaking of the gauge symmetry and are connected to the non vanishing mass of the W.
In the following, we show that their behaviour is non trivial at both low and high energies. In other words, the divergence (\ref{aiuto}) appears not only in the electroweak phase, but also at high energies.

	We consider the amplitude of the s-channel, with an off-shell W. It contains the product:
\be
\dfrac{-i}{s-m^2_W}\left[\eta_{\sigma\nu}+(\xi-1)\dfrac{(k+k^{\prime})_\sigma(k+k^{\prime})_\nu}{s-\xi m^2_W}\right]
\left\{\bar{\psi}_{\mu}^s(p)\left[\slashed{k}+\slashed{k}^{\prime},\gamma^\nu\right]
\gamma^\mu v^c_{s^\prime}(p^\prime)\right\}.
\ee
	The W boson propagator in the $\rm R_\xi$ gauge can be written as follows:
\bea
&<A_\nu(q)A_\sigma(-q)>=<A_\nu(k+k^\prime)A_\sigma(-k-k^\prime)>=\nonumber\\
&\dfrac{-i}{s-m^2_W}\left[\eta_{\sigma\nu}+(\xi-1)\dfrac{(k+k^{\prime})_\sigma(k+k^{\prime})_\nu}{s-\xi m^2_W}\right].
\eea
	This is a completely general formula, that reduces to the usual propagators in the Landau or Feynman gauge by choosing a particular value (or limit) for $\xi$. In particular, the unitarity gauge corresponds to $\xi\rightarrow\infty$:
\be
<A_\nu(k+k^\prime)A_\sigma(-k-k^\prime)>=\dfrac{-i}{s-m^2_W}\left[\eta_{\sigma\nu}+\dfrac{(k+k^{\prime})_\sigma(k+k^{\prime})_\nu}{m^2_W}\right].
\ee
	Here we see that, for both massless and massive gauge bosons, the part of the propagator which takes into account the longitudinal degrees of freedom of the particle,
\be
\dfrac{-i}{s-m^2_W}\left[(\xi-1)\dfrac{(k+k^{\prime})_\sigma(k+k^{\prime})_\nu}{s-\xi m^2_W}\right]\,,
\ee
is cancelled identically when contracted with the SUGRA vertex
\be
V(\widetilde{G},A^c_\nu,\w^c)=-\dfrac{i}{4M_{Pl}}\left[\slashed{k}+\slashed{k}^\prime,\gamma^\nu\right]\gamma^\mu\,.
\ee
This happens by virtue of the commutator. The result is thus the same in the Feynman ($\xi=1$), Landau ($\xi=0$) and unitarity gauge ($\xi\rightarrow\infty$). It is clear that this is a general result in supergravity.

	It is well known from the Standard Model that for massless gauge bosons, the longitudinal degrees of freedom are unphysical and cancel during the calculation. In contrast, if the gauge bosons are massive such cancellations should not occur, as the longitudinal polarizations of the W carry physical degrees of freedom. This might be responsible for what happens when in the squared amplitude we find terms such as
\be
\sum_{spin}|\overline{\mathcal{M}}|^2\approx-\dfrac{t(s+t)}{3\Mpl^2m^2_{\g}}\Longrightarrow \sigma_{tot}\approx \dfrac{s}{\Mpl^2\mg^2}\,.
\ee
In fact, the Feynman and the unitarity gauge are not equivalent. If the propagator is expressed in the Feynman gauge as \cite{P}
\be
<A_\nu(k+k^\prime)A_\sigma(-k-k^\prime)>=-i\frac{\eta_{\sigma\nu}}{s-m^2_W}\,,
\ee
one must recover the missing longitudinal modes by adding further diagrams with the exchange of a scalar. In the Standard Model, once the gauge symmetry is broken, the unphysical Goldstone boson is eaten by a massive gauge boson. The Goldstone controls the amplitude for emission or absorption of the gauge boson in its longitudinal polarization state \cite{P}.

	Consider now the degrees of freedom which are involved in the Higgs mechanism. We see that a massless gauge boson (two transverse polarizations) combines with a scalar Goldstone (one polarization) to produce a massive vector boson, with three polarizations. When the boson is at rest, the three polarization states are completely equivalent. At relativistic velocities the longitudinal degree of freedom "decouples" instead as a single Goldstone boson, see Fig.(3.5). This effect is called the Goldstone boson equivalence theorem and was first proven by Cornwall, Levin, Tiktopoulos and Vayonakis \cite{Gold}.
\begin{figure}
\begin{center}
\resizebox{12cm}{!}{
    \begin{picture}(471,232) (71,-54)
    \SetWidth{0.5}
    \SetColor{Black}
    \Text(272,63)[]{\Large{\Black{$=$}}}
    \Text(527,62)[]{\Large{\Black{$\times \left[ 1+\mathcal{O}\left( \dfrac{m^2_W}{E^2}\right)\right]$}}}
    \SetWidth{0.5}
    \Line(340,-48)(402,24)
    \Line(465,-54)(427,9)
    \Line(226,-54)(188,9)
    \Photon(246,165)(170,74){7.5}{8.5}
    \Text(268,170)[]{\Large{\Black{$W_L$}}}
    \DashArrowLine(419,81)(483,160){8}
    \Text(494,170)[]{\LARGE{\Black{$\phi$}}}
    \Line(71,63)(165,63)
    \Line(101,-48)(163,24)
    \GOval(170,60)(73,34)(0){0.882}
    \Line(314,62)(408,62)
    \GOval(412,60)(73,34)(0){0.882}
  \end{picture}
}
     \end{center}
  \label{fig:gold}
  \caption{Visualization of the Goldstone boson equivalence theorem. At high energy, the amplitude with absorption or emission of a massive gauge boson is equal to the amplitude with absorption or emission of the Goldstone scalar which was eaten by the gauge boson.}
\end{figure}
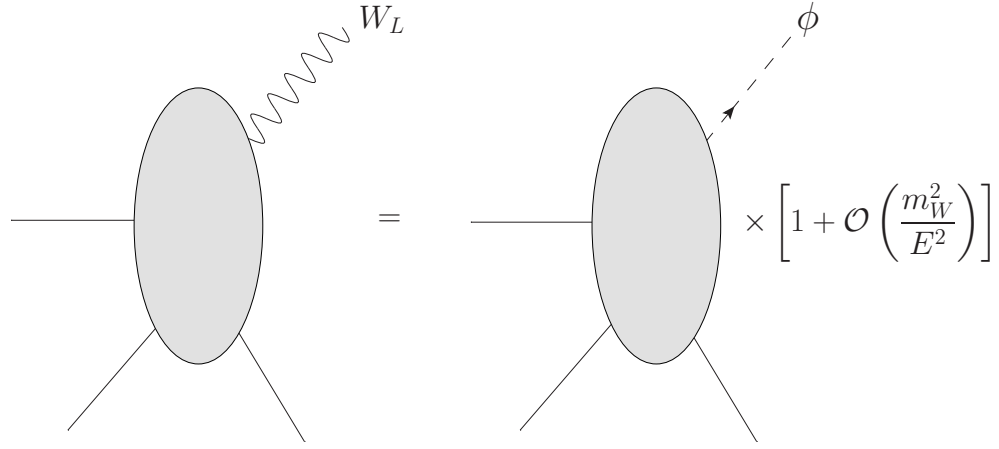
	Let us therefore consider a massive W boson. When it is at rest, the momentum is $k^\mu=(m_W,0,0,0)$ and the polarization vector is a linear combination of three orthogonal unit vectors:
\be
(0,1,0,0)\,,\qquad (0,0,1,0)\,,\qquad (0,0,0,1)\,.
\ee
	The first two give the transverse polarizations. Boosting the boson along the $\hat{z}$-axis changes the momentum to $k^\mu=(E_\mathbf{k},0,0,k)$. The polarization vectors are three unit vectors and satisfy
\be
\epsilon^\mu k_\mu=0\,, \qquad \epsilon\cdot\epsilon=-1\,.
\ee
	Clearly, since the boost is along the third direction, the transverse vectors do not change. Instead, the third one corresponds to the longitudinal degree of freedom,
\be
\epsilon_L^\alpha(k)=\left(\dfrac{k}{m_W},0,0,\dfrac{E_\mathbf{k}}{m_W} \right)\,,
\ee
which in the approximation $k\rightarrow\infty$ can be recast, component by component, as
\be
\epsilon_L^\alpha(k)=\dfrac{k^\alpha}{m_W}+\mathcal{O}\left(\dfrac{m_W}{E_\mathbf{k}}\right)\,.
\label{polvec}
\ee
The components of $k^\alpha$ are growing with $k$, in agreement with the general relations $\epsilon_L\cdot k=0$ and $k\cdot k=m^2_W$.

	By keeping in mind this last formula, we can now discuss the high energy behaviour of the result (\ref{totalm0}), and prove that the longitudinal polarizations of the two on-shell W bosons in the scattering
\be
W^a+W^b\longrightarrow \w^c+\g\,,
\ee
become strongly interacting at high centre of mass energies. This implies that the term
\be
-\dfrac{t(s+t)}{3\Mpl^2\mg^2}\,,
\ee 
holds also above the electroweak threshold, namely at any energy scale.

	In supersymmetric theories with unbroken gauge symmetry, the supercurrent $S_\mu$ is conserved in the SUSY limit. Namely, the right hand side of
\be
p^{\mu}S_{\mu}=\Delta(M)\,,
\label{conserv}
\ee
where $p^{\mu}$ is the four momentum of the gravitino and $\Delta(M)$ is the mass splitting of the supermultiplet, vanishes when all the masses are set to zero. However, in the electroweak theory the W mass cannot be set to vanish. In other words, we expect that in the process (\ref{wbosons}) the longitudinal modes of the W become strongly interacting in the limit $m_W\rightarrow 0$. Accordingly, in this case the statement (\ref{conserv}) does no longer hold. We prove this claim immediately. The matrix element can be written as 
\bea
{\cal M} = \bar{\psi}_\mu(p) S^\mu\,, 
\nonumber
\eea
therefore the squared amplitude can be recast as
\bea
 |{\cal M}|^2 = \bar{S}^{\nu} \psi_\nu(p)\cdot \bar{\psi}_\mu(p) S^\mu\,.
\eea
When summed over the gravitino polarizations, this becomes
\bea
 \sum_{\psi-spin} |{\cal M}|^2 = \bar{S}^{\nu} \left[\sum_{\psi-spin}
\psi_\nu(p)\cdot \bar{\psi}_\mu(p)\right] S^\mu\,.
\eea
	Since a term with the gravitino mass squared in the denominator can appear only from the longitudinal spin 1/2 components of $\g$ in either (\ref{gravprojtot}) or (\ref{gravprojred}), one can replace the gravitino projector with the second term of Eq.(\ref{gravprojred}), that is:
\bea
 \sum_{\psi-spin} |{\cal M}|^2 \left(\propto \dfrac{1}{\mg^2}\right) = \bar{S}^{\nu} \left[\dfrac{2}{3} \slashed{p} \dfrac{p_\nu
p_\mu}{\mg^2} \right] S^\mu\,.
\eea
	Therefore, it is always proportional to $p_\mu S^\mu = {\cal M}(\bar{\psi}_\mu \to p_\mu)$, that is to the matrix element with the
replacement $\bar{\psi}_\mu\to p_\mu$\,.

	It is now easy to prove that the scalar density $p_\mu S^\mu$ is not proportional to any SUSY breaking parameter, and it does not vanish in the SUSY limit. By explicit calculation, it can be shown that Eqs.(\ref{wwamp}) provide with the following\footnote{We have used equations like $k^\alpha \epsilon_\alpha(k) = k'^\beta
\epsilon_\beta(k') = 0$, $\slashed{p}'v^c(p') = m_{\widetilde{W}} v^c(p') =
{\cal O}(m)$, and $k^2 = k'^2 = m_W^2$.}:
\be
p^\mu S_\mu=\dfrac{g\epsilon_{abc}}{4\Mpl}\;\dfrac{\left[S^{\alpha\beta}_{ab}+\mathcal{O}(m)\right]v^c}{(s-m_W^2)(t-\mw^2)(-s-t+2m^2_W+\mg^2)}\;\epsilon^a_{\alpha}(k)\epsilon^b_{\beta}(k^{\prime})\,.
\label{curr}
\ee
	The tensor density $S^{\alpha\beta}_{ab}$ has a rather nontrivial expression:
\bea
&&S^{\alpha\beta}_{ab} =2m^2_W\left[s^2(-\gamma^\alpha\gamma^\beta\slashed{k}^{\prime}-2p^{\prime\beta}\gamma^\alpha)+t^2(\eta^{\alpha\beta}\slashed{k}-\eta^{\alpha\beta}\slashed{k}^{\prime} -2k^\beta\gamma^\alpha+2k^{\prime\alpha}\gamma^\beta)+\right.\nonumber\\
&&\left. +st(-\eta^{\alpha\beta}\slashed{k}-\eta^{\alpha\beta}\slashed{k}^{\prime}-2k^\beta\gamma^\alpha+2k^{\prime\alpha}\gamma^\beta
+\gamma^\alpha\gamma^\beta\slashed{k}-\gamma^\alpha\gamma^\beta\slashed{k}^\prime-2p^{\prime\alpha}\gamma^\beta-2p^{\prime\beta}\gamma^\alpha)     \right]\,.\nonumber\\
\eea
	At this point, one would be tempted to neglect the terms proportional to $m^2_W$ as $m_W\rightarrow 0$, and the claim of the conservation of the supercurrent would immediately follow. However, if we now recall Eq.(\ref{polvec}) and substitute it into (\ref{curr}), the result is (modulo constant factors):
\be
\dfrac{p^\mu S_\mu}{\mg}\approx\frac{(\slashed{k}^{\prime}-\slashed{k})}{\Mpl\mg}v^c+\mathcal{O}\left( \dfrac{m^2}{\mg}\right)\,,
\label{highlimit}
\ee
where now $m$ is really \emph{any} mass. We have thus seen that in the scattering of two massive gauge bosons in SUGRA, the supercurrent is not conserved in the SUSY limit. This happens since the non vanishing term in the right hand side of (\ref{highlimit}) violates the conservation of the supercurrent. It is also easy to show that the corrections $\mathcal{O}\left(m/E_\mathbf{k}\right)$ provide with terms proportional to $m_W^4$, which are absorbed into $\mathcal{O}\left(m^2/\mg\right)$. Therefore the proof given above is valid also at low energies.

	To summarise, in this section it has been shown that the structure of the supergravity vertex cancels the longitudinal degrees of freedom of \emph{any} propagating gauge boson. If the boson is massive, terms with bad high energy behaviour are generated. With an argument based on the supercurrent, we have proven that the above result holds both at low and high scales. At high centre of mass scales, the longitudinal degrees of freedom of the W become strongly interacting and violate the unitarity of the theory.

\section{Gauge eigenstates and the Higgs boson}

	In the previous sections we have understood that the terms which lead to violation of unitarity in the cross section of the process
\be
W^a+W^b\longrightarrow \w^c+\g\,,
\ee
are related to the missing longitudinal polarizations of the off-shell W in the s-channel. Accordingly, if we want to solve the problem by cancelling the divergences, we must recover those missing degrees of freedom in some way. In the case of (\ref{wbosons}), the only possibility is an annihilation channel with an off-shell scalar. We can anyway see immediately that in the basis of gauge eigenstates there might be some formal difficulties.

	First of all, in order to be consistent with the process at hand, the scalar $\phi^c$ has to be included in the adjoint representation of the gauge group. However, in Supergravity the scalars are put into the fundamental representation. In this case anyway, the gauge symmetry is broken, so the particle multiplets can pass from a representation to the other, by virtue of the mixing of gauge eigenstates. We consider the dominant couplings of the gravitino with the supercurrent in the supergravity Lagrangian (\ref{sugral}),
\be
\mathcal{L}_{\psi
J}=-\dfrac{i}{\sqrt{2}\Mpl}\left(\mathcal{\widetilde{D}_\nu}\phi^{\ast j}\bar{\psi}_{\nu}\gamma^\mu\gamma^\nu\chi^j_L-\widetilde{D}_\nu \phi
^j\bar{\chi}^j_R\gamma^\nu\gamma^\mu\psi_{\nu}\right)
-\dfrac{i}{8\Mpl}\bar{\psi}_{\mu}\left[\gamma^\nu,\gamma^\rho\right]\gamma^\mu\lambda^{(a)}F^{(a)}_{\nu\rho}\,,
\ee 
where
$\chi^j_{L,R}=P_{L,R}\chi^j=\dfrac{1}{2}\left(1\mp\gamma^5\right)\chi^j$. The coupling of the gravitino to a scalar and a gaugino does not exist in the above equation. We can anyway use the method of mass insertions, and invoke the mixing of the gaugino with a chiral fermion that couples to the gravitino and to a scalar.
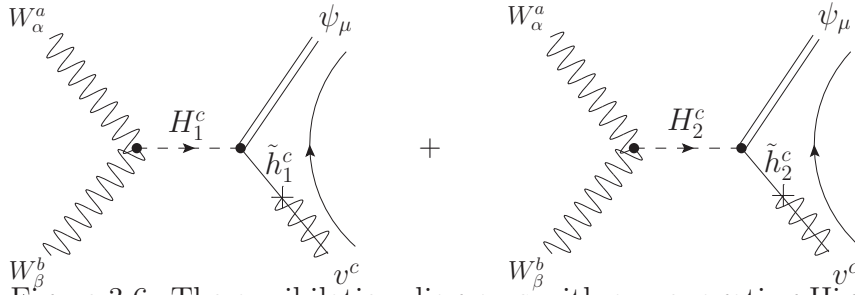
\begin{figure}[t]
\resizebox{\textwidth}{!}{
 \begin{picture}(614,131) (-40,12)
 \SetWidth{0.5}
    \SetColor{Black}
    \Text(255,71)[]{\Large{\Black{$+$}}}
    \SetWidth{0.5}
    \Vertex(425,71){2.83}
    \Photon(320,132)(366,71){7.5}{8}
    \Photon(321,13)(366,71){7.5}{8}
    \Line(423.34,72.12)(464.34,133.12)\Line(426.66,69.88)(467.66,130.88)
    \Vertex(366,71){2.83}
    \Text(308,2)[]{\large{\Black{$W^b_{\beta}$}}}
    \Text(308,143)[]{\large{\Black{$W^a_{\alpha}$}}}
    \Text(478,143)[]{\Large{\Black{$\psi_{\mu}$}}}
    \Text(396,85)[]{\Large{\Black{$H^c_2$}}}
    \Line(425,71)(473,14)
    \Photon(473,15)(448,45){7.5}{3.5}
    \Line(442,45)(454,45)\Line(448,51)(448,39)
    \Text(447,64)[]{\Large{\Black{$\tilde{h}^c_2$}}}
    \Text(484,2)[]{\Large{\Black{$v^c$}}}
    \Vertex(150,71){2.83}
    \Photon(45,131)(93,71){7.5}{8}
    \Photon(46,12)(93,71){7.5}{8}
    \Line(148.35,72.13)(189.35,132.13)\Line(151.65,69.87)(192.65,129.87)
    \Text(33,2)[]{\large{\Black{$W^b_{\beta}$}}}
    \Text(33,142)[]{\large{\Black{$W^a_{\alpha}$}}}
    \Text(203,142)[]{\Large{\Black{$\psi_{\mu}$}}}
    \Text(121,84)[]{\Large{\Black{$H^c_1$}}}
    \Line(150,71)(198,13)
    \Photon(198,14)(173,44){7.5}{3.5}
    \Line(167,44)(179,44)\Line(173,50)(173,38)
    \Text(172,63)[]{\Large{\Black{$\tilde{h}^c_1$}}}
    \Text(209,2)[]{\Large{\Black{$v^c$}}}
    \DashArrowLine(366,71)(425,71){5}
    \DashArrowLine(93,71)(150,71){5}
    \Vertex(93,71){2.83}
    \ArrowArcn(540.1,72.96)(75.11,228.16,137.19)
    \ArrowArcn(263.1,72.96)(75.11,228.16,137.19)
  \end{picture}
  }
\caption{The annihilation diagrams with a propagating Higgs.}
\label{fig:higgs}
\end{figure}

	In contrast, the supergravity couplings in (\ref{wwamp}) regard Majorana fermions which do not have definite helicity. Namely, in the interference of the Higgs scalars with the original amplitude (\ref{matrixw}) that is now labelled with $M^{(0)}$, some polarizations which might in principle contribute to the process are cancelled. Accordingly, we expect that the additional diagrams in figure (\ref{fig:higgs}) would not cancel the divergences (\ref{aiuto}).
	In the following, we consider the particle content of the MSSM \cite{HK}, that is summarised in the table at the end of this section. It contains two neutral Higgs bosons $H^{0}_1$ and $H^{0}_2$, which are suitable to our process\footnote{Throughout this section these are called $H^{c}_1$ and $H^{c}_2$, where $c$ is an index of $\ew$.}. These break the gauge symmetry by taking the vacuum expectation values $v_1$ and $v_2$,
\be
\left\langle H^0_1\right\rangle = \dfrac{v_1}{\sqrt{2}}\left( \begin{array}{c}
1\\
0 \end{array}\right), \qquad \left\langle H^0_2\right\rangle = \dfrac{v_2}{\sqrt{2}}\left( \begin{array}{c}
0\\
1 \end{array}\right),
\ee
where $v_1\neq v_2$. We label with $M^{(H)}$ the amplitude which results from the two distinct diagrams with the exchange of a Higgs, as in Fig.(\ref{fig:higgs}).

	The vertex of the Higgs with the W bosons is non trivial either, since we consider a doublet of scalars instead of a singlet. Accordingly, this is not the same coupling of the Standard Model. This leads to the following expression for the vertex \cite{R}:
\be
V(H^c_iW^a_\alpha W^b_\beta)=\dfrac{ig^2}{2}C^i_R\eta^{\alpha\beta}\epsilon^{abc}, \qquad C^i_R=v_1Z^{1i}_R+v_2Z^{2i}_R\,.
\label{mixing}
\ee
$Z_R$ is the rotation matrix transforming the gauge eigenstates $H^{c}_1$ and $H^{c}_2$ into the mass eigenstates $H$ and $h$, namely the heavy and light Higgs in the MSSM:
\be
\left( \begin{array}{c}
H\\
h\\
\end{array} \right)
=
\left( \begin{array}{cc}
\cos{\alpha} & -\sin{\alpha}  \\
\sin{\alpha} & \cos{\alpha} \end{array} \right)
\left( \begin{array}{c}
H^c_1\\
H^c_2\\
\end{array} \right).
\ee
	We are actually interested only in the gauge eigenstates, therefore we impose $\alpha\equiv0$ in (\ref{mixing}) to obtain the following:
\bea
&V(H_1^cW^a_\alpha W^b_\beta)=\dfrac{ig^2}{2}\epsilon^{abc}v_1\eta^{\alpha\beta}\,,\\
&V(H_2^cW^a_\alpha W^b_\beta)=\dfrac{ig^2}{2}\epsilon^{abc}v_2\eta^{\alpha\beta}\,.
\eea
	Recalling the mass eigenstates of the MSSM, we remember that the higgsinos and the winos mix with each other. This justifies the mass insertions. Physically, the higgsino $\tilde{h}_i$, which interacts with the scalar $H^c_i$ and with the gravitino $\g$, at a certain point becomes the wino. The mass insertion keeps trace of this mechanism. The mixing factors can be found in the neutralino mass matrix \cite{HK},
\bea
&\Delta_1&=+\dfrac{gv_1}{2}\qquad\rm for \,\tilde{h}_1\,,\\
&\Delta_2&=-\dfrac{gv_2}{2}\qquad\rm for \,\tilde{h}_2\,,
\eea
where $g$ is the coupling constant of $\ew$. Moreover, the higgsino which propagates in the external leg of each diagram provides with the factor
\be
\dfrac{1}{m_{\w}-m_{\tilde{h}_i}}\approx\frac{1}{m_{\w}}\qquad(i=1,2)\,.
\ee

	The above approximation will give a clearer form of the amplitude, and it can be safely assumed since the higgsino mass does not enter the kinematics and is not relevant for the final result. We thus obtain the amplitudes corresponding to the neutral Higgs bosons of the MSSM in the basis of gauge eigenstates,
\bea
&M^{(H_1^c)}=\dfrac{g\epsilon_{abc}}{8\sqrt{2}M_P}\left(\dfrac{g^2v_1^2}{\mw}\right)
\dfrac{\bar{\psi}_{\mu}
(\slashed{k}+\slashed{k}^{\prime})\gamma^\mu(\mathbf{1}-\gamma_5)
 v^c}{s-m_{H_1}^2}\eta^{\alpha\beta}\epsilon^a_\alpha(k)\epsilon^b_\beta(k^{\prime})\,,\\
&M^{(H_2^c)}=\dfrac{g\epsilon_{abc}}{8\sqrt{2}M_P}\left(\dfrac{g^2v_2^2}{\mw}\right)
\dfrac{\bar{\psi}_{\mu}
(\slashed{k}+\slashed{k}^{\prime})\gamma^\mu(\mathbf{1}-\gamma_5)
 v^c}{s-m_{H_2}^2}\eta^{\alpha\beta}\epsilon^a_\alpha(k)\epsilon^b_\beta(k^{\prime})\,.
\eea
By recalling the fundamental relation \cite{HK}
\be
m_W^2=\dfrac{g^2}{4}(v_1^2+v_2^2)\,,
\ee
this can be recast into a single amplitude:
\bea
&M^{(H)}\equiv M^{(H^c_1)}+M^{(H^c_2)}=\dfrac{g\epsilon_{abc}}{8\sqrt{2}\Mpl}\times
\nonumber\\
&\times\left\{ \dfrac{m_W^2}{4}s+\left[g^2(m_{H_1}^2-m_{H_2}^2)v_1^2-4m_W^2m_{H_1}^2 \right] \right\}\dfrac{\bar{\psi}_{\mu}
(\slashed{k}+\slashed{k}^{\prime})\gamma^\mu(\mathbf{1}-\gamma_5)
v^c}{\mw(s-m_{H_1}^2)(s-m_{H_2}^2)}\eta^{\alpha\beta}\epsilon^a_\alpha\epsilon^b_\beta\,.\nonumber\\
\eea
	Now we can go back to the original calculation. It can be proven that the anomaly in (\ref{aiuto}) is not cancelled by adding the above expression to (\ref{matrixw}). The dominant divergences result as:
\bea
&\sum_{spin}|M^{(0)}+M^{(H)}|^2\approx\dfrac{g^2|\epsilon^{abc}|^2}{\Mpl^2}\times
\nonumber\\
&\times\left[ -4\dfrac{t(s+t)}{3\mg^2}+\dfrac{1}{64\times48}\left( \dfrac{s^3}{\mw^2\mg^2}-\dfrac{s^2(7s+3t)}{\mw^2(s+t)} -\dfrac{64s(s+2t)}{\sqrt{2}\mw\mg}-\dfrac{3s^2}{\mg^2}\right) \right]\,.\nonumber\\
\label{m0mh}
\eea
	The interferences of $M^{(H)}$ with $M^{(0)}$ give only the term with $\sqrt{2}$. This is expected from our previous discussion, since the Higgs vertex contains the helicity projectors which cancel a number of contributions.
\begin{figure}[t]
\resizebox{\textwidth}{!}{
 \begin{picture}(614,121) (4,2)
 \SetWidth{0.5}
    \SetColor{Black}
    \Text(460.88,65.42)[]{\Large{\Black{$+$}}}
    \Text(301.06,65.42)[]{\Large{\Black{$+$}}}
    \Text(140.5,63.93)[]{\Large{\Black{$+$}}}
    \SetWidth{0.5}
    \Vertex(215.57,95.89){2.1}
    \Line(214.83,29.73)(276.53,18.58)
    \Vertex(214.83,29.73){2.1}
    \Text(152.34,8.92)[]{\large{\Black{$W^b_{\beta}$}}}
    \Text(152.88,115.22)[]{\large{\Black{$W^a_{\alpha}$}}}
    \Text(285.45,113.73)[]{\Large{\Black{$\psi_{\mu}$}}}
    \Text(285.45,11.15)[]{\Large{\Black{$v^c$}}}
    \Line(216.77,98.1)(276.99,109.99)\Line(217.35,95.18)(277.56,107.07)
    \Photon(215.57,98.12)(164.28,108.53){5.58}{8}
    \Photon(214.09,29.73)(164.28,18.58){5.58}{8}
    \SetWidth{1.0}
    \Line(241.59,19.33)(250.51,28.25)\Line(241.59,28.25)(250.51,19.33)
    \Text(225.98,14.87)[]{\Large{\Black{$\tilde{h}^c_2$}}}
    \SetWidth{0.5}
    \Vertex(373.17,95.89){2.1}
    \Line(372.42,28.25)(372.42,97.38)
    \Line(373.17,95.89)(433.38,88.46)
    \Vertex(518.12,95.89){2.1}
    \Line(517.38,28.25)(517.38,97.38)
    \Line(518.12,95.89)(578.34,88.46)
    \Vertex(372.42,29.73){2.1}
    \Text(313.93,115.22)[]{\large{\Black{$W^a_{\alpha}$}}}
    \Text(444.53,87.72)[]{\Large{\Black{$v^c$}}}
    \Text(320.65,9.66)[]{\large{\Black{$W^b_{\beta}$}}}
    \Text(445.53,38.65)[]{\Large{\Black{$\psi_{\mu}$}}}
    \Line(371.4,31.19)(433.1,43.09)\Line(371.96,28.27)(433.66,40.17)
    \Photon(324.85,110.02)(372.42,96.64){5.58}{8}
    \Photon(372.42,30.48)(327.82,19.33){5.58}{8}
    \SetWidth{1.0}
    \Line(399.19,87.72)(408.1,96.64)\Line(399.19,96.64)(408.1,87.72)
    \Text(392.49,108.53)[]{\Large{\Black{$\tilde{h}^c_1$}}}
    \SetWidth{0.5}
    \Vertex(517.38,29.73){2.1}
    \Text(458.88,115.22)[]{\large{\Black{$W^a_{\alpha}$}}}
    \Text(465.6,9.66)[]{\large{\Black{$W^b_{\beta}$}}}
    \Line(516.35,31.19)(578.05,43.09)\Line(516.92,28.27)(578.62,40.17)
    \Photon(469.8,110.02)(517.38,96.64){5.58}{8}
    \Photon(517.38,30.48)(472.78,19.33){5.58}{8}
    \SetWidth{1.0}
    \Line(544.14,87.72)(553.06,96.64)\Line(544.14,96.64)(553.06,87.72)
    \Text(537.45,108.53)[]{\Large{\Black{$\tilde{h}^c_2$}}}
    \SetWidth{0.5}
    \Line(215.57,96.64)(215.57,30.48)
    \Photon(215.57,95.15)(215.57,63.93){5.58}{4}
    \SetWidth{1.0}
    \Line(211.12,59.47)(220.03,68.39)\Line(211.12,68.39)(220.03,59.47)
    \SetWidth{0.5}
    \Photon(246.05,23.79)(277.27,19.33){5.58}{4}
    \Photon(372.42,30.48)(372.42,63.93){5.58}{4}
    \SetWidth{1.0}
    \Line(367.96,59.47)(376.88,68.39)\Line(367.96,68.39)(376.88,59.47)
    \Line(512.92,59.47)(521.84,68.39)\Line(512.92,68.39)(521.84,59.47)
    \SetWidth{0.5}
    \Photon(517.38,63.93)(518.12,30.48){5.58}{4}
    \Text(357.04,46.09)[]{\large{\Black{$\widetilde{W}^b$}}}
    \Text(202.19,46.09)[]{\Large{\Black{$\tilde{h}^a_2$}}}
    \Text(500.26,45.35)[]{\large{\Black{$\widetilde{W}^b$}}}
    \Text(200.19,78.8)[]{\large{\Black{$\widetilde{W}^a$}}}
    \Text(504.74,77.31)[]{\Large{\Black{$\tilde{h}^b_2$}}}
    \Text(360.53,78.05)[]{\Large{\Black{$\tilde{h}^b_1$}}}
    \Photon(548.6,92.18)(579.08,88.46){5.58}{4}
    \Photon(404.39,92.18)(432.64,89.2){5.58}{4}
    \Text(590.97,87.72)[]{\Large{\Black{$v^c$}}}
    \Text(591.72,38.65)[]{\Large{\Black{$\psi_{\mu}$}}}
    \Vertex(60.21,95.89){2.1}
    \Line(59.47,29.73)(121.17,18.58)
    \Line(60.21,96.64)(60.21,30.48)
    \Vertex(59.47,29.73){2.1}
    \Text(131.09,113.73)[]{\Large{\Black{$\psi_{\mu}$}}}
    \Text(130.09,11.15)[]{\Large{\Black{$v^c$}}}
    \Line(61.41,98.1)(121.62,109.99)\Line(61.99,95.18)(122.2,107.07)
    \Photon(60.21,98.12)(8.92,108.53){5.58}{8}
    \Photon(58.73,29.73)(8.92,18.58){5.58}{8}
    \SetWidth{1.0}
    \Line(55.75,59.47)(64.67,68.39)\Line(55.75,68.39)(64.67,59.47)
    \Line(86.23,19.33)(95.15,28.25)\Line(86.23,28.25)(95.15,19.33)
    \Text(70.62,14.87)[]{\Large{\Black{$\tilde{h}^c_1$}}}
    \SetWidth{0.5}
    \Photon(90.69,23.79)(121.91,18.58){5.58}{4}
    \Photon(60.21,95.89)(60.21,63.93){5.58}{4}
    \Text(47.58,46.83)[]{\Large{\Black{$\tilde{h}^a_1$}}}
    \Text(43.58,78.05)[]{\large{\Black{$\widetilde{W}^a$}}}
    \Text(-2.46,8.18)[]{\large{\Black{$W^b_{\beta}$}}}
    \Text(-4.74,115.96)[]{\large{\Black{$W^a_{\alpha}$}}}
  \end{picture}
  }
\caption{The $t-$ and $u-$ channel diagrams with the mass insertions.}
\label{fig:diatu_mi}
\end{figure}
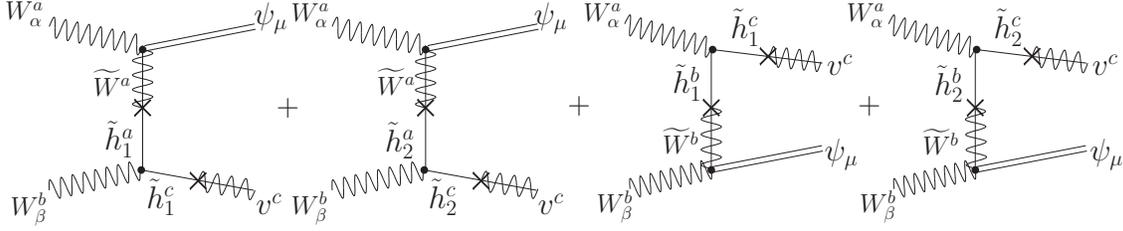
	However, by iterating the method of mass insertions, one may add further diagrams, developing a perturbation theory around the mass ratio $(m_W/\mw)$. At the first order, the matrix element $M^{(1)}$ is given by
\be
M^{(1)}=M_{t_{1,2}}+M_{u_{1,2}}\,,
\ee
where $M_{t_{1,2}}$ and $M_{u_{1,2}}$ refer to the t and u channel. The neutral higgsinos $\tilde{h}_1$ and $\tilde{h}_2$ mix both with the wino $\w$, as it is shown in figure (\ref{fig:diatu_mi}). Therefore there are two amplitudes which can be summed to give
\be
M_{t_{1,2}}=-\dfrac{1}{4\sqrt{2}}\dfrac{g\epsilon_{abc}}{M_P}\left(\dfrac{m^2_W}{m_{\w}}\right)
\left\{\bar{\psi}_{\mu}
\left[\slashed{k},\gamma^\alpha\right]\gamma^\mu\left[\dfrac{1}{t-m_{\w}^2}+\dfrac{m_{\w}(\slashed{p}-\slashed{k})}{(t-m_{\w}^2)t}\right]
\gamma^\beta v^c\right\}\epsilon^a_\alpha\epsilon^b_\beta\,,
\ee
for the diagrams with the t-channel. Regarding the u-channels, we get:
\be
M_{u_{1,2}}=-\dfrac{1}{4\sqrt{2}}\dfrac{g\epsilon_{abc}}{M_P}\left(\dfrac{m^2_W}{m_{\w}}\right)
\left\{\bar{\psi}_{\mu}
\left[\slashed{k}^{\prime},\gamma^\beta\right]\gamma^\mu\left[\dfrac{1}{u-m_{\w}^2}+\dfrac{m_{\w}(\slashed{p}-\slashed{k}^{\prime})}{(u-m_{\w}^2)u}\right]
\gamma^\alpha v^c\right\}\epsilon^a_\alpha\epsilon^b_\beta\,.
\ee
The amplitude at second order, $M^{(2)}$, contains the double mass insertion with a gaugino in the external leg. This corresponds to the diagrams in figure (\ref{fig:m2}), where $i=1,2$. The corresponding amplitudes are the following,
\bea
&&M^{(2)}=\dfrac{g\epsilon_{abc}}{4M_P}\left(\dfrac{m_W^2}{\mw^2}\right)
\Bigg(
\left[
\eta^{\alpha\beta}(k-k^{\prime})^\sigma+\eta^{\beta\sigma}(2k^{\prime}+k)^\alpha
-\eta^{\alpha\sigma}(2k+k^{\prime})^\beta\right]\times\nonumber\\
&&\times\dfrac{1}{s-m_W^2}\left(\bar{\psi}_{\mu}\left[\slashed{k}+\slashed{k}^{\prime},\gamma_\sigma\right]
\gamma^\mu
v^c\right)+\dfrac{1}
{t-\mw^2}\left\{\bar{\psi}_{\mu}
\left[\slashed{k},\gamma^\alpha\right]\gamma^\mu(\slashed{k}^{\prime}-\slashed{p}^{\prime}+\mw)
\gamma^\beta v^c\right\}\nonumber\\
&&-\dfrac{1}
{u-\mw^2}\left\{\bar{\psi}_{\mu}
[\slashed{k}^{\prime},\gamma^\beta]\gamma^\mu(\slashed{k}-\slashed{p}^{\prime}+\mw)
\gamma^\alpha v^c\right\}-\bar{\psi}_{\mu}^{(s)}
[\gamma^\alpha,\gamma^\beta]\gamma^\mu
v^c
\Bigg)
\epsilon^a_\alpha\epsilon^b_\beta\,.
\eea
The above is nothing but Eq.(\ref{matrixw}), multiplied by the factor $m_W^2/\mw^2$. The total amplitude can be finally written as the sum of the four contributions:
\be
\mathcal{M}(W^a+W^b\longrightarrow \w^c+\g) = M^{(0)}+M^{(1)}+M^{(H)}+M^{(2)}\,.
\ee
\begin{figure}[t]
\resizebox{\textwidth}{!}{
\begin{picture}(614,124) (4,2)
    \SetWidth{0.5}
    \SetColor{Black}
    \Text(14.32,120.46)[]{\large{\Black{$W^a_{\alpha}$}}}
    \Text(16.65,9.33)[]{\large{\Black{$W^b_{\beta}$}}}
    \Text(148.44,119.69)[]{\Large{\Black{$\psi_{\mu}$}}}
    \Text(81.61,83.16)[]{\large{\Black{$W^c_{\sigma}$}}}
    \Text(174.87,8.55)[]{\large{\Black{$W^b_{\beta}$}}}
    \Text(160.1,66.06)[]{\Large{\Black{$+$}}}
    \Text(607.77,115.8)[]{\Large{\Black{$\psi_{\mu}$}}}
    \SetWidth{0.5}
    \Vertex(104.92,66.84){2.2}
    \Line(103.64,66.94)(136.28,114.35)\Line(106.2,65.18)(138.84,112.59)
    \Vertex(59.07,66.06){2.2}
    \Photon(59.07,66.06)(24.09,112.69){5.83}{8}
    \Photon(58.29,66.06)(104.92,66.06){5.83}{8}
    \Photon(59.07,66.06)(24.09,20.21){5.83}{9}
    \Line(104.92,66.06)(139.89,19.43)
    \Vertex(545.59,67.62){2.2}
    \Photon(545.59,67.62)(501.29,112.69){5.83}{8}
    \Photon(545.59,67.62)(501.29,20.98){5.83}{9}
    \Text(491.96,118.91)[]{\large{\Black{$W^a_{\alpha}$}}}
    \Text(492.74,11.99)[]{\large{\Black{$W^b_{\beta}$}}}
    \Text(481.86,67.62)[]{\Large{\Black{$+$}}}
    \Text(320.2,66.84)[]{\Large{\Black{$+$}}}
    \Vertex(394.04,99.48){2.2}
    \Vertex(394.04,30.31){2.2}
    \Text(334.19,119.69)[]{\large{\Black{$W^a_{\alpha}$}}}
    \Text(338.08,9.33)[]{\large{\Black{$W^b_{\beta}$}}}
    \Text(411.91,66.06)[]{\large{\Black{$\widetilde{W}^b$}}}
    \Text(468.65,39.64)[]{\large{\Black{$\psi_{\mu}$}}}
    \Text(170.21,119.69)[]{\large{\Black{$W^a_{\alpha}$}}}
    \SetWidth{1.0}
    \Line(127.46,29.53)(132.12,37.3)\Line(125.91,35.75)(133.68,31.09)
    \SetWidth{0.5}
    \Photon(129.79,34.2)(139.89,20.21){5.83}{3}
    \Text(125.91,66.06)[]{\normalsize{\Black{$\widetilde{W}^c$}}}
    \Text(114.58,36.08)[]{\large{\Black{$\tilde{h}^c_i$}}}
    \Vertex(234.71,100.26){2.2}
    \Vertex(234.71,30.31){2.2}
    \Photon(234.71,100.26)(233.94,30.31){5.83}{8}
    \Text(307.77,118.13)[]{\Large{\Black{$\psi_{\mu}$}}}
    \Text(253.37,65.28)[]{\large{\Black{$\widetilde{W}^a$}}}
    \Line(234.42,101.78)(298.93,114.22)\Line(235.01,98.73)(299.51,111.17)
    \Photon(234.71,100.26)(181.09,112.69){5.83}{8}
    \Photon(234.71,30.31)(181.09,18.65){5.83}{8}
    \Line(234.71,100.26)(234.71,31.09)
    \SetWidth{1.0}
    \Line(412.69,93.26)(418.91,99.48)\Line(412.69,99.48)(418.91,93.26)
    \Line(439.11,88.6)(445.33,94.82)\Line(439.11,94.82)(445.33,88.6)
    \SetWidth{0.5}
    \Photon(442.22,93.26)(460.1,88.6){5.83}{3}
    \Text(427.01,83.27)[]{\large{\Black{$\tilde{h}^c_i$}}}
    \Text(410.36,110.36)[]{\normalsize{\Black{$\widetilde{W}^c$}}}
    \Text(570.35,67.51)[]{\normalsize{\Black{$\widetilde{W}^c$}}}
    \Text(605.43,15.54)[]{\Large{\Black{$v^c$}}}
    \Text(566.8,34.75)[]{\large{\Black{$\tilde{h}^c_i$}}}
    \Line(234.71,30.31)(298.44,18.65)
    \SetWidth{1.0}
    \Line(251.03,23.32)(257.25,29.53)\Line(251.03,29.53)(257.25,23.32)
    \Line(277.46,18.65)(283.68,24.87)\Line(277.46,24.87)(283.68,18.65)
    \SetWidth{0.5}
    \Photon(234.71,30.31)(254.14,26.42){5.83}{3}
    \Photon(280.57,23.32)(299.22,18.65){5.83}{3}
    \Text(243.04,15.32)[]{\normalsize{\Black{$\widetilde{W}^c$}}}
    \Text(270.91,34.64)[]{\large{\Black{$\tilde{h}^c_i$}}}
    \SetWidth{1.0}
    \Line(115.02,45.85)(119.69,53.63)\Line(113.47,52.07)(121.24,47.41)
    \SetWidth{0.5}
    \Photon(104.92,66.84)(114.25,53.63){5.83}{2}
    \Line(393.26,28.76)(393.26,101.04)
    \Photon(394.04,99.48)(415.8,96.37){5.83}{3}
    \Photon(346.63,114.25)(394.04,99.48){5.83}{8}
    \Line(394.04,99.48)(460.1,88.6)
    \Line(393.74,31.84)(456.69,44.27)\Line(394.34,28.79)(457.29,41.22)
    \Photon(394.04,30.31)(346.63,19.43){5.83}{7}
    \Photon(394.04,99.48)(394.04,30.31){5.83}{7}
    \Photon(578.23,34.97)(595.33,18.65){5.83}{3}
    \SetWidth{1.0}
    \Line(575.9,31.87)(580.56,39.64)\Line(574.35,38.08)(582.12,33.42)
    \Line(562.69,45.08)(567.35,52.85)\Line(561.13,51.29)(568.91,46.63)
    \SetWidth{0.5}
    \Line(545.59,67.62)(595.33,18.65)
    \Line(544.56,68.78)(594.3,113.08)\Line(546.62,66.46)(596.36,110.76)
    \Photon(545.59,67.62)(564.24,48.96){5.83}{3}
    \Text(146.89,12.44)[]{\Large{\Black{$v^c$}}}
    \Text(471.76,87.05)[]{\Large{\Black{$v^c$}}}
    \Text(306.99,11.66)[]{\Large{\Black{$v^c$}}}
  \end{picture}
}
\caption{The diagrams with two mass insertions in the external legs.}
\label{fig:m2}
\end{figure}
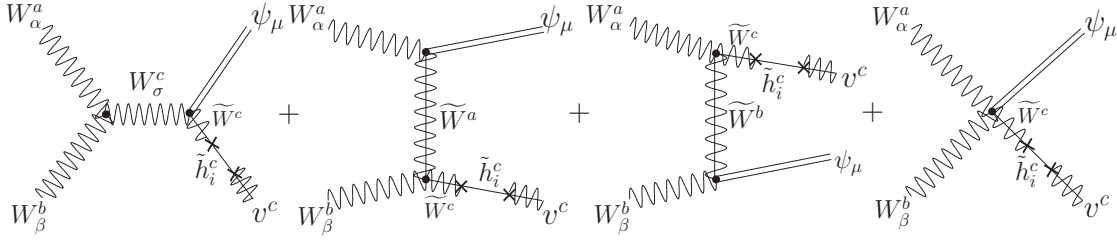
	The squares and the interference of $M^{(1)}$ and $M^{(2)}$ give subdominant terms which are proportional to for instance
\be
\left(\dfrac{m_W^2}{m_{\w}^2}\right)
\dfrac{t(s+t)}{3{m^2_{\g}}}\,.
\ee
Thus the leading divergences appear only in Eq.(\ref{m0mh}) and in
\be
\sum_{spin}|M^{(1)}+M^{(H)}|^2\approx\dfrac{g^2|\epsilon^{abc}|^2}{48\Mpl^2}\dfrac{s(s+2t)}{\sqrt{2}\mw\mg}\,.
\label{m1mh}
\ee
	Remarkably, the above factor cancels exactly the corresponding term in Eq.(\ref{m0mh}). This means that the contribution of the amplitude $M^{(1)}$ is suppressed, and that the Higgs diagrams contribute only with the square $\sum_{spin}|M^{(H)}|^2$. We can then conclude that, as expected, the neutral Higgs doublet of the MSSM gauge eigenstates do not restore unitarity.

\section{WW scattering in the basis of MSSM mass eigenstates}

\begin{center}

\begin{figure}[t]
\begin{center}
\begin{tabular}{|l|l l |l l|l|l|}
	\hline
\multicolumn{1}{|c|}
	{Ordinary particles}&
	 \multicolumn{4}{|c|} { Supersymmetric partners} \\  
	 \hline

	& \multicolumn{1}{ p{2.3cm}}{Gauge eigenstates}
	& \multicolumn{1}{ p{1.8cm}}{Name}
	& \multicolumn{1}{|p{2.2cm}}{Mass eigenstates}
	& \multicolumn{1}{ p{2cm}|}{Name}
	\\ \hline \hline
	&&& &\\
q=u,d,s,c,b,t   & $\tilde{q}_L,\tilde{q}_R$ & squark       &           $\tilde{q}_1,\tilde{q}_2$ & squark  \\

$\ell=e,\mu,\tau$  &  $\tilde{\ell}_L,\tilde{\ell}_R$ & slepton &  $\tilde{\ell}_1,\tilde{\ell}_2$ & slepton \\

$g$   &  $\tilde{g}$ &  gluino & $\tilde{g}$  & gluino  \\
&&& &\\
$\mathrm{W^\pm}$   &  $\mathrm{\widetilde{W}^\pm}$  & wino  & &\\

$\mathrm{H_u^+}$   & $\widetilde{\mathrm{H}}_u^+$  & higgsino & $\widetilde{\chi}^\pm_{1,2}$ & charginos \\

$\mathrm{H_d^-}$   & $\widetilde{\mathrm{H}}_d^-$  & higgsino && \\
 &&& &\\
$\gamma$ & $\widetilde{\gamma}$ & photino && \\
 $\mathrm{Z^0}$& $\widetilde{\mathrm{Z}}^0$ & zino &&\\
   
 $\mathrm{H_u^0}$&	$\widetilde{\mathrm{H}}_u^0$ 	& higgsino & $\neutr_i$ & neutralinos \\
   
  $\mathrm{H_d^0}$&	$\widetilde{\mathrm{H}}_d^0$ & higgsino &&\\
  
  $ \mathrm{W}^3$ & $\widetilde{\mathrm{W}}^3$ & wino&& \\
  
   $\mathrm{B} $ &	 $\widetilde{\mathrm{B}}$ & bino & &\\ 
  \hline
\end{tabular}

\end{center}
   \label{table:MSSM} 
    \caption{The particle spectrum of the MSSM.}
\end{figure}
\end{center}

	We have seen in the previous sections that the scattering
\be
W^a(k)+W^b(k^{\prime})\longrightarrow \w^c(p^{\prime})+\g(p)\,,
\ee
in the case of massive W bosons, provides terms like (\ref{aiuto}) in the squared amplitude. Such quantities can violate the unitarity of the scattering matrix above a certain scale.
The basis of gauge eigenstates is useful from a theoretical viewpoint, since it separates the contributions of the distinct gauge groups. The calculations turn out to be also simpler. However, we have seen that in this framework the anomalous terms are not cancelled.

	Moreover, the \emph{observation} of particles at colliders is based on the electroweak gauge eigenstates, while the mass eigenstates are the \emph{propagating} states, when the observation is interpreted within the theory. In the Standard Model, for an unbroken gauge symmetry such as $\rm SU(3)_C$, there is no mixing. On the other hand, when the gauge symmetry is spontaneously broken, the mass matrices are non diagonal. The eigenstates of such matrices are new physical particles, obtained after a suitable rotation that follows a change of basis in the field space.

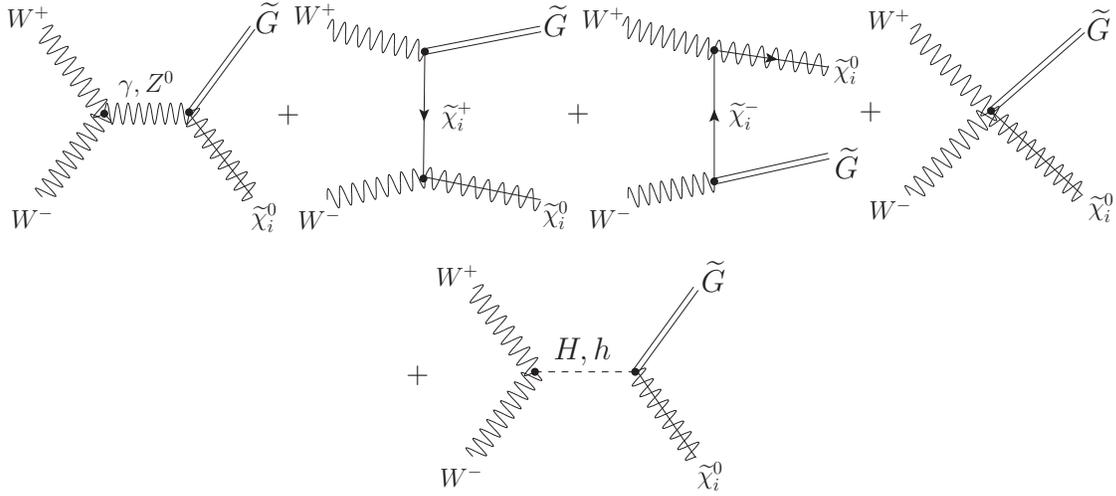
\begin{figure}[t]
\resizebox{\textwidth}{!}{
  \begin{picture}(614,267) (4,3)
    \SetWidth{0.5}
    \SetColor{Black}
    \Text(81.4,226.37)[]{\large{\Black{$\gamma,Z^0$}}}
    \Text(159.7,209.31)[]{\Large{\Black{$+$}}}
    \Text(606.23,258.93)[]{\Large{\Black{$\widetilde{G}$}}}
    \Text(607.78,158.15)[]{\large{\Black{$\widetilde{\chi}^0_i$}}}
    \SetWidth{0.5}
    \Photon(58.14,209.31)(104.66,209.31){5.81}{8}
    \Vertex(544.99,210.86){2.19}
    \Text(490.72,262.03)[]{\large{\Black{$W^+$}}}
    \Text(491.5,157.37)[]{\large{\Black{$W^-$}}}
    \Text(480.64,210.86)[]{\Large{\Black{$+$}}}
    \Text(319.4,210.09)[]{\Large{\Black{$+$}}}
    \Vertex(393.04,244.2){2.19}
    \Vertex(393.04,172.1){2.19}
    \Text(333.35,262.8)[]{\large{\Black{$W^+$}}}
    \Text(467.47,234.12)[]{\large{\Black{$\widetilde{\chi}^0_i$}}}
    \Text(337.23,152.72)[]{\large{\Black{$W^-$}}}
    \Text(410.87,209.31)[]{\large{\Black{$\widetilde{\chi}^-_i$}}}
    \Text(467.47,182.96)[]{\Large{\Black{$\widetilde{G}$}}}
    \Vertex(234.12,243.42){2.19}
    \Line(234.12,174.43)(297.69,162.02)
    \Vertex(233.35,173.65){2.19}
    \Text(306.99,261.25)[]{\Large{\Black{$\widetilde{G}$}}}
    \Text(306.99,154.27)[]{\large{\Black{$\widetilde{\chi}^0_i$}}}
    \Text(252.73,208.54)[]{\large{\Black{$\widetilde{\chi}^+_i$}}}
    \Photon(234.12,243.42)(180.63,255.83){5.81}{8}
    \Photon(234.12,174.43)(180.63,162.02){5.81}{8}
    \Vertex(349.63,67.45){2.19}
    \Line(350.41,68.22)(382.97,20.16)
    \Text(231.02,65.89)[]{\Large{\Black{$+$}}}
    \DashLine(294.59,67.45)(349.63,67.45){3.88}
    \Vertex(294.59,67.45){2.19}
    \Text(252.73,120.94)[]{\large{\Black{$W^+$}}}
    \Text(255.05,10.08)[]{\large{\Black{$W^-$}}}
    \Photon(294.59,67.45)(260.48,113.18){5.81}{8}
    \Photon(294.59,67.45)(260.48,20.93){5.81}{9}
    \Text(393.82,120.16)[]{\Large{\Black{$\widetilde{G}$}}}
    \Text(392.27,10.08)[]{\large{\Black{$\widetilde{\chi}^0_i$}}}
    \Line(348.38,68.36)(381.71,114.1)\Line(350.88,66.53)(384.22,112.27)
    \Line(104.66,209.31)(139.54,163.57)
    \Text(148.07,262.8)[]{\Large{\Black{$\widetilde{G}$}}}
    \Text(146.52,152.72)[]{\large{\Black{$\widetilde{\chi}^0_i$}}}
    \Vertex(104.66,210.09){2.19}
    \Vertex(58.14,209.31){2.19}
    \Text(16.28,263.58)[]{\large{\Black{$W^+$}}}
    \Text(18.61,152.72)[]{\large{\Black{$W^-$}}}
    \Photon(58.14,209.31)(24.03,255.83){5.81}{8}
    \Photon(58.14,209.31)(24.03,163.57){5.81}{9}
    \Line(103.41,210.23)(138.29,257.52)\Line(105.9,208.39)(140.79,255.68)
    \Line(233.82,244.95)(297.39,257.35)\Line(234.42,241.9)(297.99,254.31)
    \Photon(393.04,244.2)(455.84,235.67){5.81}{8}
    \ArrowLine(393.04,244.2)(455.84,234.9)
    \Photon(342.65,257.38)(393.04,244.2){5.81}{8}
    \Line(392.71,173.62)(455.5,187.57)\Line(393.38,170.59)(456.17,184.54)
    \Photon(393.04,172.1)(345.75,162.8){5.81}{8}
    \Photon(544.21,210.09)(594.6,164.35){5.81}{10}
    \Line(544.99,210.86)(595.38,164.35)
    \Line(543.97,212.03)(594.36,256.22)\Line(546.01,209.7)(596.4,253.89)
    \Photon(544.99,210.86)(500.8,255.83){5.81}{8}
    \Photon(544.99,210.86)(500.8,164.35){5.81}{9}
    \Text(321.72,79.07)[]{\Large{\Black{$H,h$}}}
    \Text(176.75,151.95)[]{\large{\Black{$W^-$}}}
    \Text(174.43,265.13)[]{\large{\Black{$W^+$}}}
    \Photon(104.66,210.09)(139.54,163.57){5.81}{8}
    \ArrowLine(234.12,243.42)(233.35,173.65)
    \Photon(234.12,174.43)(297.69,161.25){5.81}{8}
    \ArrowLine(393.04,172.1)(393.04,244.2)
    \Photon(349.63,67.45)(382.97,20.93){5.81}{8}
  \end{picture}
  }
\caption{WW scattering in supergravity in the basis of mass eigenstates.}
\label{fig:wmass}
\end{figure}

	This happens naturally also in supersymmetric theories, and in particular in the MSSM\footnote{A detailed discussion of the mass eigenstates of the MSSM can be found in  Ref.\cite{HK}.}. In the basis of mass eigenstates, the particle content includes the four neutralinos $\widetilde{\chi}^0_i$ ($i=1,...,4$) and the four charginos $\widetilde{\chi}^\pm_j$ ($j=1,2$). This is reported in the table in Fig.(3.9). The process
\be
W^++W^-\longrightarrow \widetilde{\chi}^0_i+\widetilde G\,,
\label{masseig}
\ee
is the counterpart of Eq.(\ref{wbosons}) and corresponds to the set of diagrams in figure (\ref{fig:wmass}). $\widetilde{\chi}^0_i$ is a neutralino with $i$ running from $1$ to $4$. This is clearly a physical process that can be observed in experiments such as the LHC \cite{WWscat}. Since the amplitudes for each single Feynman diagram are complicated, these will not be included here unless necessary.
	Before making any calculations, one might infer about the possible result by looking at the Feynman diagrams in Fig.(\ref{fig:wmass}) and at the supergravity vertices in Chapter \ref{chapt:sugra}. The vertices now factorize the mixing factors of the mass eigenstates, but do not change analytically. This means that the cancellation of the longitudinal modes in the $\rm Z^0$ boson propagator still occurs, thus we expect that also in this case the longitudinal polarizations of the on-shell W become strongly interacting in the high energy limit.

	As already discussed, the higgsinos and the electroweak gauginos mix with each other by effect of the electroweak gauge symmetry breaking. The neutral higgsinos $\widetilde{H}^0_u$ and $\widetilde{H}^0_d$, and the neutral gauginos $\widetilde{\mathrm{W}}^3$ and $\widetilde{\mathrm{B}}$ mix and form four mass eigenstates, the neutralinos. The charged higgsinos $\widetilde{H}^+_u$ and $\widetilde{H}^-_d$, and the winos $\widetilde{\mathrm{W}}^\pm$ form instead the four charginos.

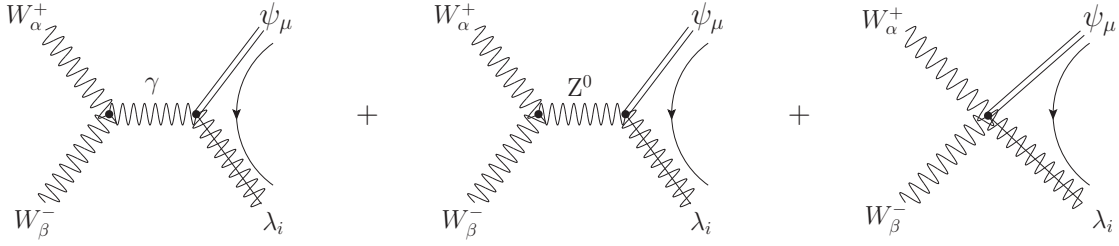
\begin{figure}[t]
\resizebox{\textwidth}{!}{
  \begin{picture}(614,126) (4,1)
    \SetWidth{0.5}
    \SetColor{Black}
    \Text(83.61,83.61)[]{\large{\Black{$\gamma$}}}
    \SetWidth{0.5}
    \Photon(59.72,67.69)(107.5,67.69){5.97}{8}
    \Text(319.32,83.61)[]{\large{\Black{$\rm Z^0$}}}
    \Photon(295.44,67.69)(343.21,67.69){5.97}{8}
    \Vertex(295.44,67.69){2.25}
    \Line(343.21,67.69)(379.05,18.32)
    \Text(389.81,121.04)[]{\Large{\Black{$\psi_{\mu}$}}}
    \Text(386.22,7.96)[]{\large{\Black{$\lambda_i$}}}
    \Vertex(343.21,67.69){2.25}
    \Line(341.95,68.65)(377.78,115.64)\Line(344.48,66.72)(380.32,113.7)
    \Photon(343.21,67.69)(379.05,19.11){5.97}{9}
    \Line(107.5,67.69)(143.34,18.32)
    \Text(152.1,121.04)[]{\Large{\Black{$\psi_{\mu}$}}}
    \Text(150.5,7.96)[]{\large{\Black{$\lambda_i$}}}
    \Vertex(107.5,67.69){2.25}
    \Line(106.24,68.65)(142.07,115.64)\Line(108.77,66.72)(144.6,113.7)
    \Photon(107.5,67.69)(143.34,19.11){5.97}{9}
    \Text(250.43,121.84)[]{\large{\Black{$W^+_{\alpha}$}}}
    \Text(254.82,7.96)[]{\large{\Black{$W^-_{\beta}$}}}
    \Photon(295.44,67.69)(260.4,113.87){5.97}{8}
    \Photon(295.44,67.69)(260.4,19.11){5.97}{9}
    \Vertex(59.72,67.69){2.25}
    \Text(14.72,121.84)[]{\large{\Black{$W^+_{\alpha}$}}}
    \Text(19.11,7.96)[]{\large{\Black{$W^-_{\beta}$}}}
    \Photon(59.72,67.69)(24.69,113.87){5.97}{8}
    \Photon(59.72,67.69)(24.69,19.11){5.97}{9}
    \Text(440.37,67.69)[]{\Large{\Black{$+$}}}
    \Text(202.27,67.69)[]{\Large{\Black{$+$}}}
    \Vertex(542.29,66.89){2.25}
    \Photon(542.29,66.89)(594.06,19.11){5.97}{10}
    \Text(606,120.26)[]{\Large{\Black{$\psi_{\mu}$}}}
    \Photon(542.29,66.89)(496.9,113.08){5.97}{8}
    \Photon(542.29,66.89)(496.9,19.11){5.97}{9}
    \Text(486.35,119.45)[]{\large{\Black{$W^+_{\alpha}$}}}
    \Text(488.14,9.94)[]{\large{\Black{$W^-_{\beta}$}}}
    \Text(607.59,9.15)[]{\large{\Black{$\lambda_i$}}}
    \SetWidth{0.25}
    \ArrowArc(177.6,67.69)(47.96,125.55,234.45)
    \ArrowArc(416.49,67.69)(47.96,125.55,234.45)
    \ArrowArc(625.93,67.69)(47.96,125.55,234.45)
    \SetWidth{0.5}
    \Line(542.29,66.89)(594.06,19.11)
    \Line(541.24,68.09)(593.01,113.48)\Line(543.34,65.69)(595.11,111.08)
  \end{picture}
  }
\caption{The diagrams which represent the amplitudes $M_s$ and $M_x$.}
\label{fig:schannels}
\end{figure}

	Let us start with the neutralinos. In two component notation, following \cite{HK} and \cite{Martin}, we can choose the basis of neutral mass eigenstates
	\be
\psi^0_j=(-i\lambda^\prime,-i\lambda_3,\psi^1_{H_u},\psi^2_{H_d})\,,
	\ee
where $\lambda^\prime$ and $\lambda_3$ are linear combinations of $\lambda_z$ and $\lambda_\gamma$. One can then write the mass matrix as follows:
\be
Y=\left( \begin{array}{cccc}
M^\prime & 0 & -m_z\sin{\theta_v}\sin{\theta_W} & m_z\cos{\theta_v}\sin{\theta_W}\\
0 & M & m_z\sin{\theta_v}\cos{\theta_W} & -m_z\cos{\theta_v}\cos{\theta_W}\\
-m_z\sin{\theta_v}\sin{\theta_W} & m_z\sin{\theta_v}\cos{\theta_W} & 0 & -\mu\\
m_z\cos{\theta_v}\sin{\theta_W} & -m_z\cos{\theta_v}\cos{\theta_W} & -\mu & 0
\end{array} \right)\,,
\\
\ee
where $\tan{\theta_v}\equiv v_1/v_2$. The mass eigenstates are defined as
\be
\chi^0_i=N_{ij}\psi^0_j\,,\qquad i=1,\cdots,4\,,
\label{Ymassmatrix}
\ee
where $N$ is a unitary matrix so that one obtains, for the diagonal matrix $N_D$ with nonnegative entries,
\be
N^*YN^{-1}=N_D\,.
\ee
Since the Feynman diagrams in Figs.(\ref{fig:schannels}) and (\ref{fig:higgdia}) contain propagating photons, $\rm Z^0$ and the heavy and light neutral Higgs bosons $H$ and $h$, we choose the basis
	\be
\psi^{0\prime}_j=(-i\lambda_\gamma,-i\lambda_z,\lambda_H,\lambda_h)\,,
	\label{basephot}
	\ee
that contains the photino $\pho$, the zino $\zi$ and the higgsinos $\widetilde{H}$ and $\tilde{h}$ as two-components spinors. This way we keep control of the gauge couplings, which are explicit in the interaction vertices. This is justified since the couplings of the gauge eigenstates are dictated by the electroweak theory, but those of mass eigenstates depend on the amount of mixing \cite{HK}. Thus there are different, yet equivalent, bases of the mass eigenstates. We can choose a basis where the photino and the other gauginos reflect the couplings. Moreover, for our purposes it is enough to define just some generic mixing coefficients. In analogy with Eq.(\ref{Ymassmatrix}), we can write
\be
\chi^0_i=N_{ij}^\prime\psi^{0\prime}_j\,,\qquad i=1,\cdots,4\,.
\label{neutr}
\ee
The non-diagonal mass matrix $Y^{\prime}$, that is analogous to (\ref{Ymassmatrix}), can be indeed diagonalized by a unitary matrix $N^{\prime}$ so that $N^{\prime *}Y^{\prime}N^{\prime -1}=N_D$, with $i,j$ running from $1$ to $4$. This needs now to be rewritten in four components. The zino, the photino and the neutral higgsinos in the basis (\ref{basephot}) are respectively represented by the matrices
\be
\pho=\left(\begin{array}{c}
-i\lambda_\gamma\\
i\overline{\lambda}_\gamma
\end{array}
\right)
, \qquad 
\zi=\left(\begin{array}{c}
-i\lambda_z\\
i\overline{\lambda}_z
\end{array}
\right)
, \qquad 
\widetilde{H}^0_u=\left(\begin{array}{c}
\psi^1_{H_u}\\
\overline{\psi}^1_{H_u}
\end{array}
\right)
, \qquad
\widetilde{H}^0_d=\left(\begin{array}{c}
\psi^2_{H_d}\\
\overline{\psi}^2_{H_d}
\end{array}
\right)\,.
\label{zphh}
\ee
	We define the four neutralinos as the Majorana fermions
\be
\neutr_i=\left(\begin{array}{c}
\chi^0_i\\\overline{\chi}^0_i
\end{array}
\right)\,.
\ee
	By abuse of notation, the mixing factors for the spinors in Eq.(\ref{zphh}) will be labelled respectively by $N^{\prime}_{\pho i}$, $N^{\prime}_{\zi i}$, $N^{\prime}_{\tilde{H}i}$ and $N^{\prime}_{\tilde{h}i}$, where the index $i$ refers to the $i$th neutralino.
	The charginos are made up of gaugino and higgsino components, and mix in analogy with the neutral mass eigenstates. These can be defined in two-component formalism as \cite{HK}
  \be
  \chi^+_i=V_{ij}\psi^+_j,\qquad\chi^-_i=U_{ij}\psi^-_j,  \qquad (i,j=1,2)\,,
\label{charg}
  \ee
where $i=1,2$. Expressing the W boson mass as
\be
m_W\equiv \dfrac{g}{2}\sqrt{v_1^2+v_2^2}\,,
\ee
the  charginos defined above are the eigenstates of the mass matrix (written in 2x2 block form)
\be
M_{\tilde{C}}=\left( \begin{array}{cc}
0 &  X^T\\
X & 0\\
\end{array} \right)\,,
\\
\ee
with the 2x2 matrix
\be
X=\left( \begin{array}{cc}
M &  m_W\sqrt{2}\cos{\theta_v}\\
m_W\sqrt{2}\sin{\theta_v}& \mu\\
\end{array} \right)\,.
\\
\ee
$M_{\tilde{C}}$ is written in the basis $\psi^{\pm}=(\widetilde{W}^+,\widetilde{H}^+_u,\widetilde{W}^-,\widetilde{H}^-_d)$, namely in components
\be
\psi^+_j=(-i\lambda^+,\psi^1_{H_d}), \qquad \psi^-_j=(-i\lambda^-,\psi^2_{H_u}), \qquad (j=1,2)\,.
\ee

In this formalism, the mass terms in the Lagrangian have the form
\be
\mathcal{L}\supset
-\half (\psi^+ \psi^-)
 \left( \begin{array}{cc} 0 & X^T \\
 X & 0 \end{array}\right)
  \left( \begin{array}{c} \psi^+ \\
 \psi^-\end{array}\right) + h.c.
\ee
and the mass eigenstates are related to the gauge eigenstates by the two unitary matrices $U$ and $V$ in Eq.(\ref{charg}):
\be
 \left( \begin{array}{c} \charg^+_1 \\
  \charg^+_2 \end{array}\right)=V\left( \begin{array}{c} \widetilde{W}^+ \\
  \widetilde{H}^+_u \end{array}\right)\,,
 \qquad
   \left( \begin{array}{c} \charg^-_1 \\
  \charg^-_2 \end{array}\right)=U\left( \begin{array}{c} \widetilde{W}^- \\
  \widetilde{H}^-_d \end{array}\right) \,.
\ee
	The mass matrix $X$ is diagonalized as follows:
	\be
	U^*XV^{-1}\equiv M_D=\left( \begin{array}{cc}
m_{\charg_1} &  0\\
0& m_{\charg_2}\\
\end{array} \right)\,.
\ee
	In four component notation, the charginos are the charged Dirac fermions
\be
\charg_1=\left(\begin{array}{c}
\chi^+_1\\\overline{\chi}^-_1
\end{array}
\right)\,,
\qquad
\charg_2=\left(\begin{array}{c}
\chi^+_2\\\overline{\chi}^-_2
\end{array}
\right)\,,
\ee
where by convention $\charg_1$ is heavier than $\charg_2$. In the following, we will label with $V_{i\chp}$ and $U_{i\chm}$ the mixing elements of, respectively, the $i$th positive chargino with the wino $\chp$ and of the $i$th negative chargino with the wino $\chm$.
\begin{figure}[t]
\resizebox{\textwidth}{!}{
 \begin{picture}(621,131) (70,10)
    \SetWidth{0.5}
    \SetColor{Black}
    \Text(366,77)[]{\Large{\Black{$+$}}}
    \SetWidth{0.3}
    \ArrowArc(313.99,71.09)(50,116.09,237.33)
    \ArrowArcn(581.9,71.5)(25.9,245.12,114.88)
    \Text(125,1)[]{\Large{\Black{$W^-_{\beta}$}}}
    \Text(125,144)[]{\Large{\Black{$W^+_{\alpha}$}}}
    \Text(302,4)[]{\Large{\Black{$\lambda_i$}}}
    \Text(232,74)[]{\Large{\Black{$\widetilde{\chi}^+_j$}}}
    \SetWidth{0.5}
    \Photon(208,121)(139,135){7.5}{8}
    \ArrowLine(208,118)(208,30)
    \Photon(208,30)(291,14){7.5}{8}
    \Vertex(511,118){2.83}
    \ArrowLine(511,28)(511,118)
    \Line(511,118)(592,107)
    \Vertex(208,118){2.83}
    \Vertex(511,28){2.83}
    \Text(434,142)[]{\Large{\Black{$W^+_{\alpha}$}}}
    \Text(607,105)[]{\Large{\Black{$\lambda_i$}}}
    \Text(434,0)[]{\Large{\Black{$W^-_{\beta}$}}}
    \Text(534,73)[]{\Large{\Black{$\widetilde{\chi}^-_j$}}}
    \Text(607,39)[]{\Large{\Black{$\psi_{\mu}$}}}
    \Line(510.64,29.97)(591.64,44.97)\Line(511.36,26.03)(592.36,41.03)
    \Photon(446,135)(511,118){7.5}{8}
    \Photon(511,28)(446,13){7.5}{8}
    \Vertex(208,30){2.83}
    \Text(302,142)[]{\Large{\Black{$\psi_{\mu}$}}}
    \Line(207.59,119.96)(289.59,136.96)\Line(208.41,116.04)(290.41,133.04)
    \Photon(208,30)(139,14){7.5}{8}
    \Line(208,30)(290,14)
    \Photon(511,118)(592,107){7.5}{8}
  \end{picture}
      }
\caption{The diagrams corresponding to the amplitudes $M_t$ and $M_u$.}
\label{fig:tuchannels}
\end{figure}
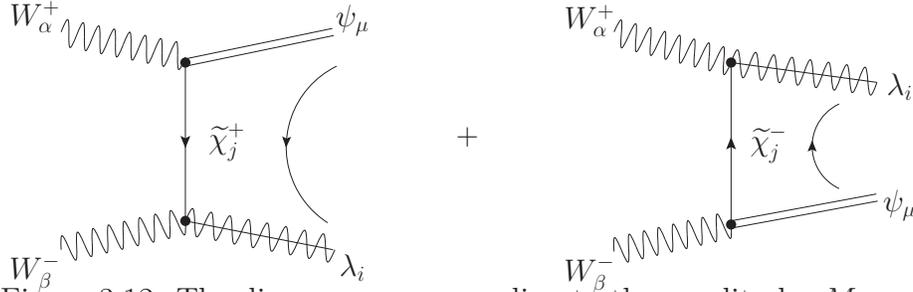

	After these preliminaries, we can now calculate the squared amplitude of the process (\ref{masseig}). First consider the exchange of a gauge boson and the four-particle vertex, where $\lambda_i$ is the wave function of the $i$-th neutralino, as in Fig.(\ref{fig:schannels}).
By using Eq.(\ref{neutr}) and the Feynman rules reported in the Appendix, we can write the amplitude of the diagram with exchange of $\rm Z^0$ as follows:
\bea
M_Z&=&N^{\prime}_{\tilde{z}i}\dfrac{g\cos{\theta_W}}{4M_P}\left[
\eta^{\alpha\beta}(k-k^{\prime})^\sigma+\eta^{\beta\sigma}(2k^{\prime}+k)^\alpha
-\eta^{\alpha\sigma}(2k+k^{\prime})^\beta\right]\times\nonumber\\
&&\times\dfrac{1}{s-m^2_Z}\left\{ \bar{\lambda}_{i}\gamma^\mu\left[\slashed{k}+\slashed{k}^{\prime},\gamma_\sigma\right]
\psi_{\mu}\right\}\epsilon_\alpha\epsilon_\beta\,.
\label{zeta}
\eea
	It is evident from the above that in $M_Z$ the longitudinal modes in the $\rm Z^0$ propagator are cancelled by the SUGRA vertex, by the mechanism studied in Section \ref{sect:theo}. Here we will not report explicitly the amplitudes of the other diagrams, which are shown in Figs.(\ref{fig:schannels}), (\ref{fig:tuchannels}) and (\ref{fig:higgdia}). Instead, we consider into detail the total squared amplitude $\mathcal{M}$ of the process $W^++W^-\longrightarrow \widetilde{\chi}^0_i+\widetilde G$, which can be written as:
\be
\mathcal{M}=M_s+M_t+M_u+M_x+M_{Higgs}\equiv M_{stux}+M_{Higgs}\,.
\label{mass_amp}
\ee
	We are interested in checking whether divergences such as (\ref{aiuto}) appear also in this case. The spin-summed square of the above depends on $N_{\widetilde{W}i}$, $O^{L}_{ij}$, and it does not have a simple form. One can anyway fix some constraints on the mixing factors, since these are free parameters until they will be determined experimentally. First we report the squared amplitude of the four diagrams corresponding to $M_{stux}$, then we consider also the Higgs bosons. By using the following equation,
\be
N_{\widetilde{W}i}=N^{\prime}_{\tilde{\gamma}i}\sin{\theta_W}+N^{\prime}_{\tilde{z}i}\cos{\theta_W}\,,
\label{1}
\ee
all the divergences which are proportional to $1/m^4_W\mg^2$ and to $1/m^4_W$ vanish. Moreover, if
\be
2|N_{\widetilde{W}i}|^2-2N_{\widetilde{W}i}(O^{L*}_{ij}+O^{R*}_{ij})+(|O^{L}_{ij}|^2+|O^{R}_{ij}|^2)=0\,,\qquad U_{j\chm}=V_{j\chp}\,,
\label{2}
\ee
also those which are proportional to $1/m_W^2$ and to $1/m_W^2\mg^2$ are cancelled.
If instead $U_{j\chm}$ and $V_{j\chp}$ are different, Eq.(\ref{2}) is rewritten as
\be
|O^{L}_{ij}|^2+|O^{R}_{ij}|^2=2N_{\widetilde{W}i}(O^{L*}_{ij}+O^{R*}_{ij})\,,\qquad U_{j\chm}\neq V_{j\chp}\,,
\ee
and no anomalies are cancelled. In any case, the terms which factorize $1/\mg^2$ still remain,
\bea
&&\sum_{spin}|M_s+M_t+M_u+M_x|^2\approx\dfrac{g^2}{M^2_P}\dfrac{1}{\cos^2{\theta_W}}\dfrac{1}{3\mg^2}\times\nonumber\\
&&\times\left\{
-|N_{\widetilde{W}i}-N^{\prime}_{\tilde{\gamma}i}\sin{\theta_W}|^2\dfrac{t(s+t)}{\cos^2{\theta_W}}
+4(|N_{\widetilde{W}i}|^2-N_{\widetilde{W}i}N^{\prime}_{\tilde{\gamma}i}\sin{\theta_W})(s^2+t^2+st)-\right.\nonumber\\
&&\left.-2N_{\widetilde{W}i}(O^{L*}_{ij}+O^{R*}_{ij})s^2+2N^{\prime}_{\tilde{\gamma}i}\sin{\theta_W}s^2-\right.\nonumber\\
&&\left.
-2\cos^2{\theta_W}\left[ 2|N_{\widetilde{W}i}|^2t(s+t)+2N_{\widetilde{W}i}(O^{L*}_{ij}+O^{R*}_{ij})s^2-4O^{L}_{ij}O^{R*}_{ij}s^2 \right]
\right.\bigg\}\,.\nonumber\\
\label{res}
\eea
	The above equation constitutes our net result. We considered Eq.(\ref{2}) for simplicity, and used $m_Z=m_W/\cos\theta_W$. $N_{\w i}$ is the mixing factor of the neutral wino $\w^3$ with the $i$th neutralino, in the basis given by
	\be
(\widetilde{B},\widetilde{W}^3,\tilde{h}_u,\tilde{h}_d)\,,
	\ee
which is alternative to (\ref{basephot}) and contains the bino $\widetilde{B}$. Both $\widetilde{W}^3$ and $\widetilde{B}$ are linear combinations of $\widetilde{\gamma}$ and $\tilde{z}$, and the unitary matrix $N$ diagonalizes the mass matrix $Y$, see Eq.(\ref{neutr}).

	Consistently with our formalism, Eq.(\ref{aiuto}) is recovered in the limit when $\theta_W\rightarrow 0$ and $N_{\widetilde{W}i},O^{L}_{ij},O^{R}_{ij}\rightarrow \epsilon^{abc}$, namely in the basis of $\rm SU(2)_L$ gauge eigenstates. We remark that during this calculation no mass has been set to zero, and that no identities other than Eqs.(\ref{1}) and (\ref{2}) have been adopted.

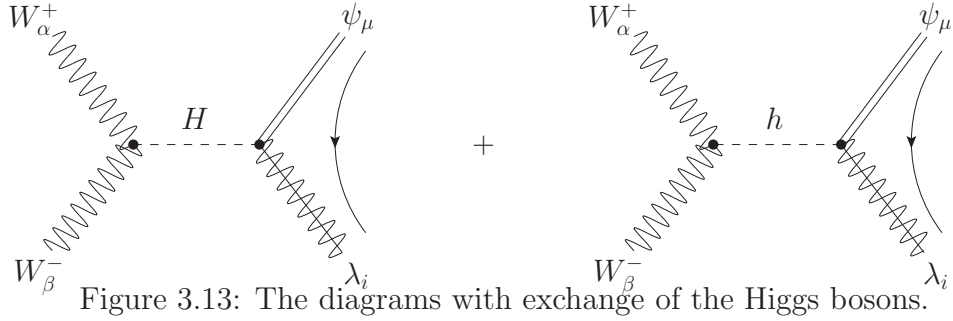
\begin{figure}[t]
\resizebox{\textwidth}{!}{
 \begin{picture}(621,131) (70,-5)
    \SetWidth{0.5}
    \SetColor{Black}
    \ArrowArc(368.5,58.5)(83.5,142.79,217.21)
    \Text(369,57)[]{\Large{\Black{$+$}}}
    \Line(243,57)(289,-3)
    \Vertex(243,57){2.83}
    \DashLine(173,57)(243,57){5}
    \Vertex(173,57){2.83}
    \Text(118,127)[]{\Large{\Black{$W^+_{\alpha}$}}}
    \Text(121,-16)[]{\Large{\Black{$W^-_{\beta}$}}}
    \Photon(173,57)(128,117){7.5}{8}
    \Photon(173,57)(128,-2){7.5}{9}
    \Text(298,-16)[]{\Large{\Black{$\lambda_i$}}}
    \Line(241.4,58.2)(287.4,119.2)\Line(244.6,55.8)(290.6,116.8)
    \Text(208,72)[]{\Large{\Black{$H$}}}
    \Line(566,57)(612,-3)
    \Vertex(566,57){2.83}
    \DashLine(496,57)(566,57){5}
    \Vertex(495,57){2.83}
    \Text(441,127)[]{\Large{\Black{$W^+_{\alpha}$}}}
    \Text(444,-16)[]{\Large{\Black{$W^-_{\beta}$}}}
    \Photon(495,57)(451,117){7.5}{8}
    \Photon(495,57)(451,-2){7.5}{9}
    \Text(621,-16)[]{\Large{\Black{$\lambda_i$}}}
    \Line(564.4,58.2)(610.4,119.2)\Line(567.6,55.8)(613.6,116.8)
    \Text(531,72)[]{\Large{\Black{$h$}}}
    \Text(299,128)[]{\Large{\Black{$\psi_{\mu}$}}}
    \Text(621,128)[]{\Large{\Black{$\psi_{\mu}$}}}
    \ArrowArc(688.5,58.5)(83.5,142.79,217.21)
    \Photon(566,57)(612,-2){7.5}{8}
    \Photon(243,57)(289,-3){7.5}{8}
  \end{picture}

      }
\caption{The diagrams with exchange of the Higgs bosons.}
\label{fig:higgdia}
\end{figure}

	The diagrams with exchange of Higgs bosons, which are reported in figure (\ref{fig:higgdia}), contribute to the leading divergences only with the square,
\bea
&\sum_{spin}|M_{Higgs}|^2\approx\dfrac{g^2}{\Mpl^2}\dfrac{1}{12}|N^{\prime}_{\tilde{h}i}\sin{(\beta-\alpha)}+N^{\prime}_{\tilde{H}i}\cos{(\beta-\alpha)}|^2\times\nonumber\\
&\times\left[ \dfrac{s^3}{m_W^2\mg^2}+ \dfrac{s^2}{m_W^2}-\dfrac{2s^2}{\cos^2{\theta_W}\mg^2}(2\cos^2{\theta_W}-1)\right]\,,
\label{higgscontrib}
\eea
where we have used the following formula for the Higgs masses \cite{HK}:
\be
m^2_{h,H}=\dfrac{1}{2}\left[ (m_A^2+m_Z^2)\mp \sqrt{(m_A^2+m_Z^2)^2-4m_A^2m_Z^2\cos^2{2\beta}} \right]\,.
\ee
	An evident analogy with Eqs.(\ref{m0mh}) and (\ref{m1mh}) has been found. The scalars contribute to the factor $1/\mg^2$ and reintroduce the divergences proportional to $1/m_W^2$ and to $1/m^2_W\mg^2$. It can be shown that the general case $U_{j\chm}\neq V_{j\chp}$ does not change the result. This is evident, since (\ref{higgscontrib}) depends only on s.
	Eqs.(\ref{res}) and (\ref{higgscontrib}) complete the calculation in the basis of the mass eigenstates, and confirm what has been obtained in the basis of gauge eigenstates.

	It is instructive to check also whether the Ward identities (\ref{ward}) hold in the present situation as well.
In the basis of mass eigenstates, suitable constraints on the mixing factors eliminate the unwanted terms in the contractions of $k_\alpha$ and $k^{\prime}_\beta$ with $M_{stux}$ in (\ref{mass_amp}). The scalar density $S$ thus appears again. 
However, the Higgs bosons maintain their peculiar role, as it is clear from the following,
\be
\left\{
\begin{tabular}{l}
$k_\alpha M^{\alpha\beta}=Sk^{\prime\beta}+S_Hk^\beta$\\
$k^{\prime}_\beta M^{\alpha\beta}=Sk^\alpha+S_Hk^{\prime\alpha}$\\
\end{tabular}
\right. \,,
\label{wardmass}
\ee
with the scalar amplitudes $S$ and $S_H$ written as
\be
S=\dfrac{gN_{\widetilde{W}i}}{4M_P}\dfrac{\bar{\lambda}_{i}\gamma^\mu\left[\slashed{k},\slashed{k}^{\prime}\right]\psi_{\mu}}{s-m_W^2}\,,
\ee
and
\bea
S_H&=&\dfrac{gm_W}{\sqrt{2}M_P}\left[N^{\prime}_{\tilde{h}i}\sin{(\beta-\alpha)}(s-m_{H}^2)+N^{\prime}_{\tilde{H}i}\cos{(\beta-\alpha)}(s-m_{h}^2)\right]\times\nonumber\\
&&\times\dfrac{\bar{\lambda}_i
(\mathbf{1}-\gamma_5)\psi_{\mu}p^{\prime\mu}}{(s-m_{h}^2)(s-m_{H}^2)}\,.
\eea
Violation of gauge invariance and therefore of unitarity is evident in Eq.(\ref{wardmass}), by effect of $S_H$. Interestingly, the contraction of the longitudinal modes of the $\rm Z^0$ in the propagator, that is
\be
\dfrac{-i}{s-m^2_W}\left[(\xi-1)\dfrac{(k+k^{\prime})_\sigma(k+k^{\prime})_\nu}{s-\xi m^2_Z}\right],
\ee
with the 3-bosons vertex
\be
\left[\eta^{\alpha\beta}(k-k^{\prime})^\sigma+\eta^{\beta\sigma}(2k^{\prime}+k)^\alpha
-\eta^{\alpha\sigma}(2k+k^{\prime})^\beta\right],
\ee
if non vanishing, would have provided with additional terms proportional to $k^{\prime\beta}$ and $k^\beta$ (or to $k^\alpha$ and $k^{\prime\alpha}$). Thus there might have been a contribution to both the scalar densities $S$ and $S_H$. These results are consistent with equation (\ref{higgscontrib}).

	To summarise, we have found that the squared amplitude of the scattering
\be
W^++W^-\longrightarrow \widetilde{\chi}^0_i+\widetilde G\,,
\nonumber
\ee
contains terms which violate the unitarity of the theory above a certain scale. The longitudinal polarizations of the $\rm Z^0$ in the propagator vanish from the amplitude (\ref{zeta}), and the Higgs bosons do not cancel the divergences. In this section, we have shown that it is not possible to restore unitarity by introducing the suitable scalar particles which are found in the observable sector of the MSSM.

\section{Conclusions and perspectives}

	We have studied into detail the WW scattering in Supergravity in the broken phase, and compared our results with the case of massless gauge bosons. If in the high-energy limit the W bosons are considered massless \emph{a priori}, as in Ref.\cite{Pradler}, the scattering amplitude does not contain any problematic terms. This is also the case of QCD \cite{BBB}.

	In contrast, the massive W provides with new structures, which could lead to the violation of unitarity in both the bases of gauge and mass eigenstates. In the annihilation diagram, the longitudinal degrees of freedom of the W boson in the propagator disappear from the amplitude by virtue of the supergravity vertex. This implies that the longitudinal polarizations of the on-shell W become strongly interacting at high energies. We show this effect in section \ref{sect:theo}, where Eq.(\ref{highlimit}) is derived. We have also shown that in both the bases of gauge and mass eigenstates of the MSSM, no scalar particle can provide with cancellation of (\ref{aiuto}).

	Let us point out that in this chapter we do not claim that the theory of supergravity has inner inconsistencies. We have simply shown that when two massive bosons scatter and produce gravitinos, standard methods lead to contradiction of well-known theorems on the conservation of the supercurrent.

	Cancellations of the molest terms would hopefully occur by adding further diagrams to the process. These can follow for instance from non trivial interaction terms in the supergravity Lagrangian (\ref{sugral}), which are not usually taken into account in this kind of calculations. Moreover, the contribution of the hidden sector might account for the missing longitudinal polarizations. The latter possibility is the topic of current investigations.

	Our results are however phenomenologically interesting, if SUGRA is regarded as the effective limit of some unified theory. It is clear from Eq.(\ref{aiuto}) that the SUSY breaking energy coincides precisely with the scale at which unitarity is violated. Accordingly, we can apply Eqs.(\ref{totalm0}), (\ref{res}) and (\ref{higgscontrib}) to particle physics and to cosmology. In fact, the process we have analysed can be observed at the LHC as a secondary reaction, for instance through gluon fusion.

	Moreover, we have shown in section (\ref{sect:theo}) that Eq.(\ref{wcs}) holds at any scale. Therefore our result contributes to the thermal gravitino production in the early universe, both at low and high reheating temperatures. How relevant the correction to the results of \cite{BBB} and \cite{Pradler} would be, is the subject of a future work.

	 We recall also that after inflation the MSSM flat directions, which are oscillating about the minimum of their potential, give a mass via their \emph{vev} to the SM gauge fields which are produced by the inflaton decay \cite{Anupammass}. Such gauge bosons scatter off each other in the primordial thermal bath, and recreate the phenomenology that is described in this chapter. Clearly, our results would affect this mechanism, providing with perhaps sizeable effects also in this particular cosmological context.

\chapter{Scalar field oscillations and gravitinos}\label{chapt:Infl}

	In this chapter we discuss the cosmological evolution of a scalar field, which at the end of inflation begins coherent oscillations about the minimum of its potential and decays into gravitinos. By demanding that the gravitinos thus produced constitute the dominant component of Cold Dark Matter (CDM), bounds on the reheating temperature $T_R$ are calculated. This is discussed in the paper \cite{Infl}.

	We begin by considering the dynamics of the scalar field, that is here identified with the SUSY breaking field in the \OR hidden sector of a model of gauge mediation with metastable vacuum. The scalar potential $V(S)$, including radiative and supergravity corrections, is calculated. Studying the decay into gravitinos, we find that the reheating temperature can be easily as low as the BBN lower bound $T_R\gsim 1\M$ and still be consistent with gravitino Dark Matter from the decay of the GMSB hidden sector.

\section{Dynamics of a scalar in an expanding background}\label{dynamics}

The cosmological evolution of the energy density associated with the oscillations of a scalar field was discussed long ago by Michael S. Turner and others \cite{T,Partprod}. Consider a scalar field $\phi$ with Lagrangian density
\be
\mathcal{L}=-\half\partial_\mu\phi\partial^\mu\phi-V(\phi)\,,
\ee
and a homogeneous and isotropic cosmology defined by the Friedman-Robertson-Walker metric Eq.(\ref{FRW}). In such models the field is homogeneous as well, with equation of motion
\be
\ddot{\phi}+3\Hub\dot{\phi}+V^{\prime}(\phi)=0\,,
\label{damped}
\ee
where an overdot denotes $d/dt$ and a prime $d/d\phi$. This is the equation of a damped harmonic oscillator, where the friction term $3\Hub\dot{\phi}$ keeps track of the expansion of the Universe. Eq.(\ref{damped}) can be recast as follows:
\bea
&&\dfrac{d}{dt}\left(\half\dot{\phi}^2+V\right)=-3\Hub\dot{\phi}^2\,,\label{eq1}\\
&&\dfrac{d\rho}{dt}=-3\Hub(\rho+p)\,,\label{eq2}\\
&&d(\rho a^3)=-pd(a^3)\,.\label{eq3}
\eea
As usual, $a(t)$ is the cosmological scale factor. The energy density and the pressure of the field $\phi$ are respectively written as
\be
\rho=\half\dot{\phi}^2+V(\phi)\,,\qquad p=\half\dot{\phi}^2-V(\phi)\,.
\ee
	The mass $m$ of the field is initially much smaller than the Hubble rate. However, while the energy density of the inflaton drives the expansion of the Universe, H decreases with time. Accordingly, Eq.(\ref{damped}) implies that when $m\approx\Hub$, $\phi$ starts coherent oscillations about the minimum of the potential $V(S)$, with frequency $\omega\approx(\dot{\phi}/\phi)$. It is clear from the above that the energy density of the scalar field $\phi$ decreases, since the kinetic term $(1/2)\dot{\phi}^2$ is red-shifted away by the expansion. Let us assume that the frequency of the oscillations $\omega$ is much larger than the Hubble rate. By defining $\gamma$ as the average of $(\rho+p)$ over an oscillation and $\gamma_p$ as the periodic part of $\dot{\phi}^2$, we obtain
\be
\dot{\phi}^2=\rho+p=(\gamma+\gamma_p)\rho\,.
\ee
Namely, the energy density of the scalar field decreases with time, since the oscillations of $\phi$ are damped. This effect turns on particle production.

	The quantum particle creation can be studied by introducing suitable couplings which determine the total decay rate $\Gamma$ of the scalar $\phi$ \cite{Partprod}. Following Turner \cite{T}, we assume for simplicity that the particles produced by the decay are photons. Let $\rho_{\gamma}$ be their energy density. We add a phenomenological term $-\dot{\phi}^2\Gamma$ to the rhs of Eq.(\ref{eq1}), to obtain the evolution equations for $\rho$ and $\rho_{\gamma}$,
\bea
&&\dot{\rho}=-(3\Hub+\Gamma)\gamma\rho\,,\\
&&\dot{\rho}_{\gamma}=-4\Hub\rho_{\gamma}+\gamma\Gamma\rho\,.
\eea
These can be easily solved to give the following:
\bea
&&\rho=\rho_0\left[\dfrac{a(t)}{a_0}\right]^{-3\gamma}e^{-\gamma\Gamma(t-t_0)}\,,\label{evo1}\\
&&\rho_{\gamma}=\rho_{\gamma_0}\left[\dfrac{a(t)}{a_0}\right]^{-4}+\rho_0\left[\dfrac{a(t)}{a_0}\right]^{-4}\int^{\gamma\Gamma t}_{\gamma\Gamma t_0}\left[\dfrac{a(t^\prime)}{a_0}\right]^{4-3\gamma}e^{u_0-u}du\,.\label{evo2}
\eea
\begin{center}
\begin{figure}[t]

\begin{center}

\resizebox{14cm}{!}{
\begin{minipage}[]{0.4\textwidth}

\begin{center}

\includegraphics[scale=1.4]{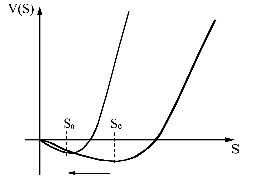}
\caption{Time-dependent shift of the minimum of the potential.}

\label{fig:shift}

\end{center}

\end{minipage}

\hspace{0.01\linewidth}
\begin{minipage}[]{0.5\textwidth}

\begin{center}

	\includegraphics[scale=1.4]{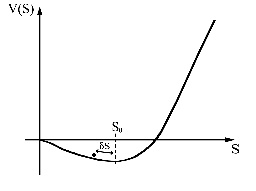}

\caption{The displacement from the minimum due to the time evolution of $S_0$.}

\label{fig:deltaS}

\end{center}

 \end{minipage} 
}
\end{center}

\end{figure}

\end{center}
	The subscript zero labels the value of that quantity at $t=t_0$, and $u\equiv \gamma\Gamma t^\prime$. It is evident from the above equations that the energy density $\rho$ during the oscillations decreases exponentially with time, and is transferred to the particles which are created through the decay\footnote{If the particles created are nonrelativistic, all the 4's in Eqs.(\ref{evo1}) and (\ref{evo2}) become 3's.}.

	In this chapter, we study the production of gravitino Dark Matter by a mechanism similar to the one sketched above. The scalar field, which we call $S$, is here included into a specific hidden sector in gauge mediation, and couples to the observable sector via loops of the messengers \cite{IK}. A similar approach has been done recently in Ref.\cite{ET}, however our study of the dynamics of the $S$-field is different. In \cite{ET}, the initial amplitude $S_c$ of the scalar field oscillations is assumed to be determined by the Hubble corrections (normalized in units of the Planck mass),
\be
\Delta V \approx e^K(3\Hub^2)\approx 3\Hub^2\left[|S|^2-\frac{5kM}{16\pi^2}(S+S^{\dagger})+\dots  \right]\,,
\label{h2corr}
\ee
which imply $|S_c|=5kM/16\pi^2$. This assumes that the inflaton dominates the energy density of the Universe when the $S$ field starts oscillating. We will also include this contribution. However, in addition to (\ref{h2corr}), the scalar potential contains terms which are not H-dependent. As we will see, this implies that the minimum $S_0(t)$ is time-dependent in the presence of Hubble corrections. The $S$ field tracks the minimum as it decreases towards the time-independent ($\Hub = 0$) minimum. This generates a displacement of $S$ from the minimum, as it is shown in Fig.\ref{fig:shift}.

\begin{center}

\begin{figure}[h]

\begin{center}

\includegraphics[scale=1.4]{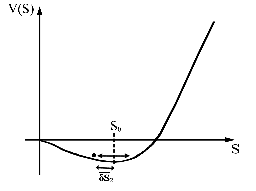}
\caption{Evolution of $\delta S$ in the potential with minimum at $\overline{\delta S_2}$.}

\label{fig:barra}

\end{center}

\end{figure}

\end{center}

	As H decreases, either the quadratic or the linear term becomes dominated by the $\Hub=0$ factor. So the minimum of $V$ is H-dependent until \emph{both} $|S|^2$ and $(S+S^\dagger)$ become H-independent. The displacement
 from the minimum at this time, $\delta S$, then determines the initial amplitude of the $S$ oscillations about the time-independent minimum $S_0$, as shown in Fig.\ref{fig:deltaS}.
	We first consider the dynamics of the GMSB sector generally, then we will apply it to the specific case of the model defined in Section \ref{gmsb}. We consider a potential of the general form
\be
V=\frac{1}{2}m_S^2S^2-aS+3\Hub^2\left(\dfrac{S^2}{2}-cS\right)\,,
\ee
where the scalar field is assumed to take real values without loss of generality (with $|S| \rightarrow S/\sqrt{2}$ for canonical normalization). When $\Hub$ decreases from a large value, either the linear or the quadratic term becomes dominated by the $\Hub$-independent term. We define $\Hub_1$ and $\Hub_2$ by 
\bea
&&\Hub=\Hub_1 \qquad \mathrm{when} \qquad a=3c\Hub^2\,,\\
&&\Hub=\Hub_2 \qquad \mathrm{when} \qquad m_S^2=3\Hub^2\,.
\eea
In the case $\Hub_2>\Hub_1$, the quadratic term becomes H-independent first. The potential when $\Hub_2 > \Hub > \Hub_1$ is therefore 
\be
V\approx\frac{1}{2}m_S^2S^2-3c\Hub^2S\,. 
\ee
The minimum is then $S_0(t)=3c\Hub^2/m_S^2$. The field $S$ will track this time-dependent minimum. If we define $S=S_0+\delta S$, where $\delta S$ is a perturbation, the scalar field equation for $S$, 
\be
\ddot{S}+3\Hub\dot{S}=-\frac{\partial V}{\partial S}\,,
\ee
becomes the following,
\be
\delta \ddot{S}+3\Hub\delta \dot{S}=-(\ddot{S}_0+3\Hub\dot{S}_0)-\left.\frac{\partial V}{\partial S}\right|_{S_0}
-\left.\frac{\partial^2 V}{\partial S^2}\right|_{S_0}\delta S.
\label{fried}
\ee
It can be shown that the higher-order terms in the expansion of the potential are negligible throughout our analysis. If the background is dominated by the inflaton, the Hubble parameter H depends on the scale factor $a(t)$ as $\Hub\propto a^{-3/2}$. Thus the above equation takes the form
\be
\delta\ddot{S}+3H\delta\dot{S}=-\frac{27}{2}\frac{c H^4}{m_S^2}-m_S^2\delta S,
\ee
and the displacement $\delta S$ evolves as if it had a potential with minimum given by
\be
\overline{\delta S_2}=-\frac{27}{2}\frac{c\Hub^4}{m^4_S}.
\label{S2}
\ee
This is illustrated in Fig.\ref{fig:barra}.
On the other hand, if $\Hub_1>\Hub_2$ then it is the linear term which becomes H-independent first. The potential when $\Hub_1 > \Hub > \Hub_2$ is then
\be
V\approx \dfrac{3}{2}\Hub^2S^2-aS\,,
\ee
and the minimum is now at $S_0(t)=a/3\Hub^2$. In this case Eq.(\ref{fried}) becomes
\be
\delta\ddot{S}+ \delta\dot{S}=-\frac{9a}{2}- 3 H^2 \delta S,
\ee
which implies 
\be
\overline{\delta S_1}=-\frac{3a}{2\Hub^2}.
\label{S1}
\ee
The shifts (\ref{S2}) and (\ref{S1}) give the initial amplitude of the oscillations of $S$ about the H-independent minimum in the two cases, where oscillations begin once $\Hub \approx \Hub_1$ and $\Hub \approx \Hub_2$ respectively. This modifies the gravitino density and the reheating temperature that were computed in \cite{ET} under the assumption that the initial amplitude is $S_c$ at $\Hub \approx m_S$.
	
	Before calculating constraints on $T_R$ according to the above dynamics, we have to include the field $S$ in a specific particle physics model, in order to make phenomenological predictions.
This is the topic of the next section.

\section{Gauge-mediated supersymmetry breaking}\label{gmsb}

	As we have seen in the previous chapters, the soft breaking terms in the matter sector must be of the order of the electroweak (EW) scale. On the other hand, the fundamental scale of spontaneous supersymmetry breaking (SSSB) can be much larger. Both with radiative corrections and at the tree level, the breaking occurs by new sets of fields and interactions, the "hidden sector". It is then transmitted to the observable sector in a variety of ways, depending on the interactions which communicate the breaking. For instance, these can be gravitational, corresponding to \emph{gravitational mediation}. The SSSB scale is of order $10^{11}\G$ \cite{LR,Dine,DineRev}. In this case, the effective Lagrangian of supergravity contains non-renormalizable mixing terms between the hidden and observable sector, which are suppressed by powers of the (reduced) Planck mass $\Mpl$. Although this formalism is elegant and suggested by string theory, it is not clear whether it can solve certain problems such as flavour changing processes \cite{DineRev}.

	Another possibility is to mediate the breaking through gauge interactions, by introducing a set of new chiral supermultiplets, the \emph{messenger sector} \cite{DR1,DR2,DFS,DF1,DS,DF2}. Gauge Mediated Supersymmetry Breaking (GMSB) is particularly attractive, since it automatically avoids the problem of flavour changing processes. Moreover, it suits well to cosmology, since in this case the gravitino is generally the lightest supersymmetric particle (LSP), that can be Dark Matter in the case of conserved R-parity \cite{GR}.
	The soft SUSY breaking terms are here introduced by the messengers. They couple directly to the hidden sector and indirectly to the matter sector through the SM interactions. In this case the gravitational interaction is automatically subdominant with respect to gauge couplings.

	If the model is renormalizable (like the MSSM), this theory is totally based on loops of the messengers. In the minimal model, these are a set of left-handed chiral supermultiplets $q$, $\bar{q}$, $\ell$, $\bar{\ell}$, which transform under $\mathbf{SU(3)}_C\times\mathbf{SU(2)}_L\times\mathbf{U(1)}_Y$. They contain messenger quarks
($\psi_q, \psi_{\bar q}$), squarks ($q, \bar q$), messenger leptons ($\psi_\ell, \psi_{\bar \ell}$) and sleptons ($\ell, \bar\ell$) \cite{Martin}. They must get very high masses in order not to have been already discovered, thus they should couple to a chiral singlet superfield $S$ as follows:
\be
W_{mess}=k_1S\ell\bar\ell+k_2Sq\bar q\,.
\label{messsuperpot}
\ee
Eq.(\ref{super}), that is discussed in the next section, has exactly this form. After that the scalar and the auxiliary components of $S$ get a \emph{vev} by supersymmetry breaking, the couplings above transmit this to the messengers, which become massive.
	It is not yet clear how the breaking occurs, however this can be achieved by either a dynamical SSB \cite{messmass} or by virtue of an \OR-type mechanism. The generation of the mass spectrum of the messengers is technically demanding and beyond the scope of this thesis. A detailed review can be found, for instance, in \cite{GR} and in \cite{Martin}.
\begin{center}

\begin{figure}[t]
\resizebox{0.75\textwidth}{!}{
\begin{minipage}[]{0.5\textwidth}
\caption{In GMSB the gaugino masses are generated by loops of virtual messengers.\label{fig:messloop}}
\end{minipage}
\hspace{0.01\linewidth}
\begin{minipage}[]{0.25\textwidth}
  \begin{picture}(204,126) (105,-69)
    \SetWidth{0.5}
    \SetColor{Black}
    \ArrowLine(105,-9)(165,-9)
    \Photon(105,-9)(165,-9){5}{6.5}
    \Text(165,36)[]{\large{\Black{$\psi_q,\psi_\ell$}}}
    \Text(120,6)[]{\large{\Black{$\lambda_{\tilde{g}}$}}}
    \Text(207,-61)[]{\large{\Black{$\langle F_S \rangle$}}}
    \Text(207,49)[]{\large{\Black{$\langle S \rangle$}}}
    \CArc(207,-6)(42,-0,180)
    \DashCArc(207,-6)(42,-180,0){10}
    \Text(207,-48)[]{\large{\Black{$\times$}}}
    \Text(207,36)[]{\large{\Black{$\times$}}}
    \Text(248,-47)[]{\large{\Black{$q,\ell$}}}
    \Photon(249,-9)(309,-9){5}{6.5}
    \ArrowLine(309,-9)(249,-9)
  \end{picture}
 \end{minipage} 
}

\end{figure}
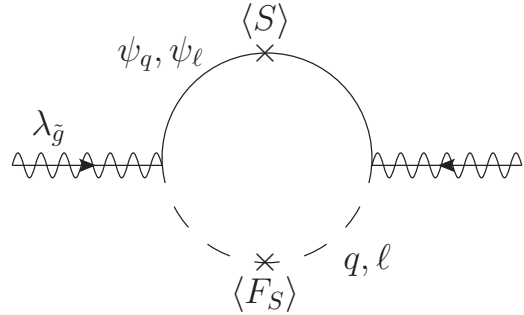
\end{center}

	Now, consider terms like (\ref{messsuperpot}) in the superpotential. After the SSSB, the scalar and fermion messengers get masses at tree level through mass terms which arise, respectively, in the Lagrangian and in the scalar potential. Loops of the messengers communicate this spectrum to the MSSM and generate the gaugino masses, as illustrated in Fig.\ref{fig:messloop}. The $q$ and $\bar q$ multiplets generate the gluino masses, while the loops of the $\ell$, $\bar{\ell}$ give masses to the wino and bino. The MSSM gaugino masses result as \cite{messmass}
\be
m_a=\dfrac{\alpha_a}{4\pi}\Lambda, \qquad (a=1,2,3)\,,
\label{gaugmass}
\ee
where the $\alpha_a$ are the couplings of the respective interactions. The cutoff $\Lambda$, which fixes the normalization, is defined as
\be
\Lambda\equiv\langle F_S \rangle/\langle S \rangle\,.
\ee
	The MSSM gauginos thus obtain their masses by virtue of spontaneous supersymmetry breaking, which has been communicated to the visible sector by one-loop processes of the messengers.

	In contrast, scalars do not get radiative corrections to their masses from one-loop diagrams, but from processes at two-loops. These are much more complicated than those in Fig.\ref{fig:messloop} and involve messenger fermions and scalars, together with ordinary gauge bosons and gauginos. The mass of each MSSM scalar results as
\be
m^2_{\phi_i}=2\Lambda^2\left[ \sum_a \left(\dfrac{\alpha_a}{4\pi}\right)^2 C_a(i) \right],\qquad (a=1,2,3)\,,
\label{scalmass}
\ee
where the $C_a(i)$ are the quadratic Casimir invariants \cite{GR}. Eqs.(\ref{gaugmass}) and (\ref{scalmass}) are valid in the limit of mass splittings within each messenger multiplet which are small with respect to the overall messenger mass scale, that is the case of the most common models.

\subsection{Gauge mediation in metastable vacua}

	We now introduce a particular class of GMSB models, which have been recently introduced by Murayama and Nomura in \cite{MN1,MN}.

	Such models are defined by three basic ingredients: i) the \K potential contains a negative quartic term in the supersymmetry breaking field, ii) an accidental $\mathrm{U(1)_R}$ symmetry in the superpotential terms which are determinant for SUSY breaking, iii) an explicit messenger mass term. One can see this by writing down the superpotential \cite{MN},
\be
W=-\mu^2S+kSf\bar{f}+Mf\bar{f}\,,
\label{supergen}
\ee
with no hidden sector yet specified\footnote{An \OR hidden sector will be included in the next section.}. $S$ is a gauge singlet chiral field, that can be either elementary or composite. The parameter $k$ can be taken real and positive without loss of generality, the scales $\mu$ and $M$ have dimension of mass, and the messengers $f$ and $\bar{f}$ belong to ${\bf 5} + {\bf 5^*}$ representations of $\rm SU(5)$ to include Standard Model interactions. The \K potential for the messengers is assumed to be canonical, whilst for the chiral field $S$ it looks as follows,
\be
K=|S|^2-\frac{|S|^4}{4\Lambda^2}+\mathcal{O}\left( \frac{|S|^6}{\Lambda^4} \right)\,,
\label{kp}
\ee
after expansion around $S=0$. $\Lambda$ is a mass scale which has to be assumed as a physical cutoff for the low-energy theory, in the absence of an explicit hidden sector. An approximate low-energy $\mathrm{U(1)_R}$ symmetry on $S$ makes it possible to obtain this form of the \K potential (this will be shown later in our particular model). For instance, the first two terms in (\ref{supergen}) may have such symmetry if $S$, $f$ and $\bar{f}$ carry respectively the charges $2$, $0$ and $0$. The third term in (\ref{supergen}) violates this symmetry anyway, but we will see that its effect on the \K potential can be neglected under certain assumptions. At the tree level the potential is given by
\be
V = \bigl|\mu^2 - k f \bar{f}\bigr|^2 
    \left[1 + \frac{|S|^2}{\Lambda^2} 
    + \mathcal{O}\left(\frac{|S|^4}{\Lambda^4}\right) \right]
    + \bigl| k S  + M \bigr|^2(|f|^2+|\bar{f}|^2),
\ee
and has a global SUSY minimum at
\be
S=-\dfrac{M}{k},\qquad f=\bar f=\dfrac{\mu}{\sqrt{k}}\,.
\ee
There is also a local SUSY breaking minimum (which is also a metastable vacuum) at $S=f=\bar f=0$, if $M^2>k\mu^2$. The latter condition is also necessary to avoid tachyonic messengers, as the scalar components of the fields have masses $m^2_S=\mu^4/\Lambda^2$ and $m_f^2=M^2\pm k\mu^2$. Since, as found in the previous section, the messenger loops generate the gaugino masses as
\be
m_a\approx \dfrac{g_a^2}{16\pi^2}\dfrac{k\mu^2}{M}\,,
\ee
by demanding these to be of order $100\G$ to $1\T$ we find that \cite{MN}
\be
\dfrac{k\mu^2}{M}\approx 100\T\,.
\ee
Moreover, the gravitino mass is given by (see Chapter \ref{chapt:sugra}),
\be
\mg\approx \frac{\langle F_S \rangle}{\Mpl}\approx \frac{\mu^2}{\Mpl}\,,
\ee
therefore we find that if gravity mediation is subdominant ($\mg\lsim 10\G$), the SUSY breaking scale $\mu$ is smaller than $10^{10}\G$.

A detailed discussion about the $\mathrm{U(1)_R}$ symmetry in the superpotential (\ref{supergen}) can be found in the paper by Murayama and Nomura Ref.\cite{MN}. Let us anyway sketch here some remarks. The SUSY breaking minimum at $S=f=\bar f=0$ might be viewed as a consequence of this symmetry, so it is clear how it plays a central role in the class of GMSB models we are considering. Nevertheless, violation of such symmetry may come from several effects. First of all, from nonlinear terms in $S$, such as $S^2$ or $S^3$, that may appear in the superpotential. In the case of an elementary $S$ field, one must simply assume that they are somehow negligible. On the other hand, if $S$ is composite, these terms can be automatically suppressed.

Another possibility of violation comes from loops of the messengers, which violate $\mathrm{U(1)_R}$ by virtue of the mass term in the superpotential. This induces a linear term in the \K potential, which makes the SUSY breaking field deviate from the origin. As we will see in the following, such effect is responsible for the decay of the field $S$ into gravitinos.

We remark that, at this point, is it quite evident that the models of gauge mediation with metastable vacua so constructed have only four parameters: $\mu$, $k$, $\Lambda$, $M$. Remarkably, the gravitino mass range allowed by this framework is very large, $1\e\lsim\mg\lsim 10\G$. We will go further in the analysis of this formalism in the next section, where a particular model with an explicit hidden sector is addressed.

\subsection{Model with an \OR hidden sector}

As we stated in the previous discussion, the form of the \K potential Eq.(\ref{kp}) is generated by the radiative corrections of the fields in the hidden sector. Moreover, in order to have a phenomenologically viable model, the cutoff $\Lambda$ should be expressed in terms of the physical parameters of such hidden sector, which then needs to be introduced explicitly. Let us then consider a field content of the \OR type, with superpotential
\be
W_{\rm h}=-\mu^2S+\lambda SX^2+mXY\,,
\ee
where the singlet fields $S$, $X$ and $Y$ have a canonical \K potential at tree level (up to terms suppressed by the Planck mass). Here $S$ is elementary, thus terms like $S^2\;,S^3\;,X^2\;,Y^2\,,SXY\;,SY^2$ are assumed to be negligible \cite{MN}. By recalling the general superpotential introduced in Eq.(\ref{supergen}), we then have:
\be
W=-\mu^2S+\lambda SX^2+mXY+kSf\bar{f}+Mf\bar{f}\equiv W_{h}+W_{m}\,.
\label{super}
\ee
Along the direction of the field space corresponding to $X=0$, $Y=0$, we still find a global SUSY minimum for the scalar potential at
\be
S=-\frac{M}{k}, \hspace{1cm} f=\bar{f}=\frac{\mu}{\sqrt{k}}\,,
\ee
and a local SUSY breaking minimum (which is a metastable vacuum) at $S=f=\bar{f}=0$, if $M^2>k\mu^2$. In fact, since at this point there is incompatibility between $F_S=0$ and $F_Y=0$, the field $S$ breaks supersymmetry.
We have seen in section \ref{section:OR} that in \OR models the SUSY breaking field $S$ is usually a flat direction of the potential at the tree level. In fact, the classical scalar potential of the hidden sector, that is obtained from the $F$-terms of the superfield multiplet, holds as follows
\bea
V^h_0(S)&=& |F_S|^2+|F_X|^2+|F_Y|^2=\mu^4+\lambda^2|X|^4-\lambda\mu^2(X^2+X^{\dagger 2})+\nonumber\\
&&+m^2(|X|^2+|Y|^2)+2\lambda m(SXY^{\dagger}+S^{\dagger} X^{\dagger} Y)+4\lambda^2|X|^2|S|^2\,.
\label{vclassic}
\eea
Thus in the direction $X=Y=0$, the above is constant (or zero, by rescaling) for any value of $S$. However, the radiative corrections originated by loops of the $X$, $Y$ fields and of the messenger fermions and scalars, have the property to lift this flat direction. We discuss this effect in the next section.

\subsection{Coleman-Weinberg potentials}\label{sect:cw}

In 1973, Sidney Coleman and Erik Weinberg published a paper, where they claimed that radiative corrections may produce spontaneous symmetry breaking in theories in which such breakdown is absent at the tree level \cite{CW}. Clearly, they could not consider supersymmetric theories at that time. However, starting from a toy model concerning massless scalar electrodynamics, they have shown that the effect is valid for both abelian and non-abelian gauge field theories, if the couplings are small.

	The extension of this effect to the Standard Model and to supersymmetric theories is naturally achieved. In several fields of research (i.e. in solid state physics, or in the study of phase transitions), this result has proven to be extremely important. In particle cosmology, there have been applications in many fields, and in particular in theories of inflation \cite{Linde,LR,LindeCW}.

	In this section, we show that the Coleman-Weinberg effective potentials for our model, which are generated by loops of the hidden sector and of the messengers, produce a series of important effects for the phenomenology. Namely, they i) lift the flat direction in the classical potential $V(S)$ by generating a mass term for the field $S$; ii) provide with an explicit form of the otherwise arbitrary cutoff $\Lambda$ in terms of the parameters of the GMSB model; iii) induce a linear term in the \K potential, which through supergravity corrections enters $V(S)$ and makes $S$ deviate from the origin.

\subsubsection{Contributions of global supersymmetry}

The mass matrix in the basis of the $X$ and $Y$ fermions $\psi_X$ and $\psi_Y$ is given by
\be
  M_\psi^h=\left( \begin{array}{cc}
    2\lambda S & m \\
    m          & 0
  \end{array} \right),
\label{fmm}
\ee
while the $F$-terms of the hidden sector, in the basis $(X,X^{\dagger},Y,Y^{\dagger})$, provide with the following mass matrix for the scalars \cite{MN}:
\be
M_\phi^h=\left( \begin{array}{cccc}
    m^2 + 4\lambda^2 |S|^2  & -2\lambda\mu^2  &  2\lambda m S^\dagger  &  0 \\
    -2\lambda\mu^2  & m^2 + 4\lambda^2 |S|^2  &  0  &  2\lambda m S         \\
    2\lambda m S     &     0   &  m^2  &  0 \\
    0  &  2\lambda m S^\dagger &  0  &  m^2
  \end{array} \right)\,.
\label{smm}
\ee
Given the mass matrices $M_\psi^h$ and $M_\phi^h$ by equations (\ref{fmm}) and (\ref{smm}) respectively, the radiative corrections to the scalar potential can be now calculated with the following formula
\be
V_{eff}^\phi=\dfrac{1}{64\pi^2}\mathrm{Tr}[M_S^2\ln M_S]\,,
\label{cws}
\ee
for the scalars, and
\be
V_{eff}^\psi=-\dfrac{1}{64\pi^2}\mathrm{Tr}[(|M_f|^2)^2\ln(|M_f|^2)]\,,
\label{cwf}
\ee
for the fermions \cite{CW}. The minus sign comes from the Fermi-Dirac statistics. The formulas (\ref{cws}) and (\ref{cwf}) give the following Coleman-Weinberg potential for the hidden sector,
\be
V^h_{eff}(S)=V^\phi_{eff}(S)+2V^\psi_{eff}(S)\approx\dfrac{\lambda^4\mu^4}{3\pi^2m^2}|S|^2-\dfrac{3\lambda^6\mu^4}{10\pi^2m^4}|S|^4+\ldots
\label{susyeff}
\ee
where we have dropped a constant, and kept the dominant terms in $\lambda\mu^2/m^2$ and in $S$. The above equation agrees with \cite{MN}. 
	Since this contribution has to be added to the tree level potential in Eq.(\ref{vclassic}), it has been verified that at the quantum level, the field $S$ is stabilized at the origin of field space (because of the expansion in $S$). Moreover, we see in (\ref{susyeff}) a positive mass term, which for small values of $S$ dominates over the quartic term. This is interesting from the point of view of the dynamics of the model.

	Also the one-loop corrections to the \K potential are important for phenomenology. They can be calculated with the following formula
\be
K^{(1)}=-\dfrac{1}{32\pi^2}\mathrm{Tr}\left[|M_\psi|^2\left(\ln\dfrac{|M_\psi|^2}{\Lambda^2}-1\right)\right],
\label{kformula}
\ee
that is a particular case of the more general expression
\bea
K^{(1)}&=&-\frac{1}{32\pi^2}\mathrm{Tr}\left[\bar{\mu}\mu\ln\frac{\bar{\mu}\mu}{\Lambda^2}-2M\ln\frac{M}{\Lambda^2}\right]\nonumber\\
&&=-\frac{1}{32\pi^2}\left\{\mathrm{Tr}\left[M^2_\phi\left(\ln\frac{M^2_\phi}{\Lambda^2}-1\right)\right]
-2\mathrm{Tr}\left[M^2_V\left(\ln\frac{M^2_V}{\Lambda^2}-1\right)\right]\right\}\,,\nonumber\\
\eea
which is valid for any renormalizable supersymmetric theory with $N=1$ \cite{Grisaru}. Here $M^2_\Phi$ and $M^2_V$ are the mass matrices of, respectively, the chiral and vector superfield sectors. In particular,
\be
(M^2_\phi)_{ij}=\overline{W}_{i\bar{k}}\delta^{\bar{k}k}W_{kj} \qquad \textrm{or} \qquad \mu_{ik}=m_{ik}+\lambda_{ijk}\Phi^j,
\ee
for the chiral sector and
\be
M_{AB}=\half\left(\overline{X}^i_A X_{Bi}+\overline{X}^i_B X_{Ai}\right),
\ee
for the vector sector (which is however absent in the model here discussed). In the case of messenger loops, we multiply the right hand side of Eq.(\ref{kformula}) by a factor of 5, to take into account for the representations of SU(5).

	In the previous section we have found that, in this class of GMSB models with metastable vacua, as long as one does not introduce an explicit hidden sector, the non-minimal \K potential for $S$ is written as:
\be
K=|S|^2-\frac{|S|^4}{4\Lambda^2}+\mathcal{O}\left( \frac{|S|^6}{\Lambda^4} \right)\,.
\ee
	In the case at hand we have defined the hidden sector, therefore the physical cutoff $\Lambda$ can be written in function of the parameters of the \OR model. By using Eq.(\ref{kformula}), for small $S$ and at the lowest order in $\lambda\mu^2/m^2$, they result as
\be
K^{(1)}_{XY}(S)\approx-\dfrac{\lambda^2}{8\pi^2}|S|^2-\dfrac{\lambda^4}{12\pi^2m^2}|S|^4\,,
\label{xykahler}
\ee
which correspond to the potential
\be
K(S)\approx|S|^2-\dfrac{|S|^4}{4\Lambda^2}+\ldots
\label{kc}
\ee
with a cutoff
\be
\Lambda^2=\frac{3\pi^2m^2}{\lambda^4},
\label{cutoff}
\ee
since $\lambda^2/8\pi^2\ll 1$ even for a strong coupling $\lambda\approx\mathcal{O}(1)$. Remarkably, quantum corrections have generated both a mass and a non-canonical \K potential for the supersymmetry breaking field.

	The contribution of the messengers is fundamental to our discussion, as they generate in the potential linear terms in $S$. These are responsible for the displacement from the origin that induces the oscillations of the field $S$, which is essential for cosmology and will be considered in the next sections.
	The scalar and fermion mass matrices $M^m_{\phi}$ and $M^m_{\psi}$, in the bases $(f,f^{\dagger},\bar{f},\bar{f}^{\dagger})$ and $(\psi_f,\psi_{\bar{f}})$, can be respectively written as
\bea
M^m_{\phi} = \left(
 \begin{array}{cccc}
    |M+kS|^2  & 0  &  0  &  -k\mu^2+k\lambda X^{\dagger 2} \\
    0  & |M+kS|^2 &  -k\mu^2+k\lambda X^2  &  0 \\
    0  & -k\mu^2+k\lambda X^{\dagger 2}   &  |M+kS|^2  &  0 \\
     -k\mu^2+k\lambda X^2 & 0 &  0  &  |M+kS|^2
  \end{array} \hspace{-1mm}\right)\,,\nonumber\\
\label{msmm}
\eea
and as
\be
  M^m_{\psi}=\left( \begin{array}{cc}
    0 &M+kS \\
    M+kS & 0
  \end{array} \right).
\label{fmmm}
\ee
	At the lowest order in $k\mu^2/M^2$ and for small values of $S$, the above matrices give
\bea
K^{(1)}_m(S)&\approx&-\dfrac{5M^2}{16\pi^2}\left[\dfrac{k}{M}(S+S^{\dagger})+2\left(\dfrac{k}{M}\right)^2|S|^2+\right.\nonumber\\
&&\left.+\dfrac{1}{2}\left(\dfrac{k}{M}\right)^3|S|^2(S+S^{\dagger})-\dfrac{1}{6}\left(\dfrac{k}{M}\right)^4|S|^2(S^2+S^{\dagger 2})\right]\,,\nonumber\\
\label{cwkahler}
\eea
that agrees with the results of \cite{ET} (notice the linear term in $S$).
The $F$-terms of the messengers give the following tree-level potential:
\be
V_0^m=k(-\mu^2+\lambda X^{\dagger 2})f\bar{f}+k(-\mu^2+\lambda X^2)f^{\dagger}\bar{f}^{\dagger}+k^2|f|^2|\bar{f}|^2\,.
\label{messpot}
\ee
The Coleman-Weinberg potential of the messengers is obtained by summing the contributions of the loops of the messenger scalars and fermions, as before. Keeping only the dominant terms in $k\mu^2/M^2$, in $\lambda\mu^2/m^2$ and in $S$, we find
\bea
V_{eff}^m(S)&\approx &
\dfrac{5k^3\mu^2}{16\pi^2}\left[\dfrac{\mu^2}{M}-\dfrac{\lambda}{M}(X^2+X^{\dagger 2})\right](S+S^{\dagger})-
\dfrac{5k^4\mu^4}{32\pi^2M^2}(S^2+S^{\dagger 2})
\nonumber\\
&&+\dfrac{5k^6\mu^4}{24\pi^2M^6}\bigg [\dfrac{\mu^4}{2}+\dfrac{\lambda^2}{2}(X^4+X^{\dagger 4}+
4|X|^4)+\lambda \mu^2(X^2+X^{\dagger 2})\bigg]|S|^2\,,\nonumber\\
\label{cwm}
\eea
where we have dropped an unimportant constant. This expression agrees with the literature \cite{ET,MN} for $X=Y=0$. Note that the messenger loops generate a mass term for the field $S$ as well.

	The radiative corrections of global SUSY to the scalar potential, i.e. Eqs.(\ref{susyeff}) and (\ref{cwm}), which are generated by loops of the supermultiplets in both the hidden and messenger sectors, can be now added to the tree-level potentials (\ref{vclassic}) and (\ref{messpot}).
	Along the direction $X=Y=f=\bar{f}=0$ in the field space, where both the SUSY and the metastable minimum are localized and where the $S$ field oscillates, the classical plus quantum potential can be rewritten as follows,
\bea
&&V_{SUSY}(S)\approx \mu^4+\dfrac{\lambda^2\mu^4}{8\pi^2}\ln\dfrac{m^2}{R^2}+\dfrac{3\lambda\mu^4}{16\pi^2}+\dfrac{15k^2\mu^4}{32\pi^2}
+\dfrac{5k^2\mu^4}{16\pi^2}\ln\dfrac{M^2}{R^2}\nonumber\\
&&+\dfrac{5k^3\mu^4}{16\pi^2M}(S+S^{\dagger})-\dfrac{5k^4\mu^4}{32\pi^2M^2}(S^2+S^{\dagger 2})
+\left(\dfrac{\lambda^4\mu^4}{3\pi^2m^2}+\dfrac{5k^6\mu^8}{48\pi^2M^6}\right)|S|^2\nonumber\\
&&+\dfrac{5k^7\mu^8}{32\pi^2M^7}(S+S^{\dagger})|S|^2-\left(\dfrac{3\lambda^6\mu^4}{10\pi^2m^4}+\dfrac{15k^4\mu^8}{64\pi^2M^8}\right)|S|^4\,,
\label{potglob}
\eea
where $R$ is a renormalization scale. This is added to the supergravity corrections which are discussed in the next section. We then obtain the full scalar potential $V(S)$ that is used to calculate the gravitino abundance.

\subsubsection{Supergravity corrections to the potential}

To embed the model of gauge mediation in supergravity, we can use the general formula \cite{Nilles}
\be
V_{SUGRA}  =  e^{\frac{K}{\Mpl^2}}  \left[ \sum_{\alpha,\beta}
\left( \dfrac{\partial^2 K}{\partial \bar{\phi}_{\alpha} \partial \phi_{\beta} }
	\right)^{-1}\times
|DW|^2
- 3 \dfrac{|W|^2}{\Mpl^2}  \right]\,,
\label{sugra}
\ee
where $\Mpl=2.43\times 10^{18}\G$ is the reduced Planck mass and
\be
DW\equiv\left( \dfrac{\partial W}{\partial \phi_{\alpha} } + \dfrac{W}{\Mpl^2}
	 \dfrac{\partial K}{\partial \phi_{\alpha} } \right)
\ee
is the \K derivative of the superpotential $W$. At this point, it might be interesting to show first the effect of the quartic term in $S$ in the \K potential at the tree level. We will then include the radiative corrections from the loops of the messengers and of the hidden sector.
Eq.(\ref{sugra}) for the SUSY breaking field is here rewritten as
\be
V_{SUGRA}(S)  =  \mu^4 e^{\frac{K}{\Mpl^2}}  \left[\left(1-\dfrac{|S|^2}{\Lambda^2}\right)^{-1}\times\left(1+2\dfrac{|S|^2}{\Mpl^2}+\dfrac{|S|^4}{\Mpl^4} \right)- 3 \dfrac{|S|^2}{\Mpl^2}  \right]\,,
\ee
with the \K potential (\ref{kc}). At the lowest order in $|S|^2$, this is expanded as
\be
V_{SUGRA}(S) \approx \mu^4\left( 1+\dfrac{|S|^2}{\Lambda^2}+\dfrac{7}{4}\dfrac{|S|^4}{\Lambda^2\Mpl^2}\right)
=\mu^4\left( 1+\dfrac{\lambda^4}{3\pi^2}\dfrac{|S|^2}{m^2}+\dfrac{7\lambda^4}{12\pi^2}\dfrac{|S|^4}{m^2\Mpl^2}\right)\,.
\label{sugracorr}
\ee
The term proportional to $|S|^2$ has a positive coefficient, consistently with the literature \cite{ET}. Using the radiative corrections (\ref{xykahler}) and (\ref{cwkahler}), expansion in $S$ gives
\be
K^{(1)}(S)=K^{(1)}_m(S)+K^{(1)}_h(S)\approx  |S|^2-\dfrac{5k}{16\pi^2}M(S+S^{\dagger})
-\dfrac{5k^3}{32\pi^2}\dfrac{|S|^2}{M}(S+S^{\dagger})-\dfrac{\lambda^4}{12\pi^2}\dfrac{|S|^4}{m^2}\,,
\ee
at the lowest order. Thus the following dominant terms are generated,
\bea
V_{SUGRA}(S) &\approx & \mu^4\dfrac{|S|^2}{\Mpl^2}\left\{\dfrac{25k^2}{64\pi^4}\left[-k^2+\left(\dfrac{M}{\Mpl}\right)^2\right]+\dfrac{\lambda^4}{3\pi^2}\left(\dfrac{\Mpl}{m}\right)^2\right\}+\nonumber\\
&&+\dfrac{5k\mu^4}{16\pi^2}(S+S^{\dagger})\left( \dfrac{k^2}{M}-2\dfrac{M}{\Mpl^2}\right)+\ldots
\eea
where we write only the terms which are linear and quadratic in $S$. This agrees with (\ref{sugracorr}) in the limit $k\rightarrow 0$. We can further assume that 
\be
\dfrac{25k^2}{64\pi^4}\left|-k^2+\left(\dfrac{M}{\Mpl}\right)^2\right|\ll\dfrac{\lambda^4}{3\pi^2}\left(\dfrac{\Mpl}{m}\right)^2,
\ee
therefore the supergravity corrections can be rewritten as
\be
V_{SUGRA}(S) \approx \mu^4\left[ \dfrac{\lambda^4}{3\pi^2m^2}|S|^2+\dfrac{5k}{16\pi^2}(S+S^{\dagger})\left( \dfrac{k^2}{M}-2\dfrac{M}{\Mpl^2}\right)+\ldots\right].
\label{vsugra}
\ee
Clearly, the sign of the term that is linear in $S$ is now determined by the coupling constant $k$. In particular, a form of the potential such as
\be
V_{SUGRA}(S) \approx \mu^4\left[ +c^2|S|^2-d^2(S+S^{\dagger})+\ldots\right]\,,
\ee
would be recovered if the following condition is satisfied:
\be
k<\sqrt{2}\dfrac{M}{\Mpl}\Rightarrow\dfrac{\mg}{m_3}>\dfrac{5.77}{\alpha_3}\,.\label{inc}
\ee
The above equation directly relates the gravitino mass $\mg$ to the gluino mass $m_3$ and to the strong coupling constant $\alpha_3$. It is calculated by using Eq.(\ref{massM}) in Section \ref{cosmo}. However, (\ref{inc}) is not consistent with gauge mediation, where the gravitino is the LSP. In particular, the framework that is used here admits gravitino masses $\mg\lsim 10 \G$, thus we conclude that also the coefficient of the linear term in (\ref{vsugra}) is identically positive.

	The supergravity corrections to the scalar potential of the model can now be written by adding (\ref{vsugra}) to the potential for $S$ which is generated by the inflaton, whose energy density dominates the Universe before the field starts oscillating \cite{ET},
\be
V(S) \approx e^{\frac{K}{\Mpl^2}} (3 \Hub^2 M_P^2)\approx 3 \Hub^2 \left[ |S|^2 - \dfrac{5 k}{16 \pi^2} M(S+S^\dagger) + \cdots\right]\,.
\ee
This provides, at the lowest order in the SUSY breaking field, with
\bea
V_{SUGRA}(S) &\approx& 3 \Hub^2 \left[ |S|^2 - \dfrac{5 k}{16 \pi^2} M(S+S^\dagger)\right]+\nonumber\\
&&+\mu^4\left[\dfrac{\lambda^4}{3\pi^2m^2}|S|^2+\dfrac{5k}{16\pi^2}(S+S^{\dagger})\left( \dfrac{k^2}{M}-2\dfrac{M}{\Mpl^2}\right)\right]\,.
\label{sugrafinal}
\eea
The above supergravity corrections, together with Eq.(\ref{potglob}), are used in the next section to calculate the gravitino abundance and the reheating temperature.

\section{Gravitino Dark Matter and reheating temperature}\label{cosmo}

      The results of Sections \ref{dynamics} and \ref{gmsb} can be now applied to gravitino production. Here we calculate the gravitino number density as a function of the reheating temperature $T_R$ and of the parameters of the GMSB model. We find bounds on $T_R$ by demanding that the gravitinos produced by the decay of the scalar field constitute the dominant component of Cold Dark Matter.
	The scalar potential of the field $S$ is obtained from the superpotential (\ref{super}) and from the \K potential (\ref{kp}), as a sum of a classical part and the corrections generated by quantum loops and by supergravity.
By summing equations (\ref{potglob}) and (\ref{sugrafinal}), we obtain
\bea
V(S)&=& V_{SUSY}+V_{SUGRA}\approx\left(3\Hub^2+\dfrac{2\lambda^4\mu^4}{3\pi^2m^2}+\dfrac{5k^6\mu^8}{48\pi^2M^6} \right)|S|^2+\nonumber\\
&&+\dfrac{5 k}{16 \pi^2} M(S+S^\dagger)\left(-3\Hub^2-\dfrac{2\mu^4}{\Mpl^2}+\dfrac{2k^2\mu^4}{M^2}\right)
\label{totalV}
~,\eea
where we have kept only the linear and quadratic terms in $S$, in the direction corresponding to $X=Y=f=\bar{f}=0$. These are dominant for sufficiently small values of the field.

	The scalar field can be treated as a modulus, with interactions given in \cite{IK}. The decay into gravitinos is not helicity suppressed, and gravitino production from $S$ decay dominates production from inflaton decay. The mass scales here considered are indeed consistent with the discussion in Ref.\cite{DineKitano}.

Dominant decay into a pair of gravitinos ($S\rightarrow\psi_\mu\psi_\nu$) occurs if the cutoff $\Lambda$ satisfies \cite{ET}
\be
\Lambda \lsim
8 \times 10^{14} \sqrt{\frac{\alpha_3}{0.1}}\left(\frac{m_3}{1\T} \right)^{-1}\left(\frac{\mg}{1\G}\right)\;\G
\label{Lambdaconstr}
~.\ee
The inclusion of the \OR sector makes it possible to relate this directly to the mass scale $m$ in the hidden sector
using Eq.(\ref{cutoff}), 
\be m \lsim 1.5 \times 10^{14} \; \lambda^2 \G\times\sqrt{\frac{\alpha_3}{0.1}}\left(\frac{m_3}{1\T} \right)^{-1}\left(\frac{\mg}{1\G} \right)\,
\label{mdecay}
~.\ee
This corresponds to an upper bound in the range $10^8\G - 10^{15}\G$ for a gravitino mass between $1\k$ and $10\G$ (the lower bound on $\mg$ is discussed in Sect.\ref{section:constr}). The messenger mass scale $M$ is related to the gluino mass $m_3$ and to the SUSY breaking scale $\mu$ by \cite{ET}
\be  M = \frac{\alpha_{3} k \mu^{2}}{4 \pi m_{3}}    ~,
\label{m3}\ee 
therefore, by using Eq.(\ref{mu2}), we find
\be
M \approx 3.3 \times 10^{13} \; k \left(\frac{\alpha_3}{0.1} \right)\left(\frac{1\T}{m_3} \right)\left(\frac{\mg}{1\G} \right) \; \G    \,.
\label{massM}
\ee
This implies $10^{7}\G\lsim M \lsim 10^{14}\G$ for the gravitino mass range reported above. Together with (\ref{mdecay}), it also gives the following approximation for the coefficient of the linear term in the scalar potential (\ref{totalV}),
\be
\left( -3 \Hub^2 -\dfrac{2 \mu^{4}}{\Mpl^2}+\dfrac{2k^2 \mu^{4}}{M^2}\right) \approx \left(-3\Hub^2 + 
\dfrac{2k^2 \mu^{4}}{M^2}\right),
\label{num}
\ee
over the entire gravitino mass range considered. Similarly, for the coefficient of the quadratic term, 
\be
\left(3\Hub^2+ \dfrac{2\lambda^4\mu^4}{3\pi^2m^2}+\dfrac{5k^6\mu^8}{48\pi^2M^6}\right)\approx\left(3\Hub^2+ \dfrac{2\lambda^4\mu^4}{3\pi^2m^2}\right).
\label{quadr}
\ee
We now briefly comment on the thermalization of the hidden sector, which was discussed for example in \cite{therm}. For $T_{R}$ smaller than the messenger mass $M$, the scattering of $S$ particles by thermal MSSM particles will be due to messenger loops, with rate $\Gamma_{s} \approx \alpha_{s} T^5/M^4$ ($\alpha_{s}$ accounts for couplings constants and numerical factors). This can be compared with the expansion rate at $T < T_{R}$, namely $H \approx T^2/M_{Pl}$. Accordingly, the hidden sector will not be thermalized as long as $T_{R} \lsim (M/\alpha_{s}M_{Pl})^{1/3}M$. Then (\ref{massM}) implies that the corresponding upper bound on the reheating temperature for which the $S$ field is not thermalized is $1 \T$ to $10^{12} \G$ (using $\alpha_{s}^{1/3} \sim 1$).

    One should also consider what happens when adding a constant term $\sim \mg\Mpl^2$ to the superpotential of the model. This is needed to tune the vacuum energy to zero in supergravity \cite{Nilles,IK}. Recalling Eq.(\ref{super}), we can write \cite{IK}
\be
W=-\mu^2S+\mg\Mpl^2+\dots
\ee
This generates both linear and quadratic terms in $S$ through the factor $|W|^2$ in the supergravity corrections (\ref{sugra}), that is
\bea
V_{SUGRA}&=& \exp(K/\Mpl^2) \left(...-3 \dfrac{|W|^2}{\Mpl^2} \right)\approx\left(1+\dfrac{K}{\Mpl^2} \right)6\mu^2\mg(S+S^{\dagger})\nonumber\\
&&\approx\left[1+\dfrac{|S|^2}{\Mpl^2}-\dfrac{5k}{16\pi^2}\dfrac{M}{\Mpl^2}(S+S^{\dagger})\right]6\mu^2\mg(S+S^{\dagger})\nonumber\\
&&\approx2\sqrt{3}\dfrac{\mu^4}{\Mpl^2}(S+S^{\dagger})-\dfrac{15k}{4\sqrt{3}\pi^2}\dfrac{M\mu^4}{\Mpl^3}|S|^2\,.
\eea
Eqs.(\ref{num}) and (\ref{quadr}) change accordingly:
\bea
&\left( 3\Hub^2 -\dfrac{2k^2\mu^4}{M^2}-\dfrac{32\sqrt{3}\pi^2\mu^4}{5k}\dfrac{1}{M \Mpl}
\right)\,,
\label{lin_corr}\\
&\left(3 \Hub^{2} + \dfrac{2\lambda^4\mu^4}{3\pi^2m^2}-\dfrac{15k\mu^4}{4\sqrt{3}\pi^2}\dfrac{M}{\Mpl^3}\right)\,.
\label{quad_corr}
\eea
For the entire gravitino mass range here considered, the contribution of the additional term in (\ref{quad_corr}) is always negligible. On the other hand, if $k\lsim 0.1$ the new term in (\ref{lin_corr}) dominates for certain values of $\mg$, thus we take it into account as well. The scalar potential of interest is therefore the following,
\be
V(S)\approx \left(3\Hub^2+\dfrac{2\lambda^4\mu^4}{3\pi^2m^2}\right)\dfrac{S^2}{2}+\dfrac{5 kM}{8 \pi^2} \dfrac{S}{\sqrt{2}}\left(-3\Hub^2+\dfrac{2k^2\mu^4}{M^2}+\dfrac{32\sqrt{3}\pi^2\mu^4}{5k}\dfrac{1}{M \Mpl}\right)\,,
\label{potential}
\ee
where it is assumed that $S$ is real for simplicity, which will be the case on minimizing the potential if all parameters are considered real. Equation (\ref{potential}) can thus be rewritten as follows,
\be
V(S)\approx \dfrac{\lambda^4\mu^4}{3\pi^2m^2}S^2+\left(\dfrac{5 k^3\mu^4}{4 \sqrt{2} \pi^2M}+2\sqrt{6}\dfrac{\mu^4}{\Mpl} \right)S+3\Hub^2\left(\dfrac{S^2}{2}-\dfrac{5 kM }{8 \sqrt{2} \pi^2}S \right)\,.
\label{recast}
\ee
Comparison of this expression with the potential
\be
V=\frac{1}{2}m_S^2S^2-aS+3\Hub^2\left(\frac{S^2}{2}-cS\right)\,,
\ee
gives the parameters which determine the dynamics of the field $S$, as discussed in Section \ref{dynamics}. It is now possible to calculate the gravitino abundance from the decay of the scalar field $S$.

	The oscillations start when the Universe is still dominated by the inflaton, thus $T>T_R$. The energy density can be written as
\be
\rho=g_\star \frac{\pi^2}{30}T^4\,,
\label{endensity}
\ee
where $T$ is the photon temperature and $g_\star$ is the number of effectively massless degrees of freedom. If the spacial curvature is $k=0$, the Hubble rate in radiation domination can be written as
\be
\Hub(T)=\sqrt{\frac{g_\star\pi^2}{90}}\frac{T^2}{\Mpl}\,.
\label{hubblet}
\ee
The entropy density of the Universe is dominated by relativistic particles, therefore $p=\rho/3$ and (\ref{endensity}) gives
\be
s=g_{\star s}\frac{4\pi^2}{90}T^3\,.
\label{entropy}
\ee
Assuming thermal equilibrium, the conservation of entropy per comoving volume, namely $s(T)a^3(T)=constant$, can be rewritten as
\be
g_\star(T)T^3a^3(T)=constant\,,
\ee
where $a(T)$ is the scale factor at temperature $T$. Let $n_{S\;osc}$ be the number density of the scalar and $a_{osc}$ the scale factor at that time. The number density $n_S(T)$ at a generic temperature is given by
\bea
n_S(T)&=&\left( \frac{a_{osc}}{a(T)} \right)^3 n_{S\;osc}  = \left( \frac{a_{osc}}{a(T_R)} \right)^3\left( \frac{a(T_R)}{a(T)} \right)^3 n_{S \; osc} =\nonumber\\
&=&\left( \frac{\Hub(T_R)}{\Hub_{osc}} \right)^2 \frac{g(T)T^3}{g(T_R)T_R^3} n_{S\;osc}\,,
\eea
since $\Hub\propto a^{-3/2}$ during inflaton domination. Recalling Eq.(\ref{hubblet}), the above becomes
\be
n_S(T)=\frac{\pi^2 }{90}\frac{g(T) T^3T_R}{\Mpl^2}\frac{n_{S\;osc}}{\Hub_{osc}^2}\,.
\ee
Therefore the number density to entropy density of $S$ is just 
\be
\frac{n_S}{s}=\frac{T_R}{4\Mpl^2}\frac{n_{S\;osc}}{\Hub_{osc}^2}\,.
\ee
From the scalar potential Eq.(\ref{recast}), $\rho_S = m_S^2S^2 /2$, which implies $n_{S \;osc}=\rho_S/m_S=(1/2)m_SS_{osc}^2$, where $S_{osc}$ is the value of the field at the beginning of the oscillations. Therefore
\be
\frac{n_S}{s} = \frac{m_S T_R S_{osc}^2}{8\Mpl^2 \Hub_{osc}^2}\,.
\label{yieldosc}
\ee
Since each $S$ scalar decays into a pair of gravitinos, the gravitino number density to entropy ratio is given by  
\be
\frac{n_{\g}}{s} = \frac{m_S T_R S_{osc}^2}{4\Mpl^2 \Hub_{osc}^2}\,
\label{yieldgrav}
~,\ee
where the initial oscillation amplitude around the minimum, $S_{osc}$, is given by $\overline{\delta S_1}$ or $\overline{\delta S_2}$ at the time when all H dependence becomes negligible.

	Consider first the case where $\Hub_2>\Hub_1$, namely where $\Hub_{osc}^2 \approx \Hub_{1}^2 = -a/3c$ and $S_{osc} \equiv \overline{\delta S_2}$. 
Therefore
\be
\frac{n_{S}}{s} = \frac{m_S T_R}{8\Mpl^2}\frac{\overline{\delta S_2}^2}{\Hub_1^2}\,.
\label{yielddelta}
\ee
The parameters $a$ and $c$ and the mass $m_S$ are fixed by the messenger and hidden sectors,
\bea
&&m_S^2=\dfrac{2\lambda^4\mu^4}{3\pi^2m^2}\,,\\
&&a= -\dfrac{5 k^3\mu^4}{4 \sqrt{2} \pi^2M}\left(1+\dfrac{16\sqrt{3}\pi^2}{5k^3}\dfrac{M}{\Mpl}\right)\,,\\
&&c = \frac{5  k M}{8 \sqrt{2}\pi^2}\,.
\eea
Using these we obtain 
\be
\dfrac{n_{S}}{s} \approx \frac{2025 \pi^7}{16 \sqrt{2}} \frac{k^4 m_{3}^4 m^{7}T_R}{\lambda^{14} \alpha_{3}^4 \mg^5 \Mpl^7}\left[1+\dfrac{12\pi\alpha_3}{5k^2}\left(\dfrac{\mg}{m_3} \right) \right]^3 \,,
\label{yieldS}
\ee
where we have used (\ref{m3}) and (\ref{mu2}) to eliminate $M$ and $\mu$. 
Therefore  
\bea
\dfrac{n_{\g}}{s} &\approx& 0.11 \times \; \dfrac{k^4}{\lambda^{14}}\left(\dfrac{\alpha_3}{0.1}\right)^{-4}\left(\dfrac{m_3}{1\T}\right)^{4}\left(\dfrac{\mg}{1\G} \right)^{-5}\nonumber\\
&&\times\left(\dfrac{m}{10^{14}\G}\right)^7\left(\dfrac{T_R}{10^8\G}\right)\left[1+\dfrac{12\pi\alpha_3}{5k^2}\left(\dfrac{\mg}{m_3} \right) \right]^3\,.
\eea
The (rescaled) density parameter of the gravitino $\Omega_{\g} h^2$ (see Chapter \ref{chapt:intro}) can be now calculated as follows. The entropy density is $s =2 \pi^2 g(T) T^3/45$, with at present $g(T) = 2$ (for the photon). Since the CMB temperature is $T=2.4\times 10^{-13}\G$, and the energy density of the gravitino is $\rho_{\g}=\mg(n_{\g}/s)s$, we obtain
\be
\Omega_{\g} h^2 =\dfrac{\rho_{\g}}{\rho_c}h^2=1.5\times10^8\left(\dfrac{\mg}{1\G} \right)\left(\dfrac{n_{\g}}{s} \right)\,,
\ee
with the present critical density $\rho_c=8.1 \times 10^{-47}h^2 \G^4$. Using the above formula we find
\bea
\Omega_{\g} h^2 &\approx& 
1.7\times 10^{7} \; \dfrac{k^4}{\lambda^{14}}\left(\dfrac{\alpha_3}{0.1}\right)^{-4}\left(\dfrac{m_3}{1\T}\right)^{4}\left(\dfrac{\mg}{1\G} \right)^{-4} \nonumber\\
&&\times\left(\dfrac{m}{10^{14}\G}\right)^7\left(\dfrac{T_R}{10^8\G}\right)\left[1+\dfrac{12\pi\alpha_3}{5k^2}\left(\dfrac{\mg}{m_3} \right) \right]^3\,.\nonumber\\
\label{gnd}
\eea
By demanding that the gravitino is the principal constituent of Cold Dark Matter,
\be
\Omega_{\g} h^2\approx\mathcal{O}(0.1)\,,
\label{cdm}
\ee
we then obtain the reheating temperature $T_R$ as a function of the parameters of the model
\bea
T_R&\approx& 0.6 \G \times \dfrac{\lambda^{14}}{k^4}\left(\dfrac{\alpha_3}{0.1}\right)^{4}
\left(\dfrac{m_3}{1\T}\right)^{-4}
\left(\dfrac{\mg}{1\G} \right)^{4}\nonumber\\
&&\times\left(\dfrac{m}{10^{14}\G}\right)^{-7}\left[1+\dfrac{12\pi\alpha_3}{5k^2}\left(\dfrac{\mg}{m_3} \right) \right]^{-3}\,.
\label{reheating21}
\eea

     In the case $\Hub_1>\Hub_2$, $\Hub_{osc}^2 \approx \Hub_{2}^{2} = m_{S}^{2}/3$ and $S_{osc} \equiv \overline{\delta S_1}$.   
We then obtain
\be
\frac{n_{S}}{s} \approx \frac{18225 \pi^3}{256 \sqrt{2}} \frac{k^4 m_{3}^2  m^{5} T_R}{\lambda^{10} \alpha_{3}^2  \mg^3 \Mpl^5}\left[1+\dfrac{12\pi\alpha_3}{5k^2}\left(\dfrac{\mg}{m_3} \right) \right]^2 \,.
\label{yieldS1}
\ee
Therefore 
\bea
 \dfrac{n_{\g}}{s}  &\approx& 3.9\times10^{-3}\dfrac{k^4}{\lambda^{10}}\left(\dfrac{\alpha_3}{0.1}\right)^{-2}\left(\dfrac{m_3}{1\T}\right)^2\left(\dfrac{\mg}{1\G}\right)^{-3}
\nonumber\\
&&\times\left(\dfrac{m}{10^{14}\G}\right)^5\left(\dfrac{T_R}{10^8\G}\right)\left[1+\dfrac{12\pi\alpha_3}{5k^2}\left(\dfrac{\mg}{m_3} \right) \right]^2\,,
\eea
and accordingly
\bea
\Omega_{\g} h^2 &\approx& 5.9\times10^{5}\dfrac{k^4}{\lambda^{10}}\left(\dfrac{\alpha_3}{0.1}\right)^{-2}\left(\dfrac{m_3}{1\T}\right)^2\left(\dfrac{\mg}{1\G} \right)^{-2}
\nonumber\\
&&\times\left(\dfrac{m}{10^{14}\G}\right)^5\left(\dfrac{T_R}{10^8\G}\right)\left[1+\dfrac{12\pi\alpha_3}{5k^2}\left(\dfrac{\mg}{m_3} \right) \right]^2\,.
\eea
The reheating temperature $T_R$ required for the correct density of gravitino Dark Matter is then 
\bea
T_R &\approx& 17 \G\times\dfrac{\lambda^{10}}{k^4}\left(\dfrac{\alpha_3}{0.1}\right)^{2}
\left(\dfrac{m_3}{1\T}\right)^{-2}
\left(\dfrac{\mg}{1\G} \right)^{2}\nonumber\\
&&\times\left(\dfrac{m}{10^{14}\G}\right)^{-5}\left[1+\dfrac{12\pi\alpha_3}{5k^2}\left(\dfrac{\mg}{m_3} \right) \right]^{-2}\,.
\label{reheating12}
\eea
Eqs.(\ref{reheating21}) and (\ref{reheating12}) are our main results. We see that the model can accommodate a very wide range of $T_R$ and still be consistent with gravitino Dark Matter from the GMSB sector.
The most striking feature of these results is their extreme sensitivity to the parameters of the model, in particular the \OR sector coupling $\lambda$. This means that the gravitino density from decay of the SUSY breaking scalar in the GMSB sector can account for Dark Matter for essentially any value of the reheating temperature above the BBN bound, $T_{R} \gsim 1 \M$. 

  Comparing the above with the previous results of \cite{ET}, which assumed that oscillations about the minimum began at $\Hub_{osc} \approx m_{S}$ with $S_{osc} \approx \sqrt{2} c$ (for real $S$), we find 
\bea
&\dfrac{n_{\g}}{s}\approx\left.\dfrac{n_{\g}}{s}\right|_{ET}\times \left(\dfrac{a}{c} \right)^3\dfrac{1}{m_S^6}\,,\qquad \left|\dfrac{a}{cm_S^2}\right|<1\,,\\
&\dfrac{n_{\g}}{s}\approx\left.\dfrac{n_{\g}}{s}\right|_{ET}\times \left(\dfrac{a}{c} \right)^2\dfrac{1}{m_S^4}\,,\qquad \left|\dfrac{a}{cm_S^2}\right|>1\,,
\eea
where $\left.n_{\g}/s\right|_{ET}$ is the value given in \cite{ET}. Namely,
\be \left.\dfrac{n_{\g}}{s}\right|_{ET} \approx 
2 \times 10^{-10} k^{4} \left(\dfrac{\alpha_{3}}{0.1}\right)^{2} \left(\dfrac{m_{3}}{1 \T}\right)^{-2} 
 \left(\dfrac{\mg}{1 \G}\right) \left(\dfrac{T_{R}}{10^{8} \G}\right)
 \left(\dfrac{\Lambda}{10^{14} \G}\right) ~.
\ee
If $\Hub_1 > \Hub_2$, $|a/c m_S^2|>1$ and so there is a strong enhancement. Similarly, if $\Hub_2 > \Hub_1$ then there is a strong suppression of the gravitino abundance.
     Note that while in the existing literature $T_R$ is necessarily written in terms of an arbitrary cutoff \cite{ET}, in our analysis there is a fully defined perturbative hidden sector, eliminating the cut-off in favour of the masses and couplings of the model. Therefore what could have been a limitation, namely focusing on a specific \OR-type sector, is in fact an advantage. Since there is no cut-off, there is no need to constrain the parameters of the model in order to make it consistent with the validity of the effective theory.

\section{Constraints on the GMSB model}\label{section:constr}

Here we derive various bounds on our model, first by demanding consistency of the series expansion of the Coleman-Weinberg potential. The coupling constants $\lambda$ and $k$ and the mass scale $m$ are indeed constrained as follows.
The scalar potential Eq.(\ref{totalV}) is obtained by expanding in terms of $\lambda\mu^2/m^2$, so the condition
\be
\dfrac{\lambda\mu^2}{m^2}\ll 1\,,
\label{mmu}
\ee
must be satisfied. Therefore 
\be m > \lambda^{1/2} \mu\,.
\label{ratio}
\ee
Now we derive further constraints on the mass scale $m$ of the hidden sector, and show that the gravitino mass range which is consistent with the model can be accordingly constrained. First, by using the relation between $\mu$ and $\mg$, 
\be
\mu^2 = \sqrt{3} \mg \Mpl   ~,
\label{mu2}
\ee
the lower bound (\ref{ratio}) can be recast as
\be
m > \lambda^{1/2} \mu=2\times10^{9} \lambda^{1/2} \G\sqrt{\dfrac{\mg}{1 \G}}\,.
\label{lowerm}
\ee
In Section \ref{cosmo}, we assume that the gravitino is the main decay product of the field $S$ when this rapidly oscillates around the minimum. Accordingly, the upper limit Eq.(\ref{mdecay}) holds:
\be
m\lsim 1.5 \times 10^{14} \; \lambda^2 \G \times\sqrt{\dfrac{\alpha_3}{0.1}}\left(\dfrac{m_3}{1\T} \right)^{-1}\left(\dfrac{\mg}{1\G} \right).
\label{upperm}
\ee
Clearly, (\ref{lowerm}) and (\ref{upperm}) must be simultaneously satisfied. This means that the formulas (\ref{reheating21}) and (\ref{reheating12}) for the reheating temperature are valid only for certain gravitino mass ranges, depending on the values of $\lambda$. In particular,
\bea
&\qquad \lambda\approx 10^{-2}\Rightarrow 1\M\lsim\mg\lsim 10\G\,,\label{mgconstr1}\\
&\qquad \lambda\approx 10^{-1}\Rightarrow 1\k\lsim\mg\lsim 10\G\,,
\label{mgconstr2}
\eea
where we have assumed only that $m_3\approx 1\T$. 
Remarkably, since these results follow from the consistency of the Coleman-Weinberg potentials and from phenomenology, they do not rely on any approximations.

	Next, let us calculate constraints on the coupling $k$ and show that it should not be too small.
In analogy with Eq.(\ref{ratio}), the effective potential (\ref{cwm}) for the messenger sector is valid if
\be
\dfrac{k\mu^2}{M^2}\ll 1\,.
\label{expk}
\ee
By recalling Eq.(\ref{massM}), i.e.
\be
M \approx 3.3 \times 10^{13} \G \times k \left(\dfrac{\alpha_3}{0.1} \right)\left(\dfrac{1\T}{m_3} \right)\left(\dfrac{\mg}{1\G} \right)\,,
\ee
we can substitute the above equation and (\ref{mu2}) in (\ref{expk}). Thus we finally obtain
\be
\dfrac{10^{-8}}{2k}
\left( \frac{0.1}{\alpha_{3}} \right)^{2} 
\left( \frac{m_{3}}{1 \T} \right)^{2}
\left(\dfrac{\mg}{1\G} \right)^{-1}\ll 1\,.
\ee 
This condition is satisfied for $k \gsim 0.01$ in the entire gravitino mass range in equations (\ref{mgconstr1}) and (\ref{mgconstr2}). 

	One should also take into account the free streaming length $\lambda_{FS}$, namely the distance that the gravitinos produced through the decay can travel until they become non-relativistic. If we want them to be Cold Dark Matter (CDM), this quantity must be smaller than $\mathcal{O}(100 \,\rm kpc)$ \cite{freestreaming}.
With the notations of Ref.\cite{ET}, the comoving free streaming length at matter-radiation equality can be defined as
\be
\lambda_{FS}\equiv\int_{t_D}^{t_{\mathrm eq}} \dfrac{\mathrm{v}_{\g}(t)}{a(t)} dt\,,
\label{freesl}
\ee
where $a(t)$ is the scale factor, $t_D$ is the time at the $S$ decay and $t_{\rm eq}\approx 2\times10^{12}\,\rm sec$ is the time at matter-radiation decoupling. The velocity of the gravitino is given by
\be
 \mathrm v_{\g}(t)=\dfrac{|{\mathbf p}_{\g}|}{E_{\g}}
 \approx \dfrac{\dfrac{m_S}{2} \left(\dfrac{a_D}{a(t)}\right)}{\sqrt{\mg^2 
 + \dfrac{ m_S^2}{4} \left(\dfrac{a_D}{a(t)}\right)^2}}\,,
\ee
where $a_D$ is the scale factor when $S$ decays, and we have assumed $m_S\gg m_{\g}$. Integrating (\ref{freesl}) and recalling that $\Lambda^2=3\pi^2m^2/\lambda^4$, one finds
\be
\lambda_{FS}\approx 1 {\rm\, kpc}\, \left(\dfrac{\sqrt{3}\pi}{\lambda^2} \right)^{\frac{3}{2}}  \sqrt{\dfrac{g_*}{100}}   \left(\dfrac{m_{\g}}{1{\rm GeV}}\right)^{-\frac{3}{2}}\left(\dfrac{m}{10^{14} {\rm GeV}}\right)^{\frac{3}{2}}\,,
\ee
where $g_*$ is the number of relativistic degrees of freedom at the time of decay.
Since the free streaming length must be smaller than 100 kpc, one obtains a lower limit on the gravitino mass,
\be
m_{\g}\gsim 5.44\times\dfrac{10^{-\frac{4}{3}}}{\lambda^2}\left(\dfrac{g_*}{100}\right)^{\frac{1}{3}}\left(\dfrac{m}{10^{14} {\rm GeV}}\right)\G\,.
\label{fslmass}
\ee
Clearly, depending on the coupling $\lambda$, the above equation determines a set of lower bounds on the gravitino mass. For $\lambda=10^{-2}$, the former constraint $m_{\g}\gsim 5\M$ still holds, whereas for $\lambda=10^{-1}$ we find a slightly different mass range. In this case, the gravitinos produced by the $S$ decay are CDM only if
\be
\lambda\approx 10^{-1}\Rightarrow 22\k\lsim\mg\lsim 10\G.
\label{lowergm}
\ee
By taking a look at Eq.(\ref{fslmass}), we see that this is not the most stringent mass bound. The lower limit $\mg\approx 22\k$ is indeed a function of $m$, which in principle can assume higher values, as soon as Eq.(\ref{upperm}) is satisfied.

\section{Conclusions and outlook}

			In this chapter we have considered the dynamics of the SUSY breaking scalar $S$ in a GMSB scenario with metastable vacua and the production of gravitino Dark Matter through its decay. Our results for the cosmological evolution of the scalar field and the resulting gravitino density are significantly different from previous investigations.
						
			We have shown that since the Universe is expanding, the minimum of the potential $S_0$ is time-dependent. The $S$ field tracks the minimum, which generates a displacement $\delta S$. Once the minimum becomes H independent, $S$ begins coherent oscillations about the minimum with initial amplitude determined by the displacement.  This produces a very different $S$ oscillation amplitude and so gravitino density as compared with previous analyses \cite{ET}. By considering a generic potential $V(S)$, we have shown that there are two possible values of $\delta S$, depending on whether the quadratic or the linear term first becomes H-independent.
The resulting gravitino density can be highly suppressed or enhanced as compared with the previous estimate of \cite{ET}. A striking feature of the gravitino density is its extreme sensitivity to the parameters of the model, with the reheating temperature having a $\lambda^{10}$ or $\lambda^{14}$ dependence on the superpotential coupling of the \OR sector.  It is therefore easy to account for gravitino Dark Matter with an arbitrarily low reheating temperature. This could be significant for the cosmology of GMSB models, since a low $T_R$ is consistent with several cosmological scenarios \cite{Lowreh}.

	One might also be interested in completing the model with a source of baryogenesis. A natural possibility would be Affleck-Dine baryogenesis. Indeed, for $\mg \sim 1 \G$, Affleck-Dine baryogenesis combined with Q-ball decay in GMSB could account for both gravitino Dark Matter and the baryon asymmetry \cite{shoe}, but for smaller $\mg$ it would provide only the baryon asymmetry. As an additional remark, we note that the \OR GMSB sector is a well-defined perturbative model. Therefore the decay of the SUSY-breaking scalar field is addressed without cutoff, nor with any stringent constraints on the mass scales nor on the coupling constants.

\chapter{Conclusions}

	In this thesis, gravitino production via different mechanisms has been discussed into details.
We have first introduced the theories of inflation and of supergravity, which provide with the background for the research papers here included. These are studied in Chapters \ref{chapt:WW} and \ref{chapt:Infl}.

	In the article \cite{WW}, the scattering of gauge bosons into gravitinos is investigated both in the gauge and mass eigenstates. If the W is massive, a term which grows with energy is produced in the cross section. This can accordingly violate the unitarity of the theory at large centre of mass scales. We show that this happens because the longitudinal degrees of freedom of the off-shell W, which is exchanged in the annihilation channel, are identically cancelled by the supergravity vertex. As proven in Section \ref{sect:theo}, this implies that the on-shell W bosons become strongly interacting at the supersymmetry breaking scale. At this point, one could wonder whether missing diagrams might cancel the unexpected terms. We show however that inclusion of processes with the neutral Higgs doublet of the MSSM does not change the result.

	There are basically two perspectives from the above. Supergravity is an effective theory, which is the limit of some unknown unified theory and is valid only under the SUSY breaking scale. Accordingly, our result can be used in phenomenology. Corrections to previous results (see \cite{Pradler}) can be thus calculated. For example, the Boltzmann collision factor of the gravitinos which are produced in the thermal bath might be significantly modified. 

	However, one might expect that the divergences would be cancelled by some yet unknown mechanism. For instance, additional diagrams might come from interaction terms in the Lagrangian (\ref{sugral}) which are usually not included. Moreover, the contribution of the hidden sector should be taken into account as well, since here both the EW gauge symmetry and supersymmetry are broken. This is a clear difference with respect to the Standard Model, where similar divergences in the scattering $W^+W^-\rightarrow W^+W^-$ are cured uniquely by diagrams with the Higgs.

	In the paper \cite{Infl} we instead discuss gravitino production from the decay of a scalar field $S$. At the end of inflation, $S$ rolls down its potential and, when the mass $m_S$ becomes comparable to the Hubble rate, it starts coherent oscillations about the minimum of the potential. This mechanism generates decay into gravitinos.
	We first consider the dynamics of the scalar field. Since the Universe is expanding, the minimum of the potential is time-dependent. Accordingly, $S$ tracks the minimum and a displacement $\delta S$ from the minimum is generated. This produces a very different oscillation amplitude with respect to previous analyses. We calculate the related corrections in a general setting, with a generic quadratic potential $V(S)$.

	$S$ is then embedded in a model of gauge mediation with metastable vacuum, that is provided with an explicit hidden sector of the \OR type. The scalar field breaks supersymmetry via a linear term in the superpotential, and the \K potential for $S$, $K(S)$, has a negative non-minimal term. We calculate the radiative and supergravity corrections to the scalar potential $V(S)$, which come from one-loop diagrams of the scalars and fermions of the hidden and messenger sector. We show that they i) lift the flat direction at the tree level by generating a mass term for $S$; ii) provide with an explicit form of the arbitrary cutoff $\Lambda$ of the theory in terms of the parameters of the hidden sector; iii) induce a linear term in $K(S)$ which makes the field deviate from the origin.

	By using the potential $V(S)$ thus obtained, we calculate the yield variable of $S$ and of the gravitino, by taking into account the correction factors previously obtained. By demanding that all the gravitinos thus produced constitute the main contribution to the amount of the observed Cold Dark Matter density, we find constraints on the reheating temperature $T_R$. A striking feature of the gravitino density and of the bounds (\ref{reheating21}) and (\ref{reheating12}) is their extreme sensitivity to the parameters of the hidden sector, especially to the coupling $\lambda$ of the \OR model. This means that it is easy to account for gravitino dark matter with an arbitrarily low reheating temperature.

	These results suggest some directions of future research. In particular, the model might be completed with a source of baryogenesis, for example the Affleck-Dine mechanism. Also the impact of other sources of gravitino dark matter, for instance the decay of MSSM flat directions, should be taken into account. Further constraints would then be imposed on a model that is a well-defined perturbative framework, by virtue of the explicit \OR hidden sector.

	All the above perspectives are object of actual investigations.

\appendix

\chapter{Notations and Feynman rules}

Here we provide with our notations and conventions, and with the Feynman rules of the MSSM and of supergravity which are used in Chapter \ref{chapt:WW}.

\section{Notations and conventions}\label{app:conv}

Regarding our conventions, we follow basically the reviews by Haber and Kane \cite{HK} and Martin \cite{Martin}.
Throughout this thesis, the main subject is particle physics, even though we deal with cosmology. Therefore, it is a good approximation to adopt the flat space-time Lorentz metric
\be
\eta_{\mu\nu}=\eta^{\mu\nu}=diag(1,-1,-1,-1)\,,
\ee
with the standard convention of particle physics for the signature. This implies $p^2=m^2$ for a particle with four-momentum $p$ and mass $m$. The spinors in both the Standard Model and in supersymmetric Lagrangians are usually written first in two-component notation, since each Dirac fermion has left- and right-handed parts with completely different electroweak interactions. This way the degrees of freedom are treated differently already from the beginning. Accordingly, we introduce the Pauli matrices
\bea
&&\sigma^0 = \left(
\begin{tabular}{ll}
1&0 \\ 0 & 1
\end{tabular} \right)\,,\qquad
\sigma^1 = \left(
\begin{tabular}{ll}
 0 & 1 \\ 1&0
\end{tabular} \right)\,, 
\nonumber\\
&&\sigma^2 = \left(
\begin{tabular}{ll}
0& $-i$ \\ $i$ & 0
\end{tabular} \right)\,,\qquad
\sigma^3=\left(
\begin{tabular}{ll}
1&0 \\ 0 & -1
\end{tabular} \right).
\eea
A two-component (left-handed) complex Weyl spinor $\xi_\alpha$ transforms under actions $M \in \mathrm{SL(2,\mathbb{C})}$, that is under the matrix representation of the Lorentz group $\mathrm{SO(1,3)}$. The right-handed spinor $\bar{\xi}_{\dot{\alpha}}$ transforms under $M^*$, and similarly for $\xi^\alpha$ and $\bar{\xi}^{\dot{\alpha}}$ one has $M^{-1}$ and $(M^{-1})^*$. The spinor indices are raised and lowered according to the antisymmetric tensor $\epsilon^{\alpha\beta}=-\epsilon^{\beta\alpha}$:
\be
\xi_\alpha = \epsilon_{\alpha\beta}
\xi^\beta,\qquad\qquad
\xi^\alpha = \epsilon^{\alpha\beta}
\xi_\beta,\qquad\qquad
\eta^\dagger_{\dot{\alpha}} = \epsilon_{\dot{\alpha}\dot{\beta}}
\eta^{\dagger\dot{\beta}},\qquad\qquad
\eta^{\dagger\dot{\alpha}} = \epsilon^{\dot{\alpha}\dot{\beta}}
\eta^\dagger_{\dot{\beta}}\,,
\ee
where $\eta_\alpha$ is another Weyl spinor, since
\be
\epsilon^{12}=-\epsilon^{21}=\epsilon_{21}=-\epsilon_{12}=1\,; \qquad \epsilon_{11}=\epsilon_{22}=\epsilon^{11}=\epsilon^{22}=0\,.
\ee
The fact that the Weyl spinors are anticommuting and the antisymmetric nature of $\epsilon^{\alpha\beta}$, compensate each other in the following remarkable result:
\be
\xi\eta = \xi^\alpha \epsilon_{\alpha\beta}
\eta^\beta = -\eta^\beta \epsilon_{\alpha\beta} \xi^\alpha =
\eta^\beta \epsilon_{\beta\alpha} \xi^\alpha =
\eta\xi\,.
\ee
The Dirac equation in two-component notation can be written as
\be
(\bar{\sigma}_\mu p^\mu)^{\dot{\alpha}\beta}\xi_\beta=m\bar{\eta}^{\dot{\alpha}}\,,\qquad
(\sigma_\mu p^\mu)_{\alpha\dot{\beta}}\bar{\eta}^{\dot{\beta}}=m\xi_\alpha\,,
\label{dirac}
\ee
where
\be
\sigma^\mu=(1,\mathbf{\sigma})\,,\qquad \bar{\sigma}^\mu=(1,-\mathbf{\sigma}).
\ee
We can now rewrite Eq.(\ref{dirac}) in four-component notation, namely
\be
(\gamma^\mu p_\mu-m)\psi=0\,,
\label{dirac4}
\ee
by defining the spinor
\be
\psi=\left(
\begin{tabular}{c}
$\xi_\alpha$ \\ $\bar{\eta}^{\dot{\alpha}}$ 
\end{tabular} \right)\,.
\ee
The Dirac (or gamma) matrices in Eq.(\ref{dirac4}) are written in the Weyl basis as
\be
\gamma^\mu=\left(
\begin{tabular}{cc}
0&$\sigma^\mu$ \\ $\bar{\sigma}^\mu$& 0 
\end{tabular} \right)\,,\qquad \gamma_5=i\gamma^0\gamma^1\gamma^2\gamma^3=\left(
\begin{tabular}{cc}
-1&0 \\ 0 & 1 
\end{tabular} \right)\,.
\ee
The above is called the chiral representation of the gamma matrices, and it is very useful in this context, where we want to separate the left- and right- polarizations of the Dirac spinor. To this aim, let us define the left- and right- handed projection operators:
\be
P_{L,R}=\dfrac{1}{2}\left(\mathbf{1}\mp\gamma^5\right)\,,
\ee
so that is it possible to write
\be
\psi=\left(
\begin{tabular}{c}
$\psi_L$ \\ $\psi_R$ 
\end{tabular} \right)\,,
\ee
where we have defined $\psi_{L,R}\equiv P_{L,R}\psi$.

\section{Feynman rules for the MSSM and SUGRA}\label{app:feyn}

The relevant Feynman rules which are used in the scattering calculations of paper \cite{WW} are here summarised. The propagators in the $R_{\xi}$ gauge and the vertices of the Minimal Supersymmetric Standard Model and of Supergravity follow Refs. \cite{M}, \cite{HK} and \cite{P}.

In the following, $\Mpl=(8\pi G_N)^{-1/2}= 2.43\times 10^{18}$ GeV is the reduced Planck mass, $g$ and $f^{abc}$ are, respectively, the coupling constant and the structure constants of a generic gauge group.
We point out that our expressions differ from those of Ref.\cite{HK} by an imaginary factor $i$, since the squared amplitude must be real. This is justified by the different normalization of the Lagrangians of Supergravity and of the MSSM.
All the momenta are considered to run into the vertex and the fermion flow is denoted by a thin line with an arrow.

\subsection{Propagators and interaction vertices}
\begin{figure}[hp]

 \begin{picture}(333,70) (0,-71)
\SetWidth{0.5}
    \SetColor{Black}
    \Vertex(85,-16){1.41}
    \Text(340,-16)[]{\large{\Black{$;$}}}
    \Text(319,-9)[]{\large{\Black{$\delta^{ab}$}}}
    \Text(240,-13)[]{\normalsize{\Black{$\left[ \eta_{\mu\nu} +(\xi-1)\dfrac{k_\mu k_\nu}{k^2-\xi m^2_A}\right]$}}}
    \Photon(20,-16)(85,-16){4}{9}
    \Vertex(20,-16){1.41}
    \Text(18,-27)[]{\normalsize{\Black{$a,\mu$}}}
    \Text(90,-27)[]{\normalsize{\Black{$b,\nu$}}}
    \Text(146,-17)[]{\Large{\Black{$\frac{-i}{k^2-m_A^2}$}}}
    \Text(107,-15)[]{\large{\Black{$=$}}}
    \Text(146,-59)[]{\Large{\Black{$\frac{-i}{k^2-m_{\Phi}^2}$}}}
    \Text(105,-56)[]{\large{\Black{$=$}}}
    \Vertex(18,-58){1.41}
    \Vertex(83,-58){1.41}
    \Text(83,-67)[]{\normalsize{\Black{$l$}}}
    \Text(15,-67)[]{\normalsize{\Black{$j$}}}
    \DashArrowLine(18,-58)(83,-58){5}
    \Text(183,-53)[]{\large{\Black{$\delta^{jl}$}}}
   
                  \end{picture}
                  
\caption{The propagators of a vector boson with mass $m_A$ and of a scalar with mass $m_{\Phi}$.}
\end{figure}

\begin{figure}[h]
\begin{center}
\resizebox{\textwidth}{!}{
\begin{picture}(394,30) (40,-78)
\SetWidth{0.3}
    \SetColor{Black}
    \Photon(40,-60)(105,-60){4}{9}
    \Vertex(105,-60){1.41}
    \Vertex(40,-60){1.41}
    \Text(42,-70)[]{\normalsize{\Black{$a,\mu$}}}
    \Text(112,-70)[]{\normalsize{\Black{$b,\nu$}}}
    \Text(176,-62)[]{\normalsize{\Black{$\dfrac{(\mp\gamma^{\rho}k_{\rho}+m_M)}{k^2-m_M^2}$}}}
    \Line(40,-60)(105,-60)
    \SetWidth{0.0}
    \ArrowLine(42,-53)(105,-53)
    \ArrowLine(105,-48)(42,-48)
    \Text(245,-61)[]{\Large{\Black{$;$}}}
    \Text(226,-57)[]{\large{\Black{$\delta^{ab}$}}}
    \Text(127,-59)[]{\large{\Black{$=i$}}}
    \Text(403,-62)[]{\normalsize{\Black{$\dfrac{(\mp\gamma^{\rho}k_{\rho}+m_D)}{k^2-m_D^2}$}}}
    \SetWidth{0.3}
    \Vertex(267,-61){1.41}
    \Vertex(332,-61){1.41}
    \Text(337,-70)[]{\normalsize{\Black{$j$}}}
    \SetWidth{0.0}
    \ArrowLine(268,-55)(331,-55)
    \Text(265,-70)[]{\normalsize{\Black{$i$}}}
    \SetWidth{0.3}
    \ArrowLine(267,-61)(332,-61)
    \SetWidth{0.0}
    \ArrowLine(331,-51)(268,-51)
    \Text(450,-57)[]{\large{\Black{$\delta^{ij}$}}}
    \Text(355,-59)[]{\large{\Black{$=i$}}}
   
                  \end{picture}
}
\end{center}
\caption{The propagators of a Majorana fermion with mass $m_{M}$ and of a Dirac fermion with mass $m_{D}$.}
\end{figure}


\begin{figure}[ht]
\begin{center}
\resizebox{14cm}{!}{

\begin{picture}(520,96) (26,10)
    \SetWidth{0.5}
    \SetColor{Black}
    \Text(71,90)[]{\large{\Black{$A^{a}_{\alpha}(k_1)$}}}
    \Text(71,8)[]{\large{\Black{$A^{b}_{\beta}(k_2)$}}}
    \Text(126,64)[]{\large{\Black{$A^{c}_{\sigma}(k_3)$}}}
    \Text(362,49)[]{\Large{\Black{$=-gf^{abc}[(k_1-k_2)^{\sigma}\eta^{\alpha\beta}+(k_2-k_3)^{\alpha}\eta^{\beta\sigma}+(k_3-k_1)^{\beta}\eta^{\sigma\alpha}]$}}}
    \SetWidth{0.5}
    \Vertex(105,49){1.41}
    \Photon(105,49)(150,49){4}{6}
    \Photon(105,49)(77,80){4}{5}
    \Photon(104,49)(77,18){4}{5}
  
  \end{picture}
  
  }
  \end{center}
\caption{Coupling of 3 gauge bosons.}
\end{figure}
\begin{figure}[ht]
\begin{center}
\resizebox{\textwidth}{!}{
   \begin{picture}(530,90) (-35,-30)
 \SetWidth{0.5}
    \SetColor{Black}
    \Text(150,9)[]{\Large{\Black{$=gf^{abc}\gamma_{\beta}$}}}
    \Text(223,9)[]{\Large{\Black{$;$}}}
    \Text(406,9)[]{\Large{\Black{$=-gf^{abc}\gamma_{\alpha}$}}}
    \SetWidth{0.5}
    \Photon(56,9)(26,-21){4}{5}
    \Photon(26,39)(56,9){4}{5}
    \Photon(56,9)(103,9){4}{6}
    \SetWidth{0.0}
    \ArrowArc(16.07,7.5)(21.98,-68.86,68.86)
    \SetWidth{0.5}
    \Vertex(56,9){1.41}
    \Text(21,48)[]{\large{\Black{$\tilde{\chi}^a$}}}
    \Text(21,-30)[]{\large{\Black{$\tilde{\chi}^c$}}}
    \Text(80,22)[]{\large{\Black{$A^b_{\beta}$}}}
    \Line(56,9)(26,-21)
    \Line(56,9)(26,39)
    \Photon(305,8)(275,-22){4}{5}
    \Photon(275,38)(305,8){4}{5}
    \Vertex(305,8){1.41}
    \Line(275,38)(305,8)
    \Photon(305,8)(352,8){4}{6}
    \SetWidth{0.0}
    \ArrowArc(265.07,6.5)(21.98,-68.86,68.86)
    \Text(330,20)[]{\large{\Black{$A^a_{\alpha}$}}}
    \Text(270,48)[]{\large{\Black{$\tilde{\chi}^b$}}}
    \Text(270,-32)[]{\large{\Black{$\tilde{\chi}^c$}}}
    \SetWidth{0.5}
    \Line(305,8)(275,-22)
   \end{picture}
                    }   
\end{center}                                        
\caption{Gauge boson-gaugino-gaugino.}
\end{figure}
\begin{figure}[ht]
\resizebox{\textwidth}{!}{
   \begin{picture}(550,116) (15,-20)
    \SetWidth{0.5}
    \SetColor{Black}
    \Text(192,33)[]{\Large{\Black{$=gm_W\cos{(\beta-\alpha)}\eta_{\alpha\beta}$}}}
    \Text(283,33)[]{\Large{\Black{$;$}}}
    \SetWidth{0.5}
    \Photon(56,33)(26,63){4}{5}
    \Photon(56,33)(26,3){4}{5}
    \DashLine(56,33)(101,33){5}
    \Vertex(56,33){1.41}
    \Text(21,75)[]{\large{\Black{$W^+_{\alpha}$}}}
    \Text(80,43)[]{\large{\Black{$H$}}}
    \Text(21,-7)[]{\large{\Black{$W^-_{\beta}$}}}
    \Text(503,34)[]{\Large{\Black{$=gm_W\sin{(\beta-\alpha)}\eta_{\alpha\beta}$}}}
    \Vertex(367,33){1.41}
    \Photon(367,33)(337,63){4}{5}
    \Photon(367,33)(337,3){4}{5}
    \DashLine(367,33)(412,33){5}
    \Text(332,75)[]{\large{\Black{$W^+_{\alpha}$}}}
   \Text(391,43)[]{\large{\Black{$h$}}}
   \Text(332,-7)[]{\large{\Black{$W^-_{\beta}$}}}
      \end{picture}
                    }                       
\caption{Higgs-gauge bosons.}
\end{figure}
\begin{figure}[ht]
\resizebox{14cm}{!}{
   \begin{picture}(557,78) (14,-38)
\SetWidth{0.5}
    \SetColor{Black}
    \Text(187,11)[]{\Large{\Black{$=g\gamma_{\beta}[O^L_{ij}P_L+O^R_{ij}P_R]$}}}
    \Text(283,11)[]{\Large{\Black{$;$}}}
    \SetWidth{0.5}
    \Photon(26,41)(56,11){4}{5}
    \Line(26,41)(56,11)
    \Photon(56,11)(103,11){4}{6}
    \SetWidth{0.0}
    \ArrowArc(16.07,9.5)(21.98,-68.86,68.86)
    \SetWidth{0.5}
    \Vertex(56,11){1.41}
    \Text(19,52)[]{\large{\Black{$\tilde{\chi}^0_i$}}}
    \Text(19,-28)[]{\large{\Black{$\tilde{\chi}^+_j$}}}
    \Text(79,26)[]{\large{\Black{$W^-_{\beta}$}}}
    \ArrowLine(27,-21)(56,11)
    \Vertex(365,10){1.41}
    \Photon(335,40)(365,10){4}{5}
    \Text(506,11)[]{\Large{\Black{$=-g\gamma_{\alpha}[O^{R*}_{ij}P_L+O^{L*}_{ij}P_R]$}}}
    \Line(335,40)(365,10)
    \Photon(365,10)(412,10){4}{6}
    \SetWidth{0.0}
    \ArrowArc(325.07,8.5)(21.98,-68.86,68.86)
    \Text(389,24)[]{\large{\Black{$W^+_{\alpha}$}}}
    \Text(328,52)[]{\large{\Black{$\tilde{\chi}^0_i$}}}
    \Text(328,-30)[]{\large{\Black{$\tilde{\chi}^-_j$}}}
    \SetWidth{0.5}
    \ArrowLine(335,-21)(365,10)
              \end{picture}
                    }                       
\caption{Gauge boson-chargino-neutralino.}
\end{figure}

\clearpage

\begin{figure}[t]
\resizebox{\textwidth}{!}{
   \begin{picture}(570,97) (13,-19)
    \SetWidth{0.5}
    \SetColor{Black}
    \Text(222,32)[]{\large{\Black{$[ \gamma^{\rho}k_{\rho},\gamma^{\sigma}]\gamma^{\mu}\delta^{ab}$}}}
    \Text(310,31)[]{\Large{\Black{$;$}}}
    \SetWidth{0.5}
    \Photon(63,30)(35,-1){4}{6}
    \Vertex(63,30){1.41}
    \Photon(63,30)(108,30){4}{6}
    \Line(33.71,60.71)(63.71,30.71)\Line(32.29,59.29)(62.29,29.29)
    \Line(33,-3)(63,30)
    \Text(28,70)[]{\normalsize{\Black{$\psi_{\mu}$}}}
    \Text(28,-11)[]{\normalsize{\Black{$\tilde{\chi}^{a}$}}}
    \Text(85,42)[]{\normalsize{\Black{$A^{b}_{\sigma}(k)$}}}
    \SetWidth{0.0}
    \ArrowArc(18.93,29.5)(25.08,-63.8,63.8)
    \Text(149,30)[]{\Large{\Black{$=-\frac{i}{4M_P}$}}}
    \SetWidth{0.5}
    \Vertex(408,30){1.41}
    \Line(378,-3)(408,30)
    \Photon(408,30)(380,-1){4}{6}
    \Photon(408,30)(453,30){4}{6}
    \Line(378.71,60.71)(408.71,30.71)\Line(377.29,59.29)(407.29,29.29)
    \Text(373,70)[]{\normalsize{\Black{$\psi_{\mu}$}}}
    \Text(373,-11)[]{\normalsize{\Black{$\tilde{\chi}^{a}$}}}
    \Text(430,42)[]{\normalsize{\Black{$A^{b}_{\sigma}(k)$}}}
    \SetWidth{0.0}
    \ArrowArcn(363.93,29.5)(25.08,63.8,-63.8)
    \Text(494,30)[]{\Large{\Black{$=-\frac{i}{4M_P}$}}}
    \Text(568,32)[]{\large{\Black{$\gamma^{\mu}[ \gamma^{\rho}k_{\rho},\gamma^{\sigma}]\delta^{ab}$}}}
  \end{picture}
}
\caption{Gauge boson-gaugino-gravitino.}
\label{fig:gvg}
\end{figure}

\begin{figure}[ht]
\resizebox{13.7cm}{!}{
 \begin{picture}(501,98) (18,-18)
    \SetWidth{0.5}
    \SetColor{Black}
    \Text(267,33)[]{\LARGE{\Black{$;$}}}
    \SetWidth{0.5}
    \Vertex(69,31){1.41}
    \Photon(69,31)(100,-1){4}{5}
    \Text(108,71)[]{\normalsize{\Black{$\psi_{\mu}$}}}
    \Text(33,71)[]{\normalsize{\Black{$A^{a}_{\alpha}$}}}
    \Text(33,-10)[]{\normalsize{\Black{$A^{b}_{\beta}$}}}
    \Text(108,-10)[]{\normalsize{\Black{$\tilde{\chi}^c$}}}
    \Text(203,35)[]{\large{\Black{$[ \gamma^{\alpha},\gamma^{\beta}]\gamma^{\mu}$}}}
    \Photon(39,62)(69,31){4}{5}
    \Photon(71,31)(39,-1){4}{5}
    \Line(69,31)(101,-1)
    \Line(68.28,31.7)(98.28,62.7)\Line(69.72,30.3)(99.72,61.3)
    \SetWidth{0.0}
    \ArrowArcn(113.07,32.5)(25.08,243.8,116.2)
    \Text(142,33)[]{\Large{\Black{$=-\frac{gf^{abc}}{4M_{P}}$}}}
    \SetWidth{0.5}
    \Vertex(365,32){1.41}
    \Photon(365,32)(396,0){4}{5}
    \Text(404,72)[]{\normalsize{\Black{$\psi_{\mu}$}}}
    \Text(329,72)[]{\normalsize{\Black{$A^{a}_{\alpha}$}}}
    \Text(329,-9)[]{\normalsize{\Black{$A^{b}_{\beta}$}}}
    \Text(404,-9)[]{\normalsize{\Black{$\tilde{\chi}^c$}}}
    \Text(504,35)[]{\large{\Black{$\gamma^{\mu}[ \gamma^{\alpha},\gamma^{\beta}]$}}}
    \Photon(335,63)(365,32){4}{5}
    \Photon(367,32)(335,0){4}{5}
    \Line(365,32)(397,0)
    \Line(364.28,32.7)(394.28,63.7)\Line(365.72,31.3)(395.72,62.3)
    \SetWidth{0.0}
    \ArrowArc(409.07,33.5)(25.08,116.2,243.8)
    \Text(444,33)[]{\Large{\Black{$=-\frac{gf^{abc}}{4M_{P}}$}}}
  \end{picture}
}
\caption{Gauge boson-gauge boson-gaugino-gravitino.}
\label{fig:4part}
\end{figure}

\begin{figure}[ht]
\resizebox{\textwidth}{!}{
 \begin{picture}(490,96) (26,-10)
    \SetWidth{0.5}
    \SetColor{Black}
    \Text(51,49)[]{\normalsize{\Black{$\Phi_i(p)$}}}
    \Text(106,78)[]{\normalsize{\Black{$\psi_{\mu}$}}}
    \Text(106,-2)[]{\normalsize{\Black{$\lambda_j$}}}
    \Text(222,41)[]{\normalsize{\Black{$(\gamma^{\rho}p_{\rho}\gamma^{\mu}P_L)\delta_{ij}$}}}
    \SetWidth{0.5}
    \Vertex(73,39){1.41}
    \DashArrowLine(26,39)(73,39){3}
    \Line(72.27,39.68)(99.27,68.68)\Line(73.73,38.32)(100.73,67.32)
    \ArrowLine(73,39)(101,8)
    \SetWidth{0.0}
    \ArrowArcn(114.89,37)(25.89,242.66,117.34)
    \Text(279,39)[]{\Large{\Black{$;$}}}
    \Text(150,38)[]{\Large{\Black{$=-\frac{i}{\sqrt{2}M_P}$}}}
    \SetWidth{0.5}
    \Vertex(347,39){1.41}
    \Text(496,41)[]{\normalsize{\Black{$(P_L\gamma^{\mu}\gamma^{\rho}p_{\rho})\delta_{ij}$}}}
    \Text(381,78)[]{\normalsize{\Black{$\psi_{\mu}$}}}
    \Text(381,-2)[]{\normalsize{\Black{$\lambda_j$}}}
    \Text(324,49)[]{\normalsize{\Black{$\Phi_i(p)$}}}
    \DashArrowLine(300,39)(347,39){3}
    \Line(346.27,39.68)(373.27,68.68)\Line(347.73,38.32)(374.73,67.32)
    \ArrowLine(347,39)(375,8)
    \SetWidth{0.0}
    \ArrowArc(388.89,37)(25.89,117.34,242.66)
    \Text(424,38)[]{\Large{\Black{$=-\frac{i}{\sqrt{2}M_P}$}}}
    \end{picture}
                        }
\caption{Scalar-Weyl fermion-gravitino.}
\label{fig:hgf}
\end{figure}

\end{document}